\documentclass[english,man]{apa}
\usepackage[T1]{fontenc}
\usepackage[latin9]{inputenc}
\usepackage{color}
\usepackage{array}
\usepackage{booktabs}
\usepackage{multirow}
\usepackage{amsbsy}
\usepackage[authoryear]{natbib}

\makeatletter

\providecommand{\tabularnewline}{\\}

\helvetica
\author{Author} 
\affiliation{Affiliation} 

\usepackage{kerkis}
\usepackage{apacite}
\usepackage{geometry}
\usepackage{multirow}
\usepackage{rotating}
\usepackage{subfig}
\usepackage{bm}
\usepackage{rotating}
\usepackage{times}
\usepackage{longtable}
\let\myEnd\endlongtable\renewcommand{\endlongtable}{\myEnd\addtocounter{table}{-1}}

\makeatother

\usepackage{babel}
\begin{document}

\title{A Bootstrap Based Between-Study Heterogeneity Test in Meta-Analysis\thispagestyle{empty}}

\threeauthors{Han Du}{Ge Jiang}{Zijun Ke}

\threeaffiliations{Department of Psychology, University of California, Los Angeles}{Department
of Educational Psychology, University of Illinois at Urbana-Champaign}{Department
of Psychology, Sun Yat-sen University \\ \vspace{1in}Correspondence
should be addressed to Han Du, Franz Hall, 502 Portola Plaza, Los
Angeles, CA 90095. Email: hdu@psych.ucla.ed}

\shorttitle{Between-study heterogeneity test}

\abstract{\textcolor{black}{Meta-analysis combines pertinent information from
existing studies to provide an overall estimate of population parameters/effect
sizes, as well as to quantify and explain the differences between
studies. However, testing the between-study heterogeneity is one of
the most troublesome topics in meta-analysis research. Additionally,
no methods have been proposed to test whether the size of the heterogeneity
is larger than a specific level. The existing methods, such as the
Q test and likelihood ratio (LR) tests, are criticized for their failure
to control the Type I error rate and/or failure to attain enough statistical
power. Although better reference distribution approximations have
been proposed in the literature, the expression is complicated and
the application is limited. In this article, we propose bootstrap
based heterogeneity tests combining the restricted maximum likelihood
(REML) ratio test or Q test with bootstrap procedures, denoted as
B-REML-LRT and B-Q respectively. Simulation studies were conducted
to examine and compare the performance of the proposed methods with
the regular LR tests, the regular Q test, and the improved Q test
in both the random-effects meta-analysis and mixed-effects meta-analysis.
Based on the results of Type I error r}ates and statistical power,
B-Q is recommended. An R package $\mathtt{boot.heterogeneity}$ is
provided to facilitate the implementation of the proposed method.\\ \vspace{0.3in}Keywords:
Meta-analysis, Between-study heterogeneity, bootstrap method}
\maketitle

\section{\textcolor{black}{1. Introduction}}

\textcolor{black}{Meta-analysis is a popular statistical technique
that combines the results across multiple studies \citep{glass1984meta,hedges1985statistical,hunter1982meta}.
Besides providing a pooled estimate of the unknown population effect
sizes, modern meta-analysis methods also have the capacity to quantify
and explain the differences between studies. Fixed-effects meta-analysis
assumes that all studies have the same true effect, whereas random-effects
and mixed-effects models assume that there is variability (i.e., between-study
heterogeneity) across study-specific population/true effects. Researchers
have suggested that one should decide whether to include the between-study
heterogeneity based on the assumption and intended generalization
of the obtained results \citep{borenstein2009,coughlin2015meta}.
More specifically, if one would like to generalize the conclusion
from a specific set of studies, the model with the between-study heterogeneity
is preferred. }

\textcolor{black}{Although we cannot rely on the heterogeneity test
to select the fixed-effects model versus random-effects model, testing
between-study heterogeneity is the research of interest in some studies
and widely used in practice. There are two types of between-study
heterogeneity tests. In the first type of test, the null hypothesis
is ``true effects are the same (no heterogeneity; $\tau^{2}=0$)''
and the alternative hypothesis is ``true effects vary ($\tau^{2}\neq0$)''.
We refer to this type of test as the }\textit{\textcolor{black}{heterogeneity
test}}\textcolor{black}{. In the second type of test, the null hypothesis
that ``the size of heterogeneity is equal to a specific level ($\tau^{2}=\lambda$)''
and the alternative hypothesis is that ``the size heterogeneity is
larger than a specific level ($\tau^{2}>\lambda$)''. We refer to
this type of test as the }\textit{\textcolor{black}{heterogeneity
magnitude test}}\textcolor{black}{. To the best of our knowledge,
no existing method can handle the latter type of hypothesis testing.
There are methods for the former type of hypothesis testing, but those
heterogeneity tests are criticized by their failure to control the
Type I error rate and low power. Hence, the goal of the current student
is to propose a more reliable and powerful test for both hypotheses.}

\textcolor{black}{The importance of the two types of between-study
heterogeneity tests is three-fold. First, ``in a meta-analysis, it
is usual to conduct a homogeneity test to determine whether the effects
measured by the included studies are sufficiently similar to justify
their combination'' (\citealp{kulinskaya2011moments}, p. 254). When
the between-study heterogeneity is large, researchers may be more
interested in the effect size of each specific study (e.g., \citealp{thalib2011combining}).
\citet{field2003problems} illustrated one example. The effect sizes
of cognitive behavior therapy for social phobia in the United Kingdom,
United States, and Netherlands are found to be different, but the
overall effect size across countries may be 0. In this case, interpreting
the overall effect size only may not provide any practical meaning,
whereas estimating the between-study heterogeneity is important and
\textquotedblleft it is sometimes useful to test the hypothesis that
the effect size variance is zero in addition to estimating this quantity\textquotedblright{}
(\citealp{hedges1985statistical}, p.197). Furthermore, in the heterogeneity
magnitude test, with a significant test result, we can claim that
the heterogeneity is larger than a specific level (e.g., a scientifically
or clinically meaningful difference). Second, one may be interested
in what factors could explain the between-study heterogeneity. To
address this research question, relevant study-level covariates (also
referred to as moderators) are included in a mixed-effects model (meta-regression)
to explain whether there are systematic reasons why effect sizes are
different \citep{field2003problems}. For example, studies on gender
differences in marital satisfaction may have discrepancy due to cohort
effects. Marriage may look very different today than it did 10 or
20 years ago. Thus, the year in which each study was initiated can
be considered as a moderator to explain why effect sizes vary across
studies. We don't anticipate that the covariates can explain the entire
residual heterogeneity, but ``the size of this residual or unexplained
heterogeneity is one consideration in evaluating the adequacy of the
moderation model'' (\citealp{card2015applied}, p. 218). The test
of (unexplained) heterogeneity in meta-regression allows for the evaluation
of more complex moderation hypothesis. \textquotedblleft For example,
one can test interactive combinations of moderators by creating product
terms. Similarly one can evaluate nonlinear moderation by the creation
of power polynomial terms\textquotedblright{} (\citealp{card2015applied},
p. 218). \citet{borenstein2009} called this test the }\textit{\textcolor{black}{goodness
of fit test}}\textcolor{black}{{} (p. 198). Third, heterogeneity tests
are particularly important in replication research. When one wants
to know whether the replication is exact (studies have identical effect
sizes), the heterogeneity test can be used \citep{hedges2018statistical}.
However, it is usually too strict to require exact replications in
scientific practice, and therefore we can test whether the heterogeneity
is negligible (approximate replication; \citealp{hedges2018statistical,schauer2020assessing}).
\textquotedblleft Small differences in the magnitude of effects may
not lead to different interpretations of a finding\textquotedblright ,
so that the replications can be treated ``almost the same'' (\citealp{hedges2018statistical},
p. 559). In this case, the heterogeneity magnitude test can be used. }

\textcolor{black}{Testing between-study heterogeneity is widely used
in real data analysis. We reviewed studies published in the flagship
journal }\textit{\textcolor{black}{Psychological Bulletin}}\textcolor{black}{.
Specifically, we searched for articles through PsycINFO using the
keywords \textquotedblleft meta-regression\textquotedblright , \textquotedblleft mixed-effects
model\textquotedblright , and \textquotedblleft meta-analysis\textquotedblright{}
from January 1, 2014 to March 15, 2019. The initial research returned
160 articles, 34 of which were identified as reply articles, comment
articles, or erratum announcements. Those 34 articles were set aside
for reference, bringing down the total number of articles to 126.
Out of the 126 articles, no article only used fixed-effect model,
3 articles (2.4\%) used both random-effects and fixed-effects, and
the remaining 123 articles (97.6\%) all chose the random-effects model
or mixed-effects models. 77 articles out of the 126 articles (61.1\%)
reported the results of testing between-study heterogeneity (with
no decline trend from 2014 to 2019), suggesting that the tests of
between-study heterogeneity are widely used in real meta-analyses.
Among the 77 articles, 75 used the Q test \citep{hedges1985statistical,hedges1983random},
one used a likelihood ratio test, and one used a Wald test.}

\textcolor{black}{Despite the theoretical and practical significance
of studying between-study heterogeneity, addressing the between-study
heterogeneity is one of the most troublesome topics in meta-analysis
research. There are three limitations. First, the existing methods,
such as the Q test \citep{hedges1985statistical,hedges1983random}
and likelihood ratio test \citep{viechtbauer2007hypothesis}, are
limited in testing for the homogeneity of effect sizes ($H_{0}:\tau^{2}=0$)
and they cannot test whether the heterogeneity is larger than a specific
level (i.e., heterogeneity magnitude test). Second, the existing methods
are criticized for their failure to control the Type I error rate
and/or failure to attain enough statistical power (e.g., \citealp{borenstein2009fixed,huedo2006assessing}).
The main reason is that the reference distributions of the Q test
and likelihood ratio test do not have exact analytic expression, and
the asymptotic approximation requires a large sample size within each
study. Although better reference distribution approximations have
been proposed in the literature, the expression is ``sufficiently
complicated'' (\citealp{hoaglin2016misunderstandings}, p. 488).
Third, few research has studied the performance of heterogeneity test
in the mixed-effect meta-analysis. To overcome these limitations,
we propose a bootstrap based heterogeneity test combining the likelihood
ratio tests and Q test with bootstrap simulation. That is, we propose
to obtain the critical value of the likelihood ratio test or Q test
using a bootstrap simulation to construct an empirical reference distribution. }

\textcolor{black}{The outline of this paper is as follows: in Section
2, an overview of the meta-analysis models and effect sizes is provided.
Section 3 presents several existing methods that test the between-study
heterogeneity, some of which motivated the current study. In Section
4, we present the proposed bootstrap based heterogeneity test and
a corresponding R package. In Section 5, the Type I error rate and
statistical power of the bootstrap based heterogeneity test with different
types of effect sizes are thoroughly examined via simulations. In
Section 6, real data examples are provided to illustrate the bootstrap
based heterogeneity test. We end the paper with some concluding remarks
in Section 7.}

\section{\textcolor{black}{2. Meta-Analysis Models and Effect Sizes}}

\textcolor{black}{Suppose $K$ unbiased effect size estimates $x_{j}$
($j=1,...,K$) are included in a meta-analysis. A typical random-effects
meta-analysis model (e.g., \citealp{hedges1985statistical}) is given
by }

\textcolor{black}{
\begin{equation}
\begin{array}{c}
x_{j}=\delta_{j}+e_{j},\:e_{j}\sim N\left(0,\:\sigma_{j}^{2}\right)\\
\delta_{j}=\mu_{\delta}+u_{j},\;u_{j}\sim N(0,\:\tau^{2})
\end{array}\label{eq:random meta analysis}
\end{equation}
where $e_{j}$ is the deviation of the observed study effect size
$x_{j}$ from the true/population study effect size $\delta_{j}$
and its variance $\sigma_{j}^{2}$ represents within-study sampling
variability of study $j$, $u_{j}$ is the deviation of the true study
effect size $\delta_{j}$ from the true/population overall effect
size $\mu_{\delta}$, and $\tau^{2}$ represents between-study heterogeneity
($\tau^{2}=0$ in a fixed-effects meta-analysis model). The calculation
of $\sigma_{j}^{2}$ varies depending on the type of effect size.
If we incorporate study-level covariates (e.g., sampling and design
characteristics), the between-study discrepancies could be explained
by the covariates to some degree. This model is referred to as a mixed-effects
model or a meta-regression model \citep{card2015applied,hedges1998fixed}.
$\tau^{2}$ in the mixed-effects meta-analysis is the between-study
heterogeneity that cannot be explained by the study-level covariates.
When all the between-study discrepancies can be explained by the covariates
and hence $\tau^{2}=0$, the model becomes a fixed-effects meta-regression.
Besides, fixed- and random-effects meta-analysis model and fixed-
and random-effects meta-regression, \citet{stanley2015neither} and
\citet{stanley2017neither} proposed an unrestricted weighted least
squares (WLS). The WLS model is neither the fixed-effects model nor
the random-effects model. It does not estimate $\tau^{2}$ and assume
the study variance $v_{j}$ can be estimated up to some unknown multiplicative
constant $\phi$ so that $v_{j}=\phi\sigma_{j}^{2}$. \citet{stanley2015neither}
and \citet{stanley2017neither} found that the WLS model outperformed
the conventional random-effects meta-analysis when there is publication
bias and outperformed the fixed-effects model when there is heterogeneity.}

\textcolor{black}{In the current article, we focus on three types
of effect sizes: standardized mean difference, Pearson's correlation,
and odds ratio. When the effects are standardized mean differences,
the effect size estimate is $g_{j}=\frac{\overline{y}_{1j}-\overline{y}_{2j}}{s_{j}}$,
where $\overline{y}_{1j}$ and $\overline{y}_{2j}$ are the sample
means of the two groups in study $j$, and $s_{j}=\sqrt{\frac{(n_{1j}-1)s_{1j}^{2}+(n_{2j}-1)s_{2j}^{2}}{n_{1j}+n_{2j}-2}}$
where $s_{1j}^{2}$ and $s_{2j}^{2}$ are the sample variances of
the two groups of study $j$ and $n_{1j}$ and $n_{2j}$ are the sample
sizes of the two groups. However, $g_{j}$ is biased especially when
the per-study sample size is small \citep{hedges1981distribution,hedges1985statistical}.
The unbiased effect size estimate is $d_{j}=(1-\frac{3}{4(n_{1j}+n_{2j})-9})g_{j}$
\citep{hedges1981distribution}.}\footnote{\textcolor{black}{We used the notations in the original work of \citet{hedges1981distribution}
and \citet{hedges1985statistical}. The notations vary across different
researchers. For example, \citet{borenstein2009} called the biased
estimate the Cohen's d and unbiased estimate the Hedge's g.}}\textcolor{black}{{} Using this effect size, $d_{j}$ asymptotically
follows $N\left(\delta_{j},\frac{n_{1j}+n_{2j}}{n_{1j}n_{2j}}+\frac{\delta_{j}^{2}}{2(n_{1j}+n_{2j})}\right)$
where $\delta_{j}$ is the study-specific true effect size (\citealp{hedges1981distribution};
\citealp{hedges1985statistical}). In practice, we usually use $d_{j}$
to replace $\delta_{j}$ to calculate the within-study sampling variance.}

\textcolor{black}{When the effects are Pearson correlations, we first
transfer correlation in each study $r_{j}$ to a Fisher's $z$ score
$z_{r,j}$ by $z_{r,j}=\frac{1}{2}ln(\frac{1+r_{j}}{1-r_{j}})$ \citep{hedges1985statistical}.
$z_{r,j}$ asymptotically follows $N\left(\delta_{j},\frac{1}{n_{j}-3}\right)$
where $n_{j}$ is the per-study sample size and $\delta_{j}$ is the
study-specific true effect size.}

\textcolor{black}{When the effects are odds ratios, the effect size
estimate is $o_{j}=\frac{n_{11j}}{n_{10j}}/\frac{n_{01j}}{n_{00j}}=\frac{n_{00j}n_{11j}}{n_{01j}n_{10j}}$
where $n_{11j}$ is the number of participants with $Y=1$ in Group
1, $n_{10j}$ is the number of participants with $Y=0$ in Group 1,
$n_{01j}$ is the number of participants with $Y=1$ in Group 2, and
$n_{00j}$ is the number of participants with $Y=0$ in Group 2. Since
odds ratios cannot be negative and hence the distribution of sample
estimates is skewed, odds ratios are usually transformed by natural
logarithm (i.e., $log\left(o_{j}\right)$; log odds ratios). The approximated
standard error of the log odds ratio is $\hat{\sigma}_{j}=\sqrt{1/n_{00j}+1/n_{10j}+1/n_{01j}+1/n_{11j}}$
(Morris \& Gardner, 1988)\nocite{morris1988calculating}.}

\section{\textcolor{black}{3. Heterogeneity Testing in Meta-Analysis}}

\textcolor{black}{To test heterogeneity of effect sizes ($H_{0}:\tau^{2}=0$
versus $H_{0}:\tau^{2}\neq0$ ), various hypothesis tests have been
proposed. We focus on the Q test, the improved Q test, likelihood
ratio test, and Bootstrap methods in the main text. There are also
other heterogeneity tests and we illustrate them in the supplemental
materials. None of the illustrated tests were proposed or have been
generalized to the heterogeneity magnitude test: $H_{0}:\tau^{2}=\lambda$
versus $H_{a}:\tau^{2}>\lambda$. To provide an overall picture of
different methods, the strengths and limitations of each method are
presented in Table 1 and will be elaborated further in the following
sections. }

\subsection{\textcolor{black}{Q test}}

\textcolor{black}{The $Q$ test is historically the most widely used
test. In our literature review on }\textit{\textcolor{black}{Psychological
Bulletin}}\textcolor{black}{, 75 articles out of 77 articles which
tested the between-study heterogeneity used the $Q$ test. The test
statistic $Q$ follows a $\chi^{2}$ distribution with $K-1$ degrees
of freedom in the random-effects meta-analysis and with $K-P-1$ ($P$
is the number of covariates) degrees of freedom in the mixed-effects
meta-analysis when the effect size is normally distributed and the
sampling variance, $\sigma_{j}^{2}$, is known \citep{hoaglin2016misunderstandings,hedges1983random,hedges1985statistical}.
When $\sigma_{j}^{2}$ is unknown and a consistent estimator of $\sigma_{j}^{2}$
is used, $Q$ asymptotically approximates the $\chi^{2}$ distribution.
The approximation is not accurate for small and medium study-level
sample sizes \citep{kulinskaya2011moments}. The Type I error rate
and power of the Q test have been widely studied with different types
of effect size, and the unified conclusion is that the Type I error
rate and power of the Q test are influenced by the number of studies
and the study-level sample sizes (e.g., \citealp{chang1993power,harwell1997empirical,hedges1985statistical,morris2000distribution,takkouche1999evaluation,viechtbauer2007hypothesis}).
More specifically, when the study-level sample sizes are small and
the number of studies is large, the Type I error rate and power (usually
not powerful enough) deviate from the theoretical values (e.g., \citealp{chang1993power,harwell1997empirical,hedges1985statistical,viechtbauer2007hypothesis}).
The direction of deviation depends on the type of effect size. Regardless
of the type of effect size, with a large number of studies and larger
study-level sample sizes, the Q test approximates the nominal $\alpha$
level (e.g., \citealp{harwell1997empirical,morris2000distribution,viechtbauer2007hypothesis}).
\citet{viechtbauer2007hypothesis} found that more studies, large
study-level sample sizes, and larger $\tau^{2}$ led to higher power
across different types of effect size, however, \citet{harwell1997empirical}
found that power values decreased with more studies, especially for
smaller study-level sample sizes for standardized mean difference.
We present the strengths and limitations of the Q test in Table 1. }

\textcolor{black}{{[}Table 1{]}}{\scriptsize{}}
\begin{table}
{\scriptsize{}\caption{Summary of the methods}
}{\scriptsize\par}

\resizebox{\textwidth}{!}{
\renewcommand{\arraystretch}{0.9}

\textcolor{blue}{}%
\begin{tabular}{lll}
\hline 
\textcolor{black}{Method} & \textcolor{black}{Strengths } & \textcolor{black}{Limitations}\tabularnewline
\hline 
\textcolor{black}{Regular Q test} & \textcolor{black}{Widely used in different software and} & \textcolor{black}{Type I error rates are not appropriately controlled; }\tabularnewline
 & \textcolor{black}{R packages} & \textcolor{black}{Power can be low; }\tabularnewline
 &  & \textcolor{black}{Cannot conduct heterogeneity magnitude test}\tabularnewline
\hline 
\textcolor{black}{Kulinskaya's improved Q test ($Q_{2}$)} & \textcolor{black}{Appropriately control Type I error rates;} & \textcolor{black}{Require raw data information such as sample means }\tabularnewline
 & \textcolor{black}{High power;} & \textcolor{black}{and sample standard deviations;}\tabularnewline
 & \textcolor{black}{R package available } & \textcolor{black}{Complicated mathematical expressions; }\tabularnewline
 &  & \textcolor{black}{Cannot handle effect size measures other than }\tabularnewline
 &  & \textcolor{black}{standardized mean difference, risk difference, and
log }\tabularnewline
 &  & \textcolor{black}{odds ratio so far; }\tabularnewline
 &  & \textcolor{black}{Cannot incorporate covariates;}\tabularnewline
 &  & \textcolor{black}{Cannot conduct heterogeneity magnitude test}\tabularnewline
\hline 
\textcolor{black}{Breslow-Day's test} & \textcolor{black}{Appropriately control Type I error rates;} & \textcolor{black}{Cannot handle effect size measures other than}\tabularnewline
 & \textcolor{black}{High power} & \textcolor{black}{{} odds ratio;}\tabularnewline
 & \textcolor{black}{R package available } & \textcolor{black}{Cannot incorporate covariates;}\tabularnewline
 &  & \textcolor{black}{Cannot conduct heterogeneity magnitude test}\tabularnewline
\hline 
\textcolor{black}{ML- and REML based LR tests } & \textcolor{black}{R package available } & \textcolor{black}{Type I error rates are not appropriately controlled;}\tabularnewline
 &  & \textcolor{black}{Power can be low;}\tabularnewline
 &  & \textcolor{black}{Cannot conduct heterogeneity magnitude test}\tabularnewline
\hline 
\textcolor{black}{Takkouche's, } & \textcolor{black}{Appropriately control Type I error rates} & \textcolor{black}{Cannot be replicated given important details missing;}\tabularnewline
\textcolor{black}{Van den Noortgate-Onghena,} & \textcolor{black}{{} in the small scale simulations} & \textcolor{black}{Performance has not been examined by large scale }\tabularnewline
\textcolor{black}{and Sinha's bootstrap tests} &  & \textcolor{black}{simulations;}\tabularnewline
 &  & \textcolor{black}{Cannot incorporate covariates;}\tabularnewline
 &  & \textcolor{black}{Cannot conduct heterogeneity magnitude test}\tabularnewline
\hline 
\textcolor{black}{Bootstrap based ML LR test } & \textcolor{black}{Control Type I error rates relatively well;} & \textcolor{black}{Can fail to control Type I error rates when the }\tabularnewline
\textcolor{black}{(B-ML-LRT),} & \textcolor{black}{Higher power compared to the regular Q test} & \textcolor{black}{study-level sample sizes were small ($SZ=24$) }\tabularnewline
\textcolor{black}{Bootstrap based REML LR test} & \textcolor{black}{Can incorporate covariates;} & \textcolor{black}{and the effect size was the log odds ratio;}\tabularnewline
\textcolor{black}{{} (B-REML-LRT)} & \textcolor{black}{Can be used in heterogeneity magnitude test;} & \textcolor{black}{Can be less powerful than the improved Q test}\tabularnewline
 & \textcolor{black}{Can be easily generalized to other types of } & \textcolor{black}{B-REML-LRT performs slightly better than}\tabularnewline
 & \textcolor{black}{effect sizes;} & \textcolor{black}{B-ML-LRT}\tabularnewline
 & \textcolor{black}{R package available } & \tabularnewline
\hline 
\textcolor{black}{Bootstrap based Q test (B-Q)} & \textcolor{black}{Appropriately control Type I error rates} & \textcolor{black}{Can be less powerful than B-REML-LRT and }\tabularnewline
 & \textcolor{black}{Higher power compared to the regular Q test} & \textcolor{black}{the improved Q test}\tabularnewline
 & \textcolor{black}{Can incorporate covariates;} & \tabularnewline
 & \textcolor{black}{Can be used in heterogeneity magnitude test;} & \tabularnewline
 & \textcolor{black}{Can be easily generalized to other types of } & \tabularnewline
 & \textcolor{black}{effect sizes;} & \tabularnewline
 & \textcolor{black}{R package available } & \tabularnewline
\hline 
\end{tabular}

}
\end{table}
{\scriptsize\par}

\textcolor{black}{Kulinskaya \citeyearpar{kulinskaya2011moments,kulinskaya2011testing,kulinskaya2015accurate}
pointed out that there are better approximations than $\chi_{K-1}^{2}$
for the Q test in meta-analysis, and the approximated distribution
of $Q$ depends on the type of effect size.}\footnote{\textcolor{black}{The improved approximations were proposed in the
cases where the effects are not normally distributed (but can be asymptotically
normally distributed) and the estimators of the weights are not independent
of the estimators of the effects. But Kulinskaya still assumed that
raw data are normally distributed. }}\textcolor{black}{{} When effect sizes are standardized mean differences,
\citet{kulinskaya2011testing} estimated the degrees of freedom of
the $\chi^{2}$ distribution using an approximation of $E\left(Q\right)$.
However, the improved Q test needs the sample mean and sample variances
information from each study, which may not be provided in meta-analysis.
When effect sizes are risk differences and log odds ratios, \citet{kulinskaya2011moments}
and \citet{kulinskaya2015accurate} approximated the reference distribution
of the Q test by matching Gamma distributions. Kulinskaya \citeyearpar{kulinskaya2011moments,kulinskaya2011testing,kulinskaya2015accurate}
found that the Type I error rate and statistical power from the improved
Q test is much better than the Q test. In order to achieve a high
approximation accuracy, mathematical expressions from Kulinskaya \citeyearpar{kulinskaya2011moments,kulinskaya2011testing,kulinskaya2015accurate}
are all quite complicated, and the generalization of Kulinskaya's
approximation to other types of effect sizes needs to be resolved
based on each specific effect size. As independent work, \citet{breslow1991statistical}
proposed a $\chi^{2}$ test for odds ratios (no logarithm transformation
needed) and the reference distribution is $\chi_{K-1}^{2}$. Interested
readers can find more details in \citet{takkouche1999evaluation},
\citet{viechtbauer2007hypothesis}, and \citet{viechtbauer2015comparison}. }

\subsection{\textcolor{black}{Likelihood ratio (LR) test}}

\textcolor{black}{The likelihood ratio (LR) test is not very widely
used but frequently mentioned in the literature probably because it
is relatively new \citep{takkouche1999evaluation,viechtbauer2007hypothesis}.
The LR test for the between-study heterogeneity is based on comparing
two models: a full model in which $\tau^{2}$ is freely estimated
(the random-effects model or the mixed-effects model) and a reduced
model in which $\tau^{2}=0$ (the fixed-effects model or the fixed-effect
meta-regression). There are two maximum likelihood based estimators:
the maximum likelihood (ML) estimator is based on the regular log-likelihood
function and the restricted maximum likelihood (REML) estimator is
based on the restricted log-likelihood function. Generally, REML is
preferred over ML for variance component estimation.}

\textcolor{black}{There is a boundary (i.e., 0) in the variances'
parameter space when only nonnegative variances are allowed (i.e.,
constrained estimation). This is the so-called boundary issue. If
we allow negative variances and the boundary issue does not exist
(i.e., unconstrained estimation), the LR test follows a $\chi_{1}^{2}$
distribution. With constrained estimation, the reference distribution
is a 0.5:0.5 mixture of chi-squared distributions, and the critical
value is 2.71 ($.5\chi_{0}^{2}+.5\chi_{1}^{2}$ at $\alpha=0.05$)
\citep{stram1994variance,stram1995variance,stoel2006likelihood}.
Similar to the Q test, the simulation in \citet{viechtbauer2007hypothesis}
showed that the performance of ML based LR test and the REML based
LR test also depended on the type of effect sizes, the number of studies,
and the study-level sample sizes. He found that a large number of
studies and large study-level sample sizes, especially the latter,
improved  the Type I error rate, except in the Fisher-transformed
correlation case \citep{viechtbauer2007hypothesis}. But even with
a large number of studies and large study-level sample sizes, the
ML based LR test still could be too conservative. The REML based LR
test was slightly better than the ML based LR test \citep{viechtbauer2007hypothesis}.
The statistical power of the Q test, the ML based LR test, and the
REML based LR test were very similar \citep{viechtbauer2007hypothesis}. }

\subsection{\textcolor{black}{Bootstrap methods}}

\textcolor{black}{Because the approximation of the LR test is sensitive
to the number of studies and study-level sample sizes, when the number
of studies is small and/or the study-level sample sizes are small,
the reference distribution may be far different from a 0.5:0.5 mixture
of $\chi_{0}^{2}$ and $\chi_{1}^{2}$. \citet{takkouche1999evaluation}
proposed a parametric bootstrap version of the LR test. The procedure
of testing the between-study heterogeneity in a random-effects model
is as follows: (1) draw with replacement from the original $K$ studies;
(2) estimate $\mu_{\delta}$ and the sampling variances, $\sigma_{j}^{2}$,
for each study within each bootstrap sample; (3) simulate $K$ observed
study effect sizes using a fixed-effects model based on the estimated
$\mu_{\delta}$ and $\sigma_{j}^{2}$s; (4) calculate the LR test
statistic, $LR^{B}$; (5) repeat steps 1 to 4 for $B$ times; (6)
the empirical $p$-value is calculated as the proportion of $LR^{B}$s
larger than the LR in the original sample. However, there are two
things unclear in \citet{takkouche1999evaluation}. First, they did
not describe which model was used to obtain $\hat{\mu}_{\delta}$
(i.e., a fixed- or random-effects model). Second and more importantly,
it is unclear which ML estimator was used and whether REML- or ML
based LR test was used. To the best of our knowledge, the ML and REML
estimators were first proposed or clearly defined by Viechtbauer in
2005 \citep{viechtbauer2005bias}, after the study of \citet{takkouche1999evaluation}.
\citet{takkouche1999evaluation} examined the performance of the regular
LR test, the bootstrap based LR test, and the Q test with the odds
ratio. They found that the Q test and the bootstrap based LR test
appropriately controlled the Type I error rate, but the regular LR
test was too conservative. The power of the Q test and the bootstrap
based LR test were similar to each other, whereas the LR test had
lower power. \citet{sinha2012bootstrap} and \citet{van2003parametric}
had independent work on parametric bootstrap tests for testing the
between-study heterogeneity. Their procedures were similar to that
of \citet{takkouche1999evaluation}. We illustrate the details in
the supplemental materials.}

\section{\textcolor{black}{4. A Bootstrap Based Heterogeneity Test}}

\textcolor{black}{Given the importance of testing the between-study
heterogeneity, researchers have worked on proposing better heterogeneity
tests. However, there are several limitations in the existing methods
as summarized in Table 1. First, the existing methods focus on testing
homogeneity ($\tau^{2}=0$ verses  $\tau^{2}\neq0$). But none of
the existing methods can test whether the heterogeneity is larger
than a specific level ($\tau^{2}=\lambda$ verses $\tau^{2}>\lambda$),
which is a quite important research question when researchers are
interested in the heterogeneity magnitude or replication. More specifically,
the reference distribution of the Q test when $\tau^{2}\neq0$ is
a weighted linear combination of chi-square distributions and this
reference distribution only can be approximated when the sampling
variances are the same across studies \citep{hedgespigott2001}. Hence,
the regular Q test is not applicable in the heterogeneity magnitude
test. The reference distribution of ML-LRT and REML-LRT (i.e., 0.5:0.5
mixture of chi-squared distributions) cannot accommodate inequality
constraints like $\tau^{2}>\lambda$. Second, despite the widespread
use of different tests of homogeneity, the existing methods are criticized
by their failure to control the Type I error rate and attain enough
statistical power (e.g., \citealp{borenstein2009fixed,huedo2006assessing}).
Third, although Kulinskaya's improved Q test perform better than the
regular Q test, the improved approximations cannot handle effect size
measures other than standardized mean difference, risk difference,
and log odds ratio, and the mathematical expressions for the proposed
approximated distributions are all quite complicated. Additionally,
the improved Q test requires the raw data information for the standardized
mean difference effect size, which may not be available in the literature.
Fourth, the bootstrap version of the LR test in \citet{takkouche1999evaluation}
cannot be replicated given important details missing. In addition,
the parametric bootstrap methods by \citet{takkouche1999evaluation},
\citet{sinha2012bootstrap}, and \citet{van2003parametric} were only
examined in small simulation studies. Fifth, the existing research
of between-study heterogeneity tests mainly focuses on random-effects
meta-analysis, and the mixed-effects meta-analysis does not get enough
attention. In a mixed-effects model, the heterogeneity test detects
the residual heterogeneity that is not accounted for by the moderators,
which can be used to search for important moderators and evaluate
more complex moderation hypothesis. }

\textcolor{black}{We now propose a parametric bootstrap based heterogeneity
test. The basic procedure of our bootstrap method is to simulate the
empirical reference distribution of a chosen test statistic (i.e.,
ML or REML based LR test or the Q test) to get a critical value for
null hypothesis testing. This method can be used with or without covariates.
When the bootstrap procedure is coupled with the ML estimator, the
REML estimator, or the Q statistic, we refer to it as B-ML-LRT, B-REML-LRT,
or B-Q, respectively. Different from \citet{sinha2012bootstrap} and
\citet{van2003parametric}, we simulate bootstrap samples based on
effect size's asymptotic sampling distributions, whereas \citet{sinha2012bootstrap}
and \citet{van2003parametric} simulated raw data first and calculate
simulated effect sizes based on the simulated data. }

\textcolor{black}{The general procedure of testing $\tau^{2}$ in
a meta-analysis with $K$ studies using the proposed method is as
follows.}

\textcolor{black}{(1) For the bootstrap based LR tests, we calculate
the ML or REML based LR test statistic based on the corresponding
log-likelihood function in the original sample $LR^{O}$. For the
bootstrap based Q test, we calculate the Q statistic in the original
sample $Q^{o}$.}

\textcolor{black}{(2) We use the original sample sizes and covariate
values, treat the REML estimates of $\mu_{\delta}$ and the influence
of covariates as the true parameters, and simulate $K$ sample study
effect sizes ($x_{j}^{B}$; $j\in[1,K]$) based on the null hypothesis
(e.g., $H_{0}:\tau^{2}=0.05$). The sampling variances used in simulation
are the sampling variances in the original study.}

\textcolor{black}{(3) With the simulated effect size $x_{j}^{B}$,
we calculate the ML or REML based LR test statistic $LR^{B}$ and
the Q statistic $Q^{B}$. Note that the heterogeneity magnitude test
is a one-sided test ($\tau^{2}=\lambda$ verses $\tau^{2}>\lambda$),
and therefore when $\hat{\tau^{2}}$ in the simulated data was smaller
than $\lambda$, the difference between the log-likelihoods was counted
as 0.}\footnote{\textcolor{black}{When the effect size is the log odds ratio, the
effect size is a function of four cell sizes. The four cell sizes
in the original data will not be consistent with the simulated $x_{j}^{B}$,
therefore there is one more step. We keep three of the simulated cell
sizes $n_{00j}^{B}$, $n_{10j}^{B}$, $n_{11j}^{B}$ and $n_{01j}^{B}$
the same as $n_{00j}$, $n_{10j}$, $n_{11j}$ and $n_{01j}$ in the
original data, and the fourth one is updated based on $x_{j}^{B}$.
For example, we fix $n_{00j}^{B}$, $n_{10j}^{B}$, and $n_{11j}^{B}$
to be the same as $n_{00j}$, $n_{10j}$ and $n_{11j}$, $n_{01j}^{B}$
is calculated as $\frac{n_{11j}^{B}n_{00j}^{B}}{n_{10j}^{B}exp\left(x_{j}^{B}\right)}$.
Each cell size has a 25\% chance to be calculated across the permutations.
Based on the calculated fourth cell size, the sampling variances of
simulated studies are updated, and then we calculate the ML- or REML
based LR test statistic and the Q statistic.}}

\textcolor{black}{(4) We repeat steps 2 and 3 $B$ times.}

\textcolor{black}{(5) The critical value $LR^{critical}$ is the $\left(1-\alpha\right)$-quantile
of the $B$ $LR^{B}$s where $\alpha$ is the pre-specified Type I
error rate, and the critical value $Q^{critical}$ is the $\left(1-\alpha\right)$-quantile
of the $B$ $Q^{B}$s.}

\textcolor{black}{(6) In the heterogeneity test, the null assumption
is rejected if $LR^{O}$ is larger than $LR^{critical}$ or $Q^{O}$
is larger than $Q^{critical}$. In the heterogeneity magnitude test,
the null assumption is rejected if $LR^{O}$ is larger than $LR^{critical}$
or $Q^{O}$ is larger than $Q^{critical}$, and $\hat{\tau^{2}}$
in the original data was larger than $\lambda$. When $\hat{\tau^{2}}$
in the original data was smaller than $\lambda$, we fail to reject
the null assumption. }

\textcolor{black}{Note that the log-likelihood function should match
the LR test. That is, when the ML estimator is used to estimate $\tau^{2}$,
$LRT=-2\left(L_{\hat{\mu}_{\delta}^{\left(ML\right)},\tau^{2}=0}-L_{\hat{\mu}_{\delta}^{\left(ML\right)},\hat{\tau}^{2\left(ML\right)}}\right)$
based on the regular log-likelihood function, and when the REML estimator
is used to estimate $\tau^{2}$, $LRT=-2\left(L_{\tau^{2}=0}^{restricted}-L_{\hat{\tau}^{2\left(REML\right)}}^{restricted}\right)$
based on the restricted log-likelihood function. }

\subsection{\textcolor{black}{R package for Implementing the Bootstrap Based
Between-Study Heterogeneity Test}}

\textcolor{black}{We provide an R package for the bootstrap based
between-study heterogeneity test, $\mathtt{boot.heterogeneity}$.
This package can be downloaded from GitHub and CRAN (The Comprehensive
R Archive Network). The code, example illustrations, and the help
manual are in the supplemental materials. $\mathtt{boot.heterogeneity}$
builds on the $\mathtt{metafor}$ package by \citet{viechtbauer2015package}.
It implements the heterogeneity test for the standardized mean difference,
Fisher-transformed Pearson correlation, and log odds ratio. For example,
$\mathtt{boot.d}$ is a function for testing the between-study heterogeneity
with the standardized mean difference.
\[
\mathtt{\mathsf{\mathrm{\mathtt{boot.d(n1,n2,est,model="random",mods=NULL,nrep=10^{4},p\_cut=0.05,adjust=FALSE,lambda=0)}}}}
\]
$\mathtt{est}$ reads in a vector of unbiased estimates of standardized
mean differences from individual studies. If the biased estimates
$g_{j}$ are read in for $\mathtt{est}$, $\mathtt{adjust=TRUE}$
must be specified to obtain the corresponding unbiased standardized
mean difference estimates $d_{j}$. $\mathtt{n1}$ and $\mathtt{n2}$
read in two vectors of sample sizes from group 1 and group 2 in each
of the included studies, respectively. $\mathtt{model}$ defines a
random- or mixed- effects model. $\mathtt{mods}$ can read in one
or more moderators in the model. $\mathtt{nrep}$ defines the number
of replications used in bootstrap simulations and the default value
is $10^{4}$. $\mathtt{p\_cut}$ defines the alpha level and the default
value is 0.05. $\mathtt{lambda}$ defines the null hypothesis ($H_{0}:\tau^{2}=\lambda$).
Bootstrap based tests of the between-study heterogeneity for Fisher-transformed
Pearson correlations and log odds ratios are implemented in functions
$\mathtt{boot.fcor()}$ and $\mathtt{boot.lnOR()}$, respectively. }

\textcolor{black}{Each function for testing the between-study heterogeneity
provides the testing results from the bootstrap based REML LR test
(B-REML-LRT), the bootstrap based Q test (B-Q), and the Q test. In
addition, for odds ratio, the standard error will be infinite if any
one of the four cells is zero. In this case, Haldane and Anscombe's
correction is used by adding 0.5 to each cell value automatically
\citep{anscombe1956estimating,haldane1940mean}. }

\section{\textcolor{black}{5. Simulation studies}}

\subsection{\textcolor{black}{5.1 Simulation Design}}

\textcolor{black}{Five factors that have been found to have an impact
on the performance of the between-study heterogeneity tests are manipulated
in the simulation study: (1) the type of effect size (standardized
mean difference, Fisher-transformed Pearson correlation, or log odds
ratio), (2) the number of studies, ($K$=10, 20, 30, 50, 100, or 300),
(3) the per-group study-level sample size ($SZ=24$, 91, or 370),
(4) the size of the between-study heterogeneity $\tau^{2}$, and (5)
the number of the study-level covariates (1 or 3) in the mixed-effects
model. The overall effect size $\mu_{\delta}$ was not manipulated
because previous research has found that its influence on the performance
of the between-study heterogeneity tests is negligible (e.g., \citealp{takkouche1999evaluation,viechtbauer2005bias}).
Our pilot simulation results confirmed this. Please see the supplemental
materials for more details. Therefore, we specified $\mu_{\delta}=0$
in the simulation. }

\textcolor{black}{Specifically, sample sizes were chosen based on
three real meta-analyses on standardized mean differences in order
to mimic the realistic setting: a 55-study meta-analysis with the
per-group sample sizes ranging from 5 to 83 and a median of 24 \citep{deci1999meta},
a 33-study meta-analysis with the per-group sample sizes ranging from
10 to 489 and a median of 91\citep{becker1986influence}, and a 186-study
meta-analysis with the per-group sample sizes ranging from 12 to 96267
and a median of 370 \citep{hyde1990gender}. These chosen study-level
sample sizes and number of studies are also representative values
based on our literature review of the 126 studies in }\textit{\textcolor{black}{Psychological
Bulletin}}\textcolor{black}{. }

\textcolor{black}{The size of the between-study heterogeneity $\tau^{2}$
is determined in the following way. \citet{pigott2012advances} provided
benchmarks for the size of a relative heterogeneity index $I^{2}$
(the proportion of total variation in the observed effect sizes that
is due to the between-study heterogeneity \citealp{higgins2002quantifying}):
small (25\%), medium (50\%) and large (75\%). We first calculated
the average within-study variance per sample size set for each type
of effect, and we compute $\tau^{2}$ based on the within-study variance
and the above-mentioned benchmarks. Hence, the size of the between-study
heterogeneity is different for different types of effect size. The
condition of SZ=370 was not considered because $\tau^{2}$ was too
small (<0.001). We illustrate different levels of $\tau^{2}$ in Table
2. Additionally, we checked heterogeneity review papers to verify
the appropriateness of $\tau^{2}$'s specification (\citealp{rhodes2015predictive}
and \citealp{van2017estimates} for the standardized mean difference;
\citealp{turner2015predictive} and \citealp{gunhan2020random} for
the log odds ratio; \citealp{van2017estimates} for the correlation).}

\textcolor{black}{{[}Table 2{]}}{\scriptsize{}}
\begin{table}
{\scriptsize{}\caption{\textcolor{black}{Levels of heterogeneity of $\tau^{2}$}}
}{\scriptsize\par}

\resizebox{\textwidth}{!}{
\renewcommand{\arraystretch}{0.8}

\begin{tabular}{c|c|ccc}
\hline 
 & \textcolor{black}{Sample size} & \multicolumn{3}{c}{\textcolor{black}{Levels of heterogeneity}}\tabularnewline
 & \textcolor{black}{(SZ)} & \textcolor{black}{Small} & \textcolor{black}{Medium} & \textcolor{black}{Large}\tabularnewline
\hline 
\multirow{2}{*}{\textcolor{black}{Standardized mean difference}} & \textcolor{black}{24} & \textcolor{black}{0.03} & \textcolor{black}{0.1} & \textcolor{black}{0.3}\tabularnewline
 & \textcolor{black}{91} & \textcolor{black}{0.006} & \textcolor{black}{0.02} & \textcolor{black}{0.05}\tabularnewline
\hline 
\multirow{2}{*}{\textcolor{black}{Fisher-transformed Pearson correlation}} & \textcolor{black}{24} & \textcolor{black}{0.01} & \textcolor{black}{0.03} & \textcolor{black}{0.1}\tabularnewline
 & \textcolor{black}{91} & \textcolor{black}{0.006} & \textcolor{black}{0.02} & \textcolor{black}{0.05}\tabularnewline
\hline 
\multirow{2}{*}{\textcolor{black}{Log odds ratio}} & \textcolor{black}{24} & \textcolor{black}{0.1} & \textcolor{black}{0.3} & \textcolor{black}{0.9}\tabularnewline
 & \textcolor{black}{91} & \textcolor{black}{0.03} & \textcolor{black}{0.1} & \textcolor{black}{0.3}\tabularnewline
\hline 
\end{tabular}

}
\end{table}
{\scriptsize\par}

In the mixed-effects meta-analyses, the study-level covariates were
simulated from a normal distribution $N(0,\,1)$ and the regression
coefficients for study-level covariates, $\boldsymbol{\beta}$, were
specified to 0.5. The number of replications was 1000 for each condition.
The study specific true effect size was simulated by $\delta_{j}\sim N(\mu_{\delta},\:\tau^{2})$
without covariates or $\delta_{j}\sim N(\mu_{\delta}+\beta_{1}Z_{1j}+...+\beta_{P}Z_{Pj},\:\tau^{2})$
with covariates. The number of bootstrap samples, $M$, was set to
10,000 to find the critical value of the LR tes\textcolor{black}{t
or Q test.}\footnote{\textcolor{black}{In our simulation and real data analyses, we found
$10^{4}$ bootstrap samples provided relatively stable critical values.}}\textcolor{black}{{} We consider a Type I error rate between 0.025 and
0.075 as satisfactory (Bradley, 1978)\nocite{bradley1978robustness}.}

\textcolor{black}{In the simulation of standardized mean differences,
we randomly drew $K$ sample sizes from the aforementioned three real
meta-analysis datasets. Based on each simulated $\delta_{j}$, the
raw data were simulated as $y_{1ij}\sim N\left(0,\:1\right)$ and
$y_{2ij}\sim N\left(\delta_{j},\:1\right)$ where $i$ indicates the
$i$th individual and $j$ indicates the $j$th study. Based on the
sample means and sample variances of each study, we calculated the
observed standardized mean differences. }

\textcolor{black}{In the simulation of Fisher-transformed Pearson
correlations, we randomly drew $K$ per-group sample sizes from one
of the two groups in the real meta-analysis data. The simulated study
specific true Fisher-transformed Pearson correlations $\delta_{j}$
were converted to Pearson correlations $\rho_{j}$. For each simulated
study, the raw data were simulated as $\left(\begin{array}{c}
y_{1ij}\\
y_{2ij}
\end{array}\right)\sim MV\left(\left(\begin{array}{c}
0\\
0
\end{array}\right),\left(\begin{array}{cc}
1 & \rho_{j}\\
\rho_{j} & 1
\end{array}\right)\right)$. Based on the simulated raw data, we calculated the observed Fisher-transformed
Pearson correlations $r_{j}$. }

\textcolor{black}{In the simulation of log odds ratio, we used the
total sample sizes of the two groups in the real meta-analysis datasets
as the total sample sizes of the odds ratios ($N_{j}$). We randomly
drew $n_{00j}$ and $n_{01j}$, and calculated $n_{10j}$ and $n_{11j}$
based on the simulated study specific true log odds ratio ($\delta_{j}$).
The current $n_{00j}$, $n_{10j}$, $n_{01j}$, and $n_{11j}$ were
the true cell sizes without sampling error. Accordingly, the estimated
proportions of the four cells $p_{00j}=\frac{n_{00j}}{N_{j}}$, $p_{01j}=\frac{n_{01j}}{N_{j}}$,
$p_{10j}=\frac{n_{10j}}{N_{j}}$, and $p_{11j}=\frac{n_{11j}}{N_{j}}$
were calculated. We used binomial distributions to simulate the raw
data, add sampling error, and update $n_{00j}$, $n_{10j}$, $n_{01j}$,
and $n_{11j}$: $n_{01j.update}\sim binomial\left(n_{00j}+n_{01j},\frac{n_{01j}}{N_{j}}\right)$,
$n_{00j.update}=n_{00j}+n_{01j}-n_{01j.update}$, $n_{11j.update}\sim binomial\left(n_{10j}+n_{11j},\frac{n_{11j}}{N_{j}}\right)$,
and $n_{10j.update}=n_{00j}+n_{01j}-n_{11j.update}$. The updated
cell sizes contained sampling errors. Based on the simulated raw data,
we calculated the observed log odds ratios. }

\textcolor{black}{We considered the bootstrap based ML LR test (B-ML-LRT),
the bootstrap based REML LR test (B-REML-LRT), and the bootstrap based
Q test (B-Q). The proposed bootstrap based methods were compared with
the regular LR tests (ML-LRT and REML-LRT), the Q test, the improved
Q test by Kulinskaya \citeyearpar{kulinskaya2011moments,kulinskaya2011testing,kulinskaya2015accurate}
($Q_{2}$ was for the standardized mean difference and log odds ratio
only), and the Breslow-Day's test by \citet{breslow1991statistical}
(BD was for the odds ratio only).}

\textcolor{black}{As pilot studies, we have tried some other tests.
The first test was the bootstrap based ML and REML LR tests with unconstrained
estimation of $\tau^{2}$. Although unconstrained estimation should
avoid the boundary issue, this method failed to control the Type I
error rates. It could lead to too small or too large Type I error
rates, especially when the study-level sample sizes were large. Second,
we examined the Type I error rates of using confidence intervals of
the heterogeneity indexes $I^{2}$ and $H$ coupled with the different
estimators of $\tau^{2}$. The Type I error rates were too small across
all combinations. Please see the supplemental materials for more details.}

\subsection{\textcolor{black}{5.2 Type I error rates}}

\subsubsection{\textcolor{black}{5.2.1 Type I error rates in meta-analyses for heterogeneity
test}}

\textcolor{black}{In this section, we evaluate the performance of
the heterogeneity tests when the true size of heterogeneity is zero,
i.e., $\tau^{2}=0$. The Type I error rates of the bootstrap based
ML LR test (Q-ML-LRT), the bootstrap based REML LR test (Q-REML-LRT),
the bootstrap based Q test (B-Q), the improved Q test by \citet{kulinskaya2015accurate}
($Q_{2}$), and the Breslow-Day's test by \citet{breslow1991statistical}
for odds ratios (BD), the regular ML based LR test (ML-LRT), the regular
REML based LR test (REML-LRT), and the regular Q test are presented
in Table 3 with their corresponding Monte Carlo standard errors. ML-LRT
produced too conservative Type I error rates in most conditions. Consistent
with the results from \citet{viechtbauer2007hypothesis}, more studies
and larger study-level sample sizes improved the Type I error rates
of ML-LRT, but even with a relatively large number of studies and
study-level sample sizes (e.g., $SZ=370$ and $K=50$), ML-LRT was
too conservative in the examined conditions. REML-LRT performed better
than ML-LRT, but it also could provide too conservative Type I error
rates, especially with the log odds ratio. The Q test could fail to
control the Type I error rates when the study-level sample sizes were
small ($SZ=24$). Especially in the case of log odds ratio, with small
study-level sample sizes, large number of studies failed to boost
the Type I error rates of the Q test; in contrast, more studies tended
to decrease the Type I error rates. The improved Q test appropriately
controlled the Type I error rates across different conditions of study-level
sample size, number of studies, and type of effect size. The Breslow-Day's
test also could appropriately control the Type I error rates for the
odds ratio. The two bootstrap LR methods, B-ML-LRT and B-REML-LRT,
could appropriately control the Type I error rates regardless of the
type of effect size, the number of studies, and the study-level sample
sizes, except when the effect size was the log odds ratio and the
study-level sample sizes were small ($SZ=24$). Their performances
were similar. The bootstrap based Q test (B-Q) could appropriately
control the Type I error rates across all examined conditions. Comparing
to the Monte Carlo standard errors which were smaller than 0.01, the
methods that could control the Type I error rates and the methods
that failed to control the Type I error rates yielded quite different
Type I error rates. The difference between their Type I error rates
should not be just due to random errors.}

\textcolor{black}{{[}Table 3{]}}{\scriptsize{}}
\begin{table}
{\scriptsize{}\caption{{\footnotesize{}Type I error rates of the bootstrap based ML LR test
(Q-ML-LRT), the bootstrap based REML LR test (Q-RE}\textcolor{black}{\footnotesize{}ML-LRT),
the bootstrap based Q test (B-Q), the improved Q test ($Q_{2}$),
the Breslow-Day's test (BD), the regular ML based LR test (ML-LRT),
the regular REML based LR test (REML-LRT), and the regular Q test
in }{\footnotesize{}meta-analyses}}
}{\scriptsize\par}

\resizebox{\textwidth}{!}{
\renewcommand{\arraystretch}{0.40}

{\scriptsize{}}%
\begin{tabular}{cccccccccc}
\toprule 
{\scriptsize{}Sample size} & {\scriptsize{}Number of} & {\scriptsize{}B-ML-LRT} & {\scriptsize{}B-REML-LRT} & {\scriptsize{}B-Q} & {\scriptsize{}$Q_{2}$} & {\scriptsize{}BD} & {\scriptsize{}ML-LRT} & {\scriptsize{}REML-LRT} & {\scriptsize{}Q}\tabularnewline
\cmidrule{3-10} 
{\scriptsize{}(SZ)} & {\scriptsize{}Studies (K)} & \multicolumn{8}{c}{{\scriptsize{}Standardized mean difference}}\tabularnewline
\midrule
\multirow{6}{*}{{\scriptsize{}24}} & {\scriptsize{}10} & {\scriptsize{}0.052 (0.007)} & {\scriptsize{}0.052 (0.007)} & {\scriptsize{}0.053 (0.007)} & {\scriptsize{}0.044 (0.006)} & {\scriptsize{}-} & \textbf{\scriptsize{}0.015 (0.004)} & {\scriptsize{}0.026 (0.005)} & {\scriptsize{}0.039 (0.006)}\tabularnewline
 & {\scriptsize{}20} & {\scriptsize{}0.052 (0.007)} & {\scriptsize{}0.052 (0.007)} & {\scriptsize{}0.055 (0.007)} & {\scriptsize{}0.051 (0.007)} & {\scriptsize{}-} & \textbf{\scriptsize{}0.013 (0.004)} & \textbf{\scriptsize{}0.022 (0.005)} & {\scriptsize{}0.041 (0.006)}\tabularnewline
 & {\scriptsize{}30} & {\scriptsize{}0.049 (0.007)} & {\scriptsize{}0.05 (0.007)} & {\scriptsize{}0.049 (0.007)} & {\scriptsize{}0.044 (0.006)} & {\scriptsize{}-} & \textbf{\scriptsize{}0.017 (0.004)} & {\scriptsize{}0.026 (0.005)} & {\scriptsize{}0.033 (0.006)}\tabularnewline
 & {\scriptsize{}50} & {\scriptsize{}0.062 (0.008)} & {\scriptsize{}0.062 (0.008)} & {\scriptsize{}0.06 (0.008)} & {\scriptsize{}0.055 (0.007)} & {\scriptsize{}-} & {\scriptsize{}0.028 (0.005)} & {\scriptsize{}0.034 (0.006)} & {\scriptsize{}0.042 (0.006)}\tabularnewline
 & {\scriptsize{}100} & {\scriptsize{}0.065 (0.008)} & {\scriptsize{}0.064 (0.008)} & {\scriptsize{}0.063 (0.008)} & {\scriptsize{}0.058 (0.007)} & {\scriptsize{}-} & {\scriptsize{}0.031 (0.005)} & {\scriptsize{}0.037 (0.006)} & {\scriptsize{}0.045 (0.007)}\tabularnewline
 & {\scriptsize{}300} & {\scriptsize{}0.058 (0.007)} & {\scriptsize{}0.058 (0.007)} & {\scriptsize{}0.057 (0.007)} & {\scriptsize{}0.058 (0.007)} & {\scriptsize{}-} & \textbf{\scriptsize{}0.019 (0.004)} & \textbf{\scriptsize{}0.022 (0.005)} & {\scriptsize{}0.028 (0.005)}\tabularnewline
\midrule
\multirow{6}{*}{{\scriptsize{}91}} & {\scriptsize{}10} & {\scriptsize{}0.054 (0.007)} & {\scriptsize{}0.056 (0.007)} & {\scriptsize{}0.049 (0.007)} & {\scriptsize{}0.046 (0.007)} & {\scriptsize{}-} & \textbf{\scriptsize{}0.012 (0.003)} & {\scriptsize{}0.032 (0.006)} & {\scriptsize{}0.046 (0.007)}\tabularnewline
 & {\scriptsize{}20} & {\scriptsize{}0.041 (0.006)} & {\scriptsize{}0.041 (0.006)} & {\scriptsize{}0.04 (0.006)} & {\scriptsize{}0.038 (0.006)} & {\scriptsize{}-} & \textbf{\scriptsize{}0.019 (0.004)} & {\scriptsize{}0.03 (0.005)} & {\scriptsize{}0.035 (0.006)}\tabularnewline
 & {\scriptsize{}30} & {\scriptsize{}0.051 (0.007)} & {\scriptsize{}0.052 (0.007)} & {\scriptsize{}0.053 (0.007)} & {\scriptsize{}0.048 (0.007)} & {\scriptsize{}-} & {\scriptsize{}0.03 (0.005)} & {\scriptsize{}0.038 (0.006)} & {\scriptsize{}0.045 (0.007)}\tabularnewline
 & {\scriptsize{}50} & {\scriptsize{}0.054 (0.007)} & {\scriptsize{}0.055 (0.007)} & {\scriptsize{}0.052 (0.007)} & {\scriptsize{}0.05 (0.007)} & {\scriptsize{}-} & {\scriptsize{}0.028 (0.005)} & {\scriptsize{}0.037 (0.006)} & {\scriptsize{}0.046 (0.007)}\tabularnewline
 & {\scriptsize{}100} & {\scriptsize{}0.055 (0.007)} & {\scriptsize{}0.056 (0.007)} & {\scriptsize{}0.052 (0.007)} & {\scriptsize{}0.049 (0.007)} & {\scriptsize{}-} & {\scriptsize{}0.03 (0.005)} & {\scriptsize{}0.035 (0.006)} & {\scriptsize{}0.046 (0.007)}\tabularnewline
 & {\scriptsize{}300} & {\scriptsize{}0.043 (0.006)} & {\scriptsize{}0.043 (0.006)} & {\scriptsize{}0.045 (0.007)} & {\scriptsize{}0.04 (0.006)} & {\scriptsize{}-} & {\scriptsize{}0.032 (0.006)} & {\scriptsize{}0.032 (0.006)} & {\scriptsize{}0.032 (0.006)}\tabularnewline
\midrule
\multirow{6}{*}{{\scriptsize{}370}} & {\scriptsize{}10} & {\scriptsize{}0.043 (0.006)} & {\scriptsize{}0.049 (0.007)} & {\scriptsize{}0.051 (0.007)} & {\scriptsize{}0.05 (0.007)} & {\scriptsize{}-} & \textbf{\scriptsize{}0.007 (0.003)} & \textbf{\scriptsize{}0.022 (0.005)} & {\scriptsize{}0.05 (0.007)}\tabularnewline
 & {\scriptsize{}20} & {\scriptsize{}0.048 (0.007)} & {\scriptsize{}0.051 (0.007)} & {\scriptsize{}0.041 (0.006)} & {\scriptsize{}0.04 (0.006)} & {\scriptsize{}-} & \textbf{\scriptsize{}0.017 (0.004)} & {\scriptsize{}0.032 (0.006)} & {\scriptsize{}0.04 (0.006)}\tabularnewline
 & {\scriptsize{}30} & {\scriptsize{}0.052 (0.007)} & {\scriptsize{}0.054 (0.007)} & {\scriptsize{}0.053 (0.007)} & {\scriptsize{}0.053 (0.007)} & {\scriptsize{}-} & \textbf{\scriptsize{}0.021 (0.005)} & {\scriptsize{}0.037 (0.006)} & {\scriptsize{}0.051 (0.007)}\tabularnewline
 & {\scriptsize{}50} & {\scriptsize{}0.053 (0.007)} & {\scriptsize{}0.055 (0.007)} & {\scriptsize{}0.048 (0.007)} & {\scriptsize{}0.047 (0.007)} & {\scriptsize{}-} & \textbf{\scriptsize{}0.021 (0.005)} & {\scriptsize{}0.04 (0.006)} & {\scriptsize{}0.047 (0.007)}\tabularnewline
 & {\scriptsize{}100} & {\scriptsize{}0.058 (0.007)} & {\scriptsize{}0.056 (0.007)} & {\scriptsize{}0.059 (0.007)} & {\scriptsize{}0.057 (0.007)} & {\scriptsize{}-} & {\scriptsize{}0.029 (0.005)} & {\scriptsize{}0.044 (0.006)} & {\scriptsize{}0.057 (0.007)}\tabularnewline
 & {\scriptsize{}300} & {\scriptsize{}0.055 (0.007)} & {\scriptsize{}0.054 (0.007)} & {\scriptsize{}0.051 (0.007)} & {\scriptsize{}0.049 (0.007)} & {\scriptsize{}-} & {\scriptsize{}0.037 (0.006)} & {\scriptsize{}0.046 (0.007)} & {\scriptsize{}0.046 (0.007)}\tabularnewline
\midrule 
 &  & \multicolumn{8}{c}{{\scriptsize{}Fisher-transformed Pearson correlation}}\tabularnewline
\midrule
\multirow{6}{*}{{\scriptsize{}24}} & {\scriptsize{}10} & {\scriptsize{}0.073 (0.008)} & {\scriptsize{}0.074 (0.008)} & {\scriptsize{}0.074 (0.008)} & {\scriptsize{}-} & {\scriptsize{}-} & {\scriptsize{}0.032 (0.006)} & {\scriptsize{}0.05 (0.007)} & \textbf{\scriptsize{}0.077 (0.008)}\tabularnewline
 & {\scriptsize{}20} & {\scriptsize{}0.054 (0.007)} & {\scriptsize{}0.053 (0.007)} & {\scriptsize{}0.053 (0.007)} & {\scriptsize{}-} & {\scriptsize{}-} & {\scriptsize{}0.029 (0.005)} & {\scriptsize{}0.043 (0.006)} & {\scriptsize{}0.053 (0.007)}\tabularnewline
 & {\scriptsize{}30} & {\scriptsize{}0.046 (0.007)} & {\scriptsize{}0.048 (0.007)} & {\scriptsize{}0.048 (0.007)} & {\scriptsize{}-} & {\scriptsize{}-} & {\scriptsize{}0.031 (0.005)} & {\scriptsize{}0.037 (0.006)} & {\scriptsize{}0.048 (0.007)}\tabularnewline
 & {\scriptsize{}50} & {\scriptsize{}0.06 (0.008)} & {\scriptsize{}0.061 (0.008)} & {\scriptsize{}0.064 (0.008)} & {\scriptsize{}-} & {\scriptsize{}-} & {\scriptsize{}0.043 (0.006)} & {\scriptsize{}0.057 (0.007)} & {\scriptsize{}0.063 (0.008)}\tabularnewline
 & {\scriptsize{}100} & {\scriptsize{}0.047 (0.007)} & {\scriptsize{}0.047 (0.007)} & {\scriptsize{}0.052 (0.007)} & {\scriptsize{}-} & {\scriptsize{}-} & {\scriptsize{}0.031 (0.005)} & {\scriptsize{}0.04 (0.006)} & {\scriptsize{}0.053 (0.007)}\tabularnewline
 & {\scriptsize{}300} & {\scriptsize{}0.056 (0.007)} & {\scriptsize{}0.056 (0.007)} & {\scriptsize{}0.043 (0.006)} & {\scriptsize{}-} & {\scriptsize{}-} & {\scriptsize{}0.044 (0.006)} & {\scriptsize{}0.051 (0.007)} & {\scriptsize{}0.043 (0.006)}\tabularnewline
\midrule
\multirow{6}{*}{{\scriptsize{}91}} & {\scriptsize{}10} & {\scriptsize{}0.053 (0.007)} & {\scriptsize{}0.053 (0.007)} & {\scriptsize{}0.05 (0.007)} & {\scriptsize{}-} & {\scriptsize{}-} & \textbf{\scriptsize{}0.024 (0.005)} & {\scriptsize{}0.038 (0.006)} & {\scriptsize{}0.049 (0.007)}\tabularnewline
 & {\scriptsize{}20} & {\scriptsize{}0.055 (0.007)} & {\scriptsize{}0.055 (0.007)} & {\scriptsize{}0.053 (0.007)} & {\scriptsize{}-} & {\scriptsize{}-} & \textbf{\scriptsize{}0.023 (0.005)} & {\scriptsize{}0.04 (0.006)} & {\scriptsize{}0.055 (0.007)}\tabularnewline
 & {\scriptsize{}30} & {\scriptsize{}0.038 (0.006)} & {\scriptsize{}0.037 (0.006)} & {\scriptsize{}0.047 (0.007)} & {\scriptsize{}-} & {\scriptsize{}-} & \textbf{\scriptsize{}0.024 (0.005)} & {\scriptsize{}0.032 (0.006)} & {\scriptsize{}0.047 (0.007)}\tabularnewline
 & {\scriptsize{}50} & {\scriptsize{}0.047 (0.007)} & {\scriptsize{}0.048 (0.007)} & {\scriptsize{}0.061 (0.008)} & {\scriptsize{}-} & {\scriptsize{}-} & {\scriptsize{}0.03 (0.005)} & {\scriptsize{}0.037 (0.006)} & {\scriptsize{}0.062 (0.008)}\tabularnewline
 & {\scriptsize{}100} & {\scriptsize{}0.046 (0.007)} & {\scriptsize{}0.046 (0.007)} & {\scriptsize{}0.046 (0.007)} & {\scriptsize{}-} & {\scriptsize{}-} & {\scriptsize{}0.041 (0.006)} & {\scriptsize{}0.045 (0.007)} & {\scriptsize{}0.044 (0.006)}\tabularnewline
 & {\scriptsize{}300} & {\scriptsize{}0.044 (0.006)} & {\scriptsize{}0.044 (0.006)} & {\scriptsize{}0.043 (0.006)} & {\scriptsize{}-} & {\scriptsize{}-} & {\scriptsize{}0.037 (0.006)} & {\scriptsize{}0.039 (0.006)} & {\scriptsize{}0.044 (0.006)}\tabularnewline
\midrule
\multirow{6}{*}{{\scriptsize{}370}} & {\scriptsize{}10} & {\scriptsize{}0.041 (0.006)} & {\scriptsize{}0.042 (0.006)} & {\scriptsize{}0.052 (0.007)} & {\scriptsize{}-} & {\scriptsize{}-} & \textbf{\scriptsize{}0.008 (0.003)} & {\scriptsize{}0.026 (0.005)} & {\scriptsize{}0.052 (0.007)}\tabularnewline
 & {\scriptsize{}20} & {\scriptsize{}0.049 (0.007)} & {\scriptsize{}0.045 (0.007)} & {\scriptsize{}0.047 (0.007)} & {\scriptsize{}-} & {\scriptsize{}-} & \textbf{\scriptsize{}0.016 (0.004)} & {\scriptsize{}0.026 (0.005)} & {\scriptsize{}0.049 (0.007)}\tabularnewline
 & {\scriptsize{}30} & {\scriptsize{}0.046 (0.007)} & {\scriptsize{}0.053 (0.007)} & {\scriptsize{}0.048 (0.007)} & {\scriptsize{}-} & {\scriptsize{}-} & \textbf{\scriptsize{}0.018 (0.004)} & {\scriptsize{}0.032 (0.006)} & {\scriptsize{}0.048 (0.007)}\tabularnewline
 & {\scriptsize{}50} & {\scriptsize{}0.05 (0.007)} & {\scriptsize{}0.052 (0.007)} & {\scriptsize{}0.049 (0.007)} & {\scriptsize{}-} & {\scriptsize{}-} & \textbf{\scriptsize{}0.022 (0.005)} & {\scriptsize{}0.034 (0.006)} & {\scriptsize{}0.048 (0.007)}\tabularnewline
 & {\scriptsize{}100} & {\scriptsize{}0.047 (0.007)} & {\scriptsize{}0.047 (0.007)} & {\scriptsize{}0.047 (0.007)} & {\scriptsize{}-} & {\scriptsize{}-} & {\scriptsize{}0.026 (0.005)} & {\scriptsize{}0.035 (0.006)} & {\scriptsize{}0.047 (0.007)}\tabularnewline
 & {\scriptsize{}300} & {\scriptsize{}0.048 (0.007)} & {\scriptsize{}0.047 (0.007)} & {\scriptsize{}0.056 (0.007)} & {\scriptsize{}-} & {\scriptsize{}-} & {\scriptsize{}0.031 (0.005)} & {\scriptsize{}0.037 (0.006)} & {\scriptsize{}0.055 (0.007)}\tabularnewline
\midrule 
 &  & \multicolumn{8}{c}{{\scriptsize{}Log odds ratio}}\tabularnewline
\midrule
\multirow{6}{*}{{\scriptsize{}24}} & {\scriptsize{}10} & {\scriptsize{}0.037 (0.006)} & {\scriptsize{}0.036 (0.006)} & {\scriptsize{}0.041 (0.006)} & {\scriptsize{}0.04 (0.006)} & {\scriptsize{}0.037 (0.006)} & \textbf{\scriptsize{}0.009 (0.003)} & \textbf{\scriptsize{}0.017 (0.004)} & \textbf{\scriptsize{}0.024 (0.005)}\tabularnewline
 & {\scriptsize{}20} & {\scriptsize{}0.03 (0.005)} & {\scriptsize{}0.03 (0.005)} & {\scriptsize{}0.036 (0.006)} & {\scriptsize{}0.042 (0.006)} & {\scriptsize{}0.03 (0.005)} & \textbf{\scriptsize{}0.003 (0.002)} & \textbf{\scriptsize{}0.005 (0.002)} & \textbf{\scriptsize{}0.008 (0.003)}\tabularnewline
 & {\scriptsize{}30} & {\scriptsize{}0.032 (0.006)} & {\scriptsize{}0.033 (0.006)} & {\scriptsize{}0.031 (0.005)} & {\scriptsize{}0.044 (0.006)} & {\scriptsize{}0.037 (0.006)} & \textbf{\scriptsize{}0.007 (0.003)} & \textbf{\scriptsize{}0.012 (0.003)} & \textbf{\scriptsize{}0.016 (0.004)}\tabularnewline
 & {\scriptsize{}50} & {\scriptsize{}0.027 (0.005)} & {\scriptsize{}0.028 (0.005)} & {\scriptsize{}0.039 (0.006)} & {\scriptsize{}0.039 (0.006)} & {\scriptsize{}0.037 (0.006)} & \textbf{\scriptsize{}0.007 (0.003)} & \textbf{\scriptsize{}0.008 (0.003)} & \textbf{\scriptsize{}0.011 (0.003)}\tabularnewline
 & {\scriptsize{}100} & \textbf{\scriptsize{}0.015 (0.004)} & \textbf{\scriptsize{}0.015 (0.004)} & {\scriptsize{}0.043 (0.006)} & {\scriptsize{}0.036 (0.006)} & {\scriptsize{}0.03 (0.005)} & \textbf{\scriptsize{}0.006 (0.002)} & \textbf{\scriptsize{}0.009 (0.003)} & \textbf{\scriptsize{}0.009 (0.003)}\tabularnewline
 & {\scriptsize{}300} & \textbf{\scriptsize{}0.012 (0.003)} & \textbf{\scriptsize{}0.012 (0.003)} & {\scriptsize{}0.04 (0.006)} & {\scriptsize{}0.034 (0.006)} & {\scriptsize{}0.035 (0.006)} & \textbf{\scriptsize{}0.001 (0.001)} & \textbf{\scriptsize{}0.001 (0.001)} & \textbf{\scriptsize{}0.002 (0.001)}\tabularnewline
\midrule
\multirow{6}{*}{{\scriptsize{}91}} & {\scriptsize{}10} & {\scriptsize{}0.039 (0.006)} & {\scriptsize{}0.039 (0.006)} & {\scriptsize{}0.045 (0.007)} & {\scriptsize{}0.05 (0.007)} & {\scriptsize{}0.05 (0.007)} & \textbf{\scriptsize{}0.015 (0.004)} & {\scriptsize{}0.028 (0.005)} & {\scriptsize{}0.039 (0.006)}\tabularnewline
 & {\scriptsize{}20} & {\scriptsize{}0.047 (0.007)} & {\scriptsize{}0.048 (0.007)} & {\scriptsize{}0.048 (0.007)} & {\scriptsize{}0.052 (0.007)} & {\scriptsize{}0.052 (0.007)} & \textbf{\scriptsize{}0.02 (0.004)} & {\scriptsize{}0.03 (0.005)} & {\scriptsize{}0.036 (0.006)}\tabularnewline
 & {\scriptsize{}30} & {\scriptsize{}0.041 (0.006)} & {\scriptsize{}0.039 (0.006)} & {\scriptsize{}0.042 (0.006)} & {\scriptsize{}0.041 (0.006)} & {\scriptsize{}0.043 (0.006)} & \textbf{\scriptsize{}0.019 (0.004)} & \textbf{\scriptsize{}0.023 (0.005)} & {\scriptsize{}0.032 (0.006)}\tabularnewline
 & {\scriptsize{}50} & {\scriptsize{}0.029 (0.005)} & {\scriptsize{}0.03 (0.005)} & {\scriptsize{}0.047 (0.007)} & {\scriptsize{}0.048 (0.007)} & {\scriptsize{}0.056 (0.007)} & \textbf{\scriptsize{}0.015 (0.004)} & \textbf{\scriptsize{}0.021 (0.005)} & {\scriptsize{}0.027 (0.005)}\tabularnewline
 & {\scriptsize{}100} & {\scriptsize{}0.038 (0.006)} & {\scriptsize{}0.039 (0.006)} & {\scriptsize{}0.058 (0.007)} & {\scriptsize{}0.053 (0.007)} & {\scriptsize{}0.056 (0.007)} & \textbf{\scriptsize{}0.02 (0.004)} & \textbf{\scriptsize{}0.022 (0.005)} & {\scriptsize{}0.034 (0.006)}\tabularnewline
 & {\scriptsize{}300} & {\scriptsize{}0.033 (0.006)} & {\scriptsize{}0.033 (0.006)} & {\scriptsize{}0.049 (0.007)} & {\scriptsize{}0.038 (0.006)} & {\scriptsize{}0.05 (0.007)} & \textbf{\scriptsize{}0.022 (0.005)} & \textbf{\scriptsize{}0.022 (0.005)} & {\scriptsize{}0.023 (0.005)}\tabularnewline
\midrule
\multirow{6}{*}{{\scriptsize{}370}} & {\scriptsize{}10} & {\scriptsize{}0.035 (0.006)} & {\scriptsize{}0.039 (0.006)} & {\scriptsize{}0.054 (0.007)} & {\scriptsize{}0.054 (0.007)} & {\scriptsize{}0.051 (0.007)} & \textbf{\scriptsize{}0.009 (0.003)} & {\scriptsize{}0.028 (0.005)} & {\scriptsize{}0.05 (0.007)}\tabularnewline
 & {\scriptsize{}20} & {\scriptsize{}0.033 (0.006)} & {\scriptsize{}0.036 (0.006)} & {\scriptsize{}0.045 (0.007)} & {\scriptsize{}0.046 (0.007)} & {\scriptsize{}0.045 (0.007)} & \textbf{\scriptsize{}0.011 (0.003)} & \textbf{\scriptsize{}0.021 (0.005)} & {\scriptsize{}0.044 (0.006)}\tabularnewline
 & {\scriptsize{}30} & {\scriptsize{}0.049 (0.007)} & {\scriptsize{}0.053 (0.007)} & {\scriptsize{}0.048 (0.007)} & {\scriptsize{}0.05 (0.007)} & {\scriptsize{}0.052 (0.007)} & \textbf{\scriptsize{}0.015 (0.004)} & {\scriptsize{}0.031 (0.005)} & {\scriptsize{}0.043 (0.006)}\tabularnewline
 & {\scriptsize{}50} & {\scriptsize{}0.051 (0.007)} & {\scriptsize{}0.049 (0.007)} & {\scriptsize{}0.034 (0.006)} & {\scriptsize{}0.035 (0.006)} & {\scriptsize{}0.035 (0.006)} & \textbf{\scriptsize{}0.019 (0.004)} & {\scriptsize{}0.034 (0.006)} & {\scriptsize{}0.033 (0.006)}\tabularnewline
 & {\scriptsize{}100} & {\scriptsize{}0.039 (0.006)} & {\scriptsize{}0.04 (0.006)} & {\scriptsize{}0.044 (0.006)} & {\scriptsize{}0.049 (0.007)} & {\scriptsize{}0.048 (0.007)} & \textbf{\scriptsize{}0.021 (0.005)} & {\scriptsize{}0.032 (0.006)} & {\scriptsize{}0.039 (0.006)}\tabularnewline
 & {\scriptsize{}300} & {\scriptsize{}0.054 (0.007)} & {\scriptsize{}0.054 (0.007)} & {\scriptsize{}0.053 (0.007)} & {\scriptsize{}0.055 (0.007)} & {\scriptsize{}0.054 (0.007)} & {\scriptsize{}0.035 (0.006)} & {\scriptsize{}0.043 (0.006)} & {\scriptsize{}0.042 (0.006)}\tabularnewline
\bottomrule
\end{tabular}{\scriptsize\par}

}

{\scriptsize{}No}\textcolor{black}{\scriptsize{}te: Monte Carlo standard
errors are presented in the parentheses. A}{\scriptsize{}ll unacceptable
Type I error rates (i.e., <0.025 or >0.075) are bold. ML or REML estimation
sometimes did not have converged results, but the nonconvergence rates
across all conditions were within 1\%. ``-'' indicate not available.}{\scriptsize\par}
\end{table}
{\scriptsize\par}

\subsubsection{5.\textcolor{black}{2.2 Type I error rates in mixed-effects models
for heterogeneity test}}

\textcolor{black}{Here we evaluate the performance of heterogeneity
tests when the true size of residual heterogeneity is zero, i.e.,
$\tau^{2}=0$. The Type I error rates of B-ML-LRT, B-REML-LRT, B-Q,
ML-LRT, REML-LRT, and the Q test (Q) with one and three covariates
are presented in Tables 4 and 5 with their corresponding Monte Carlo
standard errors. The improved Q test and Breslow-Day's test are not
applicable in mixed-effects models. Similar to the model without covariates,
ML-LRT was too conservative in most examined conditions, regardless
of which type of effect size was of interest and the number of covariates.
REML-LRT and the Q test also could be too conservative, especially
when the effect size of interest was the log odds ratio. A larger
number of studies could aggravate the conservative Type I error rates
of ML-LRT, REML-LRT, and the Q test when the effect size was the log
odds ratio and the study-level sample sizes were small ($SZ=24$).
B-REML-LRT generally performed well, except when the study-level sample
sizes were small ($SZ=24$) and the effect size was the log odds ratio.
B-Q outperformed B-ML-LRT and B-REML-LRT, and appropriately controlled
the Type I error rates regardless of the type of effect size, the
number of studies, and the study-level sample sizes. Unlike the cases
without covariates, B-ML-LRT could fail to control the Type I error
rates when the study-level sample sizes were large ($SZ=370$). Since
the Monte Carlo standard errors were all smaller than 0.01, the difference
between the methods that could control the Type I error rates and
the methods that failed to control the Type I error rates was not
due to random errors.}

\textcolor{black}{{[}Table 4{]}}{\scriptsize{}}
\begin{table}
{\scriptsize{}\caption{{\footnotesize{}With one covariate, Type I error rates of the bootstrap
based ML LR test (Q-ML-}\textcolor{black}{\footnotesize{}LRT), the
bootstrap based REML LR test (Q-REML-LRT), the bootstrap based Q test
(B-Q), the regular ML based LR test (ML-LRT), the regular REML based
LR test (REML-LRT), and the regular Q test in }{\footnotesize{}mixed-effects
models}}
}{\scriptsize\par}

\resizebox{\textwidth}{!}{
\renewcommand{\arraystretch}{0.09}

{\scriptsize{}}%
\begin{tabular}{cccccccc}
\toprule 
{\scriptsize{}Sample size} & {\scriptsize{}Number of} & {\scriptsize{}B-ML-LRT} & {\scriptsize{}B-REML-LRT} & {\scriptsize{}B-Q} & {\scriptsize{}ML-LRT} & {\scriptsize{}REML-LRT} & {\scriptsize{}Q}\tabularnewline
\cmidrule{3-8} 
{\scriptsize{}(SZ)} & {\scriptsize{}Studies (K)} & \multicolumn{6}{c}{{\scriptsize{}Standardized mean difference}}\tabularnewline
\midrule
\multirow{6}{*}{{\scriptsize{}24}} & {\scriptsize{}10} & {\scriptsize{}0.063 (0.008)} & {\scriptsize{}0.059 (0.007)} & {\scriptsize{}0.059 (0.007)} & \textbf{\scriptsize{}0.015 (0.004)} & {\scriptsize{}0.046 (0.007)} & {\scriptsize{}0.059 (0.007)}\tabularnewline
 & {\scriptsize{}20} & {\scriptsize{}0.06 (0.008)} & {\scriptsize{}0.055 (0.007)} & {\scriptsize{}0.047 (0.007)} & \textbf{\scriptsize{}0.017 (0.004)} & {\scriptsize{}0.04 (0.006)} & {\scriptsize{}0.046 (0.007)}\tabularnewline
 & {\scriptsize{}30} & {\scriptsize{}0.051 (0.007)} & {\scriptsize{}0.05 (0.007)} & {\scriptsize{}0.05 (0.007)} & {\scriptsize{}0.026 (0.005)} & {\scriptsize{}0.042 (0.006)} & {\scriptsize{}0.049 (0.007)}\tabularnewline
 & {\scriptsize{}50} & {\scriptsize{}0.059 (0.007)} & {\scriptsize{}0.059 (0.007)} & {\scriptsize{}0.055 (0.007)} & {\scriptsize{}0.033 (0.006)} & {\scriptsize{}0.045 (0.007)} & {\scriptsize{}0.055 (0.007)}\tabularnewline
 & {\scriptsize{}100} & {\scriptsize{}0.053 (0.007)} & {\scriptsize{}0.052 (0.007)} & {\scriptsize{}0.043 (0.006)} & {\scriptsize{}0.036 (0.006)} & {\scriptsize{}0.045 (0.007)} & {\scriptsize{}0.043 (0.006)}\tabularnewline
 & {\scriptsize{}300} & {\scriptsize{}0.047 (0.007)} & {\scriptsize{}0.047 (0.007)} & {\scriptsize{}0.044 (0.006)} & {\scriptsize{}0.028 (0.005)} & {\scriptsize{}0.042 (0.006)} & {\scriptsize{}0.043 (0.006)}\tabularnewline
\midrule
\multirow{6}{*}{{\scriptsize{}91}} & {\scriptsize{}10} & {\scriptsize{}0.048 (0.007)} & {\scriptsize{}0.05 (0.007)} & {\scriptsize{}0.05 (0.007)} & \textbf{\scriptsize{}0.009 (0.003)} & {\scriptsize{}0.03 (0.005)} & {\scriptsize{}0.05 (0.007)}\tabularnewline
 & {\scriptsize{}20} & {\scriptsize{}0.047 (0.007)} & {\scriptsize{}0.044 (0.006)} & {\scriptsize{}0.039 (0.006)} & \textbf{\scriptsize{}0.016 (0.004)} & {\scriptsize{}0.032 (0.006)} & {\scriptsize{}0.036 (0.006)}\tabularnewline
 & {\scriptsize{}30} & {\scriptsize{}0.055 (0.007)} & {\scriptsize{}0.053 (0.007)} & {\scriptsize{}0.049 (0.007)} & \textbf{\scriptsize{}0.018 (0.004)} & {\scriptsize{}0.039 (0.006)} & {\scriptsize{}0.046 (0.007)}\tabularnewline
 & {\scriptsize{}50} & {\scriptsize{}0.059 (0.007)} & {\scriptsize{}0.059 (0.007)} & {\scriptsize{}0.052 (0.007)} & {\scriptsize{}0.026 (0.005)} & {\scriptsize{}0.046 (0.007)} & {\scriptsize{}0.044 (0.006)}\tabularnewline
 & {\scriptsize{}100} & {\scriptsize{}0.049 (0.007)} & {\scriptsize{}0.048 (0.007)} & {\scriptsize{}0.044 (0.006)} & {\scriptsize{}0.027 (0.005)} & {\scriptsize{}0.035 (0.006)} & {\scriptsize{}0.04 (0.006)}\tabularnewline
 & {\scriptsize{}300} & {\scriptsize{}0.044 (0.006)} & {\scriptsize{}0.044 (0.006)} & {\scriptsize{}0.034 (0.006)} & {\scriptsize{}0.033 (0.006)} & {\scriptsize{}0.042 (0.006)} & {\scriptsize{}0.034 (0.006)}\tabularnewline
\midrule
\multirow{6}{*}{{\scriptsize{}370}} & {\scriptsize{}10} & {\scriptsize{}0.03 (0.005)} & {\scriptsize{}0.053 (0.007)} & {\scriptsize{}0.051 (0.007)} & \textbf{\scriptsize{}0.006 (0.002)} & {\scriptsize{}0.032 (0.006)} & {\scriptsize{}0.048 (0.007)}\tabularnewline
 & {\scriptsize{}20} & {\scriptsize{}0.041 (0.006)} & {\scriptsize{}0.046 (0.007)} & {\scriptsize{}0.05 (0.007)} & \textbf{\scriptsize{}0.008 (0.003)} & {\scriptsize{}0.037 (0.006)} & {\scriptsize{}0.051 (0.007)}\tabularnewline
 & {\scriptsize{}30} & {\scriptsize{}0.058 (0.007)} & {\scriptsize{}0.053 (0.007)} & {\scriptsize{}0.064 (0.008)} & \textbf{\scriptsize{}0.012 (0.003)} & {\scriptsize{}0.035 (0.006)} & {\scriptsize{}0.063 (0.008)}\tabularnewline
 & {\scriptsize{}50} & {\scriptsize{}0.051 (0.007)} & {\scriptsize{}0.052 (0.007)} & {\scriptsize{}0.047 (0.007)} & \textbf{\scriptsize{}0.012 (0.003)} & {\scriptsize{}0.037 (0.006)} & {\scriptsize{}0.045 (0.007)}\tabularnewline
 & {\scriptsize{}100} & {\scriptsize{}0.05 (0.007)} & {\scriptsize{}0.05 (0.007)} & {\scriptsize{}0.051 (0.007)} & {\scriptsize{}0.025 (0.005)} & {\scriptsize{}0.039 (0.006)} & {\scriptsize{}0.051 (0.007)}\tabularnewline
 & {\scriptsize{}300} & {\scriptsize{}0.053 (0.007)} & {\scriptsize{}0.054 (0.007)} & {\scriptsize{}0.052 (0.007)} & {\scriptsize{}0.032 (0.006)} & {\scriptsize{}0.044 (0.006)} & {\scriptsize{}0.048 (0.007)}\tabularnewline
\midrule 
 &  & \multicolumn{6}{c}{{\scriptsize{}Fisher-transformed Pearson correlation}}\tabularnewline
\midrule
\multirow{6}{*}{{\scriptsize{}24}} & {\scriptsize{}10} & {\scriptsize{}0.063 (0.008)} & {\scriptsize{}0.06 (0.008)} & {\scriptsize{}0.058 (0.007)} & \textbf{\scriptsize{}0.015 (0.004)} & {\scriptsize{}0.046 (0.007)} & {\scriptsize{}0.059 (0.007)}\tabularnewline
 & {\scriptsize{}20} & {\scriptsize{}0.06 (0.008)} & {\scriptsize{}0.056 (0.007)} & {\scriptsize{}0.049 (0.007)} & \textbf{\scriptsize{}0.017 (0.004)} & {\scriptsize{}0.04 (0.006)} & {\scriptsize{}0.046 (0.007)}\tabularnewline
 & {\scriptsize{}30} & {\scriptsize{}0.048 (0.007)} & {\scriptsize{}0.048 (0.007)} & {\scriptsize{}0.05 (0.007)} & {\scriptsize{}0.026 (0.005)} & {\scriptsize{}0.042 (0.006)} & {\scriptsize{}0.049 (0.007)}\tabularnewline
 & {\scriptsize{}50} & {\scriptsize{}0.059 (0.007)} & {\scriptsize{}0.058 (0.007)} & {\scriptsize{}0.055 (0.007)} & {\scriptsize{}0.033 (0.006)} & {\scriptsize{}0.045 (0.007)} & {\scriptsize{}0.055 (0.007)}\tabularnewline
 & {\scriptsize{}100} & {\scriptsize{}0.053 (0.007)} & {\scriptsize{}0.053 (0.007)} & {\scriptsize{}0.042 (0.006)} & {\scriptsize{}0.036 (0.006)} & {\scriptsize{}0.045 (0.007)} & {\scriptsize{}0.043 (0.006)}\tabularnewline
 & {\scriptsize{}300} & {\scriptsize{}0.048 (0.007)} & {\scriptsize{}0.048 (0.007)} & {\scriptsize{}0.044 (0.006)} & {\scriptsize{}0.028 (0.005)} & {\scriptsize{}0.042 (0.006)} & {\scriptsize{}0.043 (0.006)}\tabularnewline
\midrule
\multirow{6}{*}{{\scriptsize{}91}} & {\scriptsize{}10} & {\scriptsize{}0.057 (0.007)} & {\scriptsize{}0.06 (0.008)} & {\scriptsize{}0.063 (0.008)} & \textbf{\scriptsize{}0.015 (0.004)} & {\scriptsize{}0.038 (0.006)} & {\scriptsize{}0.062 (0.008)}\tabularnewline
 & {\scriptsize{}20} & {\scriptsize{}0.051 (0.007)} & {\scriptsize{}0.051 (0.007)} & {\scriptsize{}0.056 (0.007)} & \textbf{\scriptsize{}0.016 (0.004)} & {\scriptsize{}0.038 (0.006)} & {\scriptsize{}0.054 (0.007)}\tabularnewline
 & {\scriptsize{}30} & {\scriptsize{}0.049 (0.007)} & {\scriptsize{}0.045 (0.007)} & {\scriptsize{}0.046 (0.007)} & \textbf{\scriptsize{}0.017 (0.004)} & {\scriptsize{}0.038 (0.006)} & {\scriptsize{}0.046 (0.007)}\tabularnewline
 & {\scriptsize{}50} & {\scriptsize{}0.052 (0.007)} & {\scriptsize{}0.053 (0.007)} & {\scriptsize{}0.054 (0.007)} & \textbf{\scriptsize{}0.024 (0.005)} & {\scriptsize{}0.045 (0.007)} & {\scriptsize{}0.056 (0.007)}\tabularnewline
 & {\scriptsize{}100} & {\scriptsize{}0.051 (0.007)} & {\scriptsize{}0.05 (0.007)} & {\scriptsize{}0.048 (0.007)} & {\scriptsize{}0.026 (0.005)} & {\scriptsize{}0.047 (0.007)} & {\scriptsize{}0.047 (0.007)}\tabularnewline
 & {\scriptsize{}300} & {\scriptsize{}0.044 (0.006)} & {\scriptsize{}0.044 (0.006)} & {\scriptsize{}0.033 (0.006)} & {\scriptsize{}0.033 (0.006)} & {\scriptsize{}0.042 (0.006)} & {\scriptsize{}0.034 (0.006)}\tabularnewline
\midrule
\multirow{6}{*}{{\scriptsize{}370}} & {\scriptsize{}10} & {\scriptsize{}0.032 (0.006)} & {\scriptsize{}0.051 (0.007)} & {\scriptsize{}0.06 (0.008)} & \textbf{\scriptsize{}0.004 (0.002)} & {\scriptsize{}0.033 (0.006)} & {\scriptsize{}0.06 (0.008)}\tabularnewline
 & {\scriptsize{}20} & {\scriptsize{}0.048 (0.007)} & {\scriptsize{}0.051 (0.007)} & {\scriptsize{}0.047 (0.007)} & \textbf{\scriptsize{}0.008 (0.003)} & {\scriptsize{}0.03 (0.005)} & {\scriptsize{}0.048 (0.007)}\tabularnewline
 & {\scriptsize{}30} & {\scriptsize{}0.048 (0.007)} & {\scriptsize{}0.047 (0.007)} & {\scriptsize{}0.055 (0.007)} & \textbf{\scriptsize{}0.007 (0.003)} & {\scriptsize{}0.031 (0.005)} & {\scriptsize{}0.057 (0.007)}\tabularnewline
 & {\scriptsize{}50} & {\scriptsize{}0.049 (0.007)} & {\scriptsize{}0.049 (0.007)} & {\scriptsize{}0.047 (0.007)} & \textbf{\scriptsize{}0.013 (0.004)} & {\scriptsize{}0.036 (0.006)} & {\scriptsize{}0.047 (0.007)}\tabularnewline
 & {\scriptsize{}100} & {\scriptsize{}0.049 (0.007)} & {\scriptsize{}0.048 (0.007)} & {\scriptsize{}0.044 (0.006)} & \textbf{\scriptsize{}0.017 (0.004)} & {\scriptsize{}0.037 (0.006)} & {\scriptsize{}0.047 (0.007)}\tabularnewline
 & {\scriptsize{}300} & {\scriptsize{}0.052 (0.007)} & {\scriptsize{}0.053 (0.007)} & {\scriptsize{}0.045 (0.007)} & {\scriptsize{}0.027 (0.005)} & {\scriptsize{}0.042 (0.006)} & {\scriptsize{}0.044 (0.006)}\tabularnewline
\midrule 
 &  & \multicolumn{6}{c}{{\scriptsize{}Log odds ratio}}\tabularnewline
\midrule
\multirow{6}{*}{{\scriptsize{}24}} & {\scriptsize{}10} & {\scriptsize{}0.033 (0.006)} & {\scriptsize{}0.033 (0.006)} & {\scriptsize{}0.04 (0.006)} & \textbf{\scriptsize{}0.005 (0.002)} & \textbf{\scriptsize{}0.013 (0.004)} & \textbf{\scriptsize{}0.018 (0.004)}\tabularnewline
 & {\scriptsize{}20} & {\scriptsize{}0.028 (0.005)} & {\scriptsize{}0.026 (0.005)} & {\scriptsize{}0.033 (0.006)} & \textbf{\scriptsize{}0.009 (0.003)} & \textbf{\scriptsize{}0.014 (0.004)} & \textbf{\scriptsize{}0.017 (0.004)}\tabularnewline
 & {\scriptsize{}30} & {\scriptsize{}0.03 (0.005)} & {\scriptsize{}0.031 (0.005)} & {\scriptsize{}0.045 (0.007)} & \textbf{\scriptsize{}0.008 (0.003)} & \textbf{\scriptsize{}0.014 (0.004)} & \textbf{\scriptsize{}0.012 (0.003)}\tabularnewline
 & {\scriptsize{}50} & {\scriptsize{}0.032 (0.006)} & {\scriptsize{}0.03 (0.005)} & {\scriptsize{}0.033 (0.006)} & \textbf{\scriptsize{}0.009 (0.003)} & \textbf{\scriptsize{}0.016 (0.004)} & \textbf{\scriptsize{}0.015 (0.004)}\tabularnewline
 & {\scriptsize{}100} & \textbf{\scriptsize{}0.017 (0.004)} & \textbf{\scriptsize{}0.016 (0.004)} & {\scriptsize{}0.034 (0.006)} & \textbf{\scriptsize{}0.004 (0.002)} & \textbf{\scriptsize{}0.007 (0.003)} & \textbf{\scriptsize{}0.007 (0.003)}\tabularnewline
 & {\scriptsize{}300} & \textbf{\scriptsize{}0.013 (0.004)} & \textbf{\scriptsize{}0.013 (0.004)} & {\scriptsize{}0.04 (0.006)} & \textbf{\scriptsize{}0 (0<0.001)} & \textbf{\scriptsize{}0.001 (0.001)} & \textbf{\scriptsize{}0.004 (0.002)}\tabularnewline
\midrule
\multirow{6}{*}{{\scriptsize{}91}} & {\scriptsize{}10} & {\scriptsize{}0.046 (0.007)} & {\scriptsize{}0.048 (0.007)} & {\scriptsize{}0.049 (0.007)} & \textbf{\scriptsize{}0.009 (0.003)} & \textbf{\scriptsize{}0.028 (0.005)} & \textbf{\scriptsize{}0.037 (0.006)}\tabularnewline
 & {\scriptsize{}20} & {\scriptsize{}0.04 (0.006)} & {\scriptsize{}0.041 (0.006)} & {\scriptsize{}0.046 (0.007)} & \textbf{\scriptsize{}0.01 (0.003)} & \textbf{\scriptsize{}0.02 (0.004)} & {\scriptsize{}0.036 (0.006)}\tabularnewline
 & {\scriptsize{}30} & {\scriptsize{}0.046 (0.007)} & {\scriptsize{}0.046 (0.007)} & {\scriptsize{}0.051 (0.007)} & \textbf{\scriptsize{}0.021 (0.005)} & {\scriptsize{}0.029 (0.005)} & {\scriptsize{}0.033 (0.006)}\tabularnewline
 & {\scriptsize{}50} & {\scriptsize{}0.044 (0.006)} & {\scriptsize{}0.041 (0.006)} & {\scriptsize{}0.044 (0.006)} & \textbf{\scriptsize{}0.011 (0.003)} & \textbf{\scriptsize{}0.02 (0.004)} & \textbf{\scriptsize{}0.024 (0.005)}\tabularnewline
 & {\scriptsize{}100} & {\scriptsize{}0.036 (0.006)} & {\scriptsize{}0.035 (0.006)} & {\scriptsize{}0.048 (0.007)} & \textbf{\scriptsize{}0.015 (0.004)} & {\scriptsize{}0.026 (0.005)} & {\scriptsize{}0.026 (0.005)}\tabularnewline
 & {\scriptsize{}300} & {\scriptsize{}0.047 (0.007)} & {\scriptsize{}0.046 (0.007)} & {\scriptsize{}0.051 (0.007)} & \textbf{\scriptsize{}0.024 (0.005)} & {\scriptsize{}0.032 (0.006)} & {\scriptsize{}0.025 (0.005)}\tabularnewline
\midrule
\multirow{6}{*}{{\scriptsize{}370}} & {\scriptsize{}10} & {\scriptsize{}0.023 (0.005)} & {\scriptsize{}0.042 (0.006)} & {\scriptsize{}0.039 (0.006)} & \textbf{\scriptsize{}0.002 (0.001)} & \textbf{\scriptsize{}0.02 (0.004)} & {\scriptsize{}0.036 (0.006)}\tabularnewline
 & {\scriptsize{}20} & {\scriptsize{}0.042 (0.006)} & {\scriptsize{}0.044 (0.006)} & {\scriptsize{}0.047 (0.007)} & \textbf{\scriptsize{}0.009 (0.003)} & {\scriptsize{}0.028 (0.005)} & {\scriptsize{}0.041 (0.006)}\tabularnewline
 & {\scriptsize{}30} & {\scriptsize{}0.04 (0.006)} & {\scriptsize{}0.047 (0.007)} & {\scriptsize{}0.05 (0.007)} & \textbf{\scriptsize{}0.01 (0.003)} & {\scriptsize{}0.03 (0.005)} & {\scriptsize{}0.045 (0.007)}\tabularnewline
 & {\scriptsize{}50} & {\scriptsize{}0.052 (0.007)} & {\scriptsize{}0.053 (0.007)} & {\scriptsize{}0.059 (0.007)} & \textbf{\scriptsize{}0.013 (0.004)} & {\scriptsize{}0.043 (0.006)} & {\scriptsize{}0.051 (0.007)}\tabularnewline
 & {\scriptsize{}100} & {\scriptsize{}0.045 (0.007)} & {\scriptsize{}0.044 (0.006)} & {\scriptsize{}0.044 (0.006)} & \textbf{\scriptsize{}0.012 (0.003)} & {\scriptsize{}0.029 (0.005)} & {\scriptsize{}0.036 (0.006)}\tabularnewline
 & {\scriptsize{}300} & {\scriptsize{}0.061 (0.008)} & {\scriptsize{}0.057 (0.007)} & {\scriptsize{}0.047 (0.007)} & {\scriptsize{}0.033 (0.006)} & {\scriptsize{}0.047 (0.007)} & {\scriptsize{}0.026 (0.005)}\tabularnewline
\bottomrule
\end{tabular}{\scriptsize\par}

}

\textcolor{black}{\scriptsize{}Note: Monte Carlo standard errors are
presented in the parentheses. All unacceptable Type}{\scriptsize{}
I error rates (i.e., <0.025 or >0.075) are bold. ML or REML estimation
sometimes did not have converged results, but the nonconvergence rates
across all conditions were within 0.8\%.}{\scriptsize\par}
\end{table}
{\scriptsize\par}

{[}Table 5{]}{\scriptsize{}}
\begin{table}
{\scriptsize{}\caption{{\footnotesize{}With three covariates, Type I error rates of the bootstrap
ML-based LR test (Q}\textcolor{black}{\footnotesize{}-ML-LRT), the
bootstrap REML-based LR test (Q-REML-LRT), the bootstrap Q test (B-Q),
the regular ML-based LR test (ML-LRT), the regular REML-based LR test
(REML-LRT), and the regular Q test i}{\footnotesize{}n mixed-effects
models}}
}{\scriptsize\par}

\resizebox{\textwidth}{!}{
\renewcommand{\arraystretch}{0.09}

{\scriptsize{}}%
\begin{tabular}{cccccccc}
\toprule 
{\scriptsize{}Sample size} & {\scriptsize{}Number of} & {\scriptsize{}B-ML-LRT} & {\scriptsize{}B-REML-LRT} & {\scriptsize{}B-Q} & {\scriptsize{}ML-LRT} & {\scriptsize{}REML-LRT} & {\scriptsize{}Q}\tabularnewline
\cmidrule{3-8} 
{\scriptsize{}(SZ)} & {\scriptsize{}Studies (K)} & \multicolumn{6}{c}{{\scriptsize{}Standardized mean difference}}\tabularnewline
\midrule
\multirow{6}{*}{{\scriptsize{}24}} & {\scriptsize{}10} & {\scriptsize{}0.052 (0.007)} & {\scriptsize{}0.058 (0.007)} & {\scriptsize{}0.054 (0.007)} & \textbf{\scriptsize{}0.002 (0.001)} & {\scriptsize{}0.032 (0.006)} & {\scriptsize{}0.049 (0.007)}\tabularnewline
 & {\scriptsize{}20} & {\scriptsize{}0.037 (0.006)} & {\scriptsize{}0.035 (0.006)} & {\scriptsize{}0.036 (0.006)} & \textbf{\scriptsize{}0.006 (0.002)} & \textbf{\scriptsize{}0.024 (0.005)} & {\scriptsize{}0.031 (0.005)}\tabularnewline
 & {\scriptsize{}30} & {\scriptsize{}0.054 (0.007)} & {\scriptsize{}0.055 (0.007)} & {\scriptsize{}0.052 (0.007)} & \textbf{\scriptsize{}0.008 (0.003)} & {\scriptsize{}0.032 (0.006)} & {\scriptsize{}0.046 (0.007)}\tabularnewline
 & {\scriptsize{}50} & {\scriptsize{}0.045 (0.007)} & {\scriptsize{}0.047 (0.007)} & {\scriptsize{}0.049 (0.007)} & \textbf{\scriptsize{}0.01 (0.003)} & {\scriptsize{}0.029 (0.005)} & {\scriptsize{}0.04 (0.006)}\tabularnewline
 & {\scriptsize{}100} & {\scriptsize{}0.045 (0.007)} & {\scriptsize{}0.043 (0.006)} & {\scriptsize{}0.049 (0.007)} & \textbf{\scriptsize{}0.015 (0.004)} & {\scriptsize{}0.031 (0.005)} & {\scriptsize{}0.041 (0.006)}\tabularnewline
 & {\scriptsize{}300} & {\scriptsize{}0.034 (0.006)} & {\scriptsize{}0.034 (0.006)} & {\scriptsize{}0.039 (0.006)} & \textbf{\scriptsize{}0.012 (0.003)} & \textbf{\scriptsize{}0.019 (0.004)} & {\scriptsize{}0.028 (0.005)}\tabularnewline
\midrule
\multirow{6}{*}{{\scriptsize{}91}} & {\scriptsize{}10} & {\scriptsize{}0.044 (0.006)} & {\scriptsize{}0.05 (0.007)} & {\scriptsize{}0.049 (0.007)} & \textbf{\scriptsize{}0.001 (0.001)} & {\scriptsize{}0.025 (0.005)} & {\scriptsize{}0.048 (0.007)}\tabularnewline
 & {\scriptsize{}20} & {\scriptsize{}0.063 (0.008)} & {\scriptsize{}0.061 (0.008)} & {\scriptsize{}0.058 (0.007)} & \textbf{\scriptsize{}0.007 (0.003)} & {\scriptsize{}0.047 (0.007)} & {\scriptsize{}0.058 (0.007)}\tabularnewline
 & {\scriptsize{}30} & {\scriptsize{}0.054 (0.007)} & {\scriptsize{}0.051 (0.007)} & {\scriptsize{}0.051 (0.007)} & \textbf{\scriptsize{}0.009 (0.003)} & {\scriptsize{}0.038 (0.006)} & {\scriptsize{}0.048 (0.007)}\tabularnewline
 & {\scriptsize{}50} & {\scriptsize{}0.055 (0.007)} & {\scriptsize{}0.054 (0.007)} & {\scriptsize{}0.059 (0.007)} & \textbf{\scriptsize{}0.02 (0.004)} & {\scriptsize{}0.042 (0.006)} & {\scriptsize{}0.054 (0.007)}\tabularnewline
 & {\scriptsize{}100} & {\scriptsize{}0.041 (0.006)} & {\scriptsize{}0.042 (0.006)} & {\scriptsize{}0.044 (0.006)} & \textbf{\scriptsize{}0.014 (0.004)} & {\scriptsize{}0.03 (0.005)} & {\scriptsize{}0.04 (0.006)}\tabularnewline
 & {\scriptsize{}300} & {\scriptsize{}0.053 (0.007)} & {\scriptsize{}0.055 (0.007)} & {\scriptsize{}0.049 (0.007)} & \textbf{\scriptsize{}0.023 (0.005)} & {\scriptsize{}0.047 (0.007)} & {\scriptsize{}0.043 (0.006)}\tabularnewline
\midrule
\multirow{6}{*}{{\scriptsize{}370}} & {\scriptsize{}10} & {\scriptsize{}0.01 (0.003)} & {\scriptsize{}0.063 (0.008)} & {\scriptsize{}0.051 (0.007)} & \textbf{\scriptsize{}0.001 (0.001)} & {\scriptsize{}0.036 (0.006)} & {\scriptsize{}0.053 (0.007)}\tabularnewline
 & {\scriptsize{}20} & {\scriptsize{}0.021 (0.005)} & {\scriptsize{}0.053 (0.007)} & {\scriptsize{}0.047 (0.007)} & \textbf{\scriptsize{}0.002 (0.001)} & {\scriptsize{}0.032 (0.006)} & {\scriptsize{}0.046 (0.007)}\tabularnewline
 & {\scriptsize{}30} & {\scriptsize{}0.046 (0.007)} & {\scriptsize{}0.053 (0.007)} & {\scriptsize{}0.045 (0.007)} & \textbf{\scriptsize{}0.003 (0.002)} & {\scriptsize{}0.034 (0.006)} & {\scriptsize{}0.046 (0.007)}\tabularnewline
 & {\scriptsize{}50} & {\scriptsize{}0.048 (0.007)} & {\scriptsize{}0.051 (0.007)} & {\scriptsize{}0.042 (0.006)} & \textbf{\scriptsize{}0.003 (0.002)} & {\scriptsize{}0.033 (0.006)} & {\scriptsize{}0.038 (0.006)}\tabularnewline
 & {\scriptsize{}100} & {\scriptsize{}0.054 (0.007)} & {\scriptsize{}0.053 (0.007)} & {\scriptsize{}0.056 (0.007)} & \textbf{\scriptsize{}0.011 (0.003)} & {\scriptsize{}0.043 (0.006)} & {\scriptsize{}0.053 (0.007)}\tabularnewline
 & {\scriptsize{}300} & {\scriptsize{}0.049 (0.007)} & {\scriptsize{}0.051 (0.007)} & {\scriptsize{}0.043 (0.006)} & \textbf{\scriptsize{}0.013 (0.004)} & {\scriptsize{}0.039 (0.006)} & {\scriptsize{}0.041 (0.006)}\tabularnewline
\midrule 
 &  & \multicolumn{6}{c}{{\scriptsize{}Fisher-transformed Pearson correlation}}\tabularnewline
\midrule
\multirow{6}{*}{{\scriptsize{}24}} & {\scriptsize{}10} & {\scriptsize{}0.054 (0.007)} & {\scriptsize{}0.057 (0.007)} & {\scriptsize{}0.06 (0.008)} & \textbf{\scriptsize{}0.004 (0.002)} & {\scriptsize{}0.04 (0.006)} & {\scriptsize{}0.061 (0.008)}\tabularnewline
 & {\scriptsize{}20} & {\scriptsize{}0.047 (0.007)} & {\scriptsize{}0.047 (0.007)} & {\scriptsize{}0.045 (0.007)} & \textbf{\scriptsize{}0.006 (0.002)} & {\scriptsize{}0.035 (0.006)} & {\scriptsize{}0.045 (0.007)}\tabularnewline
 & {\scriptsize{}30} & {\scriptsize{}0.059 (0.007)} & {\scriptsize{}0.058 (0.007)} & {\scriptsize{}0.059 (0.007)} & \textbf{\scriptsize{}0.013 (0.004)} & {\scriptsize{}0.047 (0.007)} & {\scriptsize{}0.059 (0.007)}\tabularnewline
 & {\scriptsize{}50} & {\scriptsize{}0.05 (0.007)} & {\scriptsize{}0.051 (0.007)} & {\scriptsize{}0.054 (0.007)} & \textbf{\scriptsize{}0.018 (0.004)} & {\scriptsize{}0.041 (0.006)} & {\scriptsize{}0.054 (0.007)}\tabularnewline
 & {\scriptsize{}100} & {\scriptsize{}0.053 (0.007)} & {\scriptsize{}0.053 (0.007)} & {\scriptsize{}0.049 (0.007)} & \textbf{\scriptsize{}0.021 (0.005)} & {\scriptsize{}0.048 (0.007)} & {\scriptsize{}0.047 (0.007)}\tabularnewline
 & {\scriptsize{}300} & {\scriptsize{}0.05 (0.007)} & {\scriptsize{}0.05 (0.007)} & {\scriptsize{}0.046 (0.007)} & {\scriptsize{}0.04 (0.006)} & {\scriptsize{}0.048 (0.007)} & {\scriptsize{}0.045 (0.007)}\tabularnewline
\midrule
\multirow{6}{*}{{\scriptsize{}91}} & {\scriptsize{}10} & {\scriptsize{}0.044 (0.006)} & {\scriptsize{}0.049 (0.007)} & {\scriptsize{}0.048 (0.007)} & \textbf{\scriptsize{}0.006 (0.002)} & {\scriptsize{}0.033 (0.006)} & {\scriptsize{}0.047 (0.007)}\tabularnewline
 & {\scriptsize{}20} & {\scriptsize{}0.036 (0.006)} & {\scriptsize{}0.035 (0.006)} & {\scriptsize{}0.048 (0.007)} & \textbf{\scriptsize{}0.005 (0.002)} & {\scriptsize{}0.027 (0.005)} & {\scriptsize{}0.046 (0.007)}\tabularnewline
 & {\scriptsize{}30} & {\scriptsize{}0.046 (0.007)} & {\scriptsize{}0.043 (0.006)} & {\scriptsize{}0.049 (0.007)} & \textbf{\scriptsize{}0.007 (0.003)} & {\scriptsize{}0.037 (0.006)} & {\scriptsize{}0.046 (0.007)}\tabularnewline
 & {\scriptsize{}50} & {\scriptsize{}0.049 (0.007)} & {\scriptsize{}0.047 (0.007)} & {\scriptsize{}0.059 (0.007)} & \textbf{\scriptsize{}0.013 (0.004)} & {\scriptsize{}0.034 (0.006)} & {\scriptsize{}0.059 (0.007)}\tabularnewline
 & {\scriptsize{}100} & {\scriptsize{}0.047 (0.007)} & {\scriptsize{}0.046 (0.007)} & {\scriptsize{}0.046 (0.007)} & \textbf{\scriptsize{}0.017 (0.004)} & {\scriptsize{}0.044 (0.006)} & {\scriptsize{}0.046 (0.007)}\tabularnewline
 & {\scriptsize{}300} & {\scriptsize{}0.053 (0.007)} & {\scriptsize{}0.053 (0.007)} & {\scriptsize{}0.055 (0.007)} & {\scriptsize{}0.038 (0.006)} & {\scriptsize{}0.049 (0.007)} & {\scriptsize{}0.054 (0.007)}\tabularnewline
\midrule
\multirow{6}{*}{{\scriptsize{}370}} & {\scriptsize{}10} & \textbf{\scriptsize{}0.009 (0.003)} & {\scriptsize{}0.052 (0.007)} & {\scriptsize{}0.048 (0.007)} & \textbf{\scriptsize{}0.001 (0.001)} & \textbf{\scriptsize{}0.023 (0.005)} & {\scriptsize{}0.048 (0.007)}\tabularnewline
 & {\scriptsize{}20} & \textbf{\scriptsize{}0.021 (0.005)} & {\scriptsize{}0.051 (0.007)} & {\scriptsize{}0.049 (0.007)} & \textbf{\scriptsize{}0.001 (0.001)} & {\scriptsize{}0.034 (0.006)} & {\scriptsize{}0.049 (0.007)}\tabularnewline
 & {\scriptsize{}30} & {\scriptsize{}0.037 (0.006)} & {\scriptsize{}0.055 (0.007)} & {\scriptsize{}0.047 (0.007)} & \textbf{\scriptsize{}0.003 (0.002)} & {\scriptsize{}0.035 (0.006)} & {\scriptsize{}0.047 (0.007)}\tabularnewline
 & {\scriptsize{}50} & {\scriptsize{}0.041 (0.006)} & {\scriptsize{}0.046 (0.007)} & {\scriptsize{}0.054 (0.007)} & \textbf{\scriptsize{}0.004 (0.002)} & {\scriptsize{}0.026 (0.005)} & {\scriptsize{}0.054 (0.007)}\tabularnewline
 & {\scriptsize{}100} & {\scriptsize{}0.059 (0.007)} & {\scriptsize{}0.055 (0.007)} & {\scriptsize{}0.044 (0.006)} & \textbf{\scriptsize{}0.012 (0.003)} & {\scriptsize{}0.041 (0.006)} & {\scriptsize{}0.045 (0.007)}\tabularnewline
 & {\scriptsize{}300} & {\scriptsize{}0.044 (0.006)} & {\scriptsize{}0.044 (0.006)} & {\scriptsize{}0.047 (0.007)} & \textbf{\scriptsize{}0.02 (0.004)} & {\scriptsize{}0.037 (0.006)} & {\scriptsize{}0.048 (0.007)}\tabularnewline
\midrule 
 &  & \multicolumn{6}{c}{{\scriptsize{}Log odds ratio}}\tabularnewline
\midrule
\multirow{6}{*}{{\scriptsize{}24}} & {\scriptsize{}10} & {\scriptsize{}0.038 (0.006)} & {\scriptsize{}0.035 (0.006)} & {\scriptsize{}0.035 (0.006)} & \textbf{\scriptsize{}0.002 (0.001)} & \textbf{\scriptsize{}0.017 (0.004)} & {\scriptsize{}0.027 (0.005)}\tabularnewline
 & {\scriptsize{}20} & {\scriptsize{}0.029 (0.005)} & {\scriptsize{}0.028 (0.005)} & {\scriptsize{}0.03 (0.005)} & \textbf{\scriptsize{}0.001 (0.001)} & \textbf{\scriptsize{}0.014 (0.004)} & \textbf{\scriptsize{}0.016 (0.004)}\tabularnewline
 & {\scriptsize{}30} & \textbf{\scriptsize{}0.022 (0.005)} & \textbf{\scriptsize{}0.023 (0.005)} & {\scriptsize{}0.032 (0.006)} & \textbf{\scriptsize{}0.001 (0.001)} & \textbf{\scriptsize{}0.007 (0.003)} & \textbf{\scriptsize{}0.009 (0.003)}\tabularnewline
 & {\scriptsize{}50} & \textbf{\scriptsize{}0.021 (0.005)} & \textbf{\scriptsize{}0.02 (0.004)} & {\scriptsize{}0.028 (0.005)} & \textbf{\scriptsize{}0.001 (0.001)} & \textbf{\scriptsize{}0.006 (0.002)} & \textbf{\scriptsize{}0.012 (0.003)}\tabularnewline
 & {\scriptsize{}100} & \textbf{\scriptsize{}0.017 (0.004)} & \textbf{\scriptsize{}0.018 (0.004)} & {\scriptsize{}0.034 (0.006)} & \textbf{\scriptsize{}0 (0<0.001)} & \textbf{\scriptsize{}0.001 (0.001)} & \textbf{\scriptsize{}0.005 (0.002)}\tabularnewline
 & {\scriptsize{}300} & \textbf{\scriptsize{}0.012 (0.003)} & \textbf{\scriptsize{}0.012 (0.003)} & {\scriptsize{}0.033 (0.006)} & \textbf{\scriptsize{}0 (0<0.001)} & \textbf{\scriptsize{}0 (0<0.001)} & \textbf{\scriptsize{}0 (0<0.001)}\tabularnewline
\midrule
\multirow{6}{*}{{\scriptsize{}91}} & {\scriptsize{}10} & {\scriptsize{}0.039 (0.006)} & {\scriptsize{}0.044 (0.006)} & {\scriptsize{}0.042 (0.006)} & \textbf{\scriptsize{}0.001 (0.001)} & \textbf{\scriptsize{}0.024 (0.005)} & {\scriptsize{}0.037 (0.006)}\tabularnewline
 & {\scriptsize{}20} & {\scriptsize{}0.049 (0.007)} & {\scriptsize{}0.048 (0.007)} & {\scriptsize{}0.042 (0.006)} & \textbf{\scriptsize{}0.005 (0.002)} & {\scriptsize{}0.034 (0.006)} & {\scriptsize{}0.035 (0.006)}\tabularnewline
 & {\scriptsize{}30} & {\scriptsize{}0.043 (0.006)} & {\scriptsize{}0.041 (0.006)} & {\scriptsize{}0.046 (0.007)} & \textbf{\scriptsize{}0.004 (0.002)} & \textbf{\scriptsize{}0.024 (0.005)} & {\scriptsize{}0.034 (0.006)}\tabularnewline
 & {\scriptsize{}50} & {\scriptsize{}0.048 (0.007)} & {\scriptsize{}0.05 (0.007)} & {\scriptsize{}0.054 (0.007)} & \textbf{\scriptsize{}0.01 (0.003)} & {\scriptsize{}0.029 (0.005)} & {\scriptsize{}0.033 (0.006)}\tabularnewline
 & {\scriptsize{}100} & {\scriptsize{}0.042 (0.006)} & {\scriptsize{}0.041 (0.006)} & {\scriptsize{}0.043 (0.006)} & \textbf{\scriptsize{}0.013 (0.004)} & {\scriptsize{}0.028 (0.005)} & {\scriptsize{}0.024 (0.005)}\tabularnewline
 & {\scriptsize{}300} & {\scriptsize{}0.035 (0.006)} & {\scriptsize{}0.034 (0.006)} & {\scriptsize{}0.04 (0.006)} & \textbf{\scriptsize{}0.007 (0.003)} & \textbf{\scriptsize{}0.013 (0.004)} & \textbf{\scriptsize{}0.013 (0.004)}\tabularnewline
\midrule
\multirow{6}{*}{{\scriptsize{}370}} & {\scriptsize{}10} & \textbf{\scriptsize{}0.009 (0.003)} & {\scriptsize{}0.052 (0.007)} & {\scriptsize{}0.049 (0.007)} & \textbf{\scriptsize{}0.001 (0.001)} & \textbf{\scriptsize{}0.022 (0.005)} & {\scriptsize{}0.043 (0.006)}\tabularnewline
 & {\scriptsize{}20} & \textbf{\scriptsize{}0.019 (0.004)} & {\scriptsize{}0.046 (0.007)} & {\scriptsize{}0.051 (0.007)} & \textbf{\scriptsize{}0 (0<0.001)} & {\scriptsize{}0.034 (0.006)} & {\scriptsize{}0.046 (0.007)}\tabularnewline
 & {\scriptsize{}30} & {\scriptsize{}0.033 (0.006)} & {\scriptsize{}0.046 (0.007)} & {\scriptsize{}0.044 (0.006)} & \textbf{\scriptsize{}0.004 (0.002)} & {\scriptsize{}0.029 (0.005)} & {\scriptsize{}0.036 (0.006)}\tabularnewline
 & {\scriptsize{}50} & {\scriptsize{}0.045 (0.007)} & {\scriptsize{}0.045 (0.007)} & {\scriptsize{}0.047 (0.007)} & \textbf{\scriptsize{}0.004 (0.002)} & {\scriptsize{}0.034 (0.006)} & {\scriptsize{}0.04 (0.006)}\tabularnewline
 & {\scriptsize{}100} & {\scriptsize{}0.042 (0.006)} & {\scriptsize{}0.041 (0.006)} & {\scriptsize{}0.049 (0.007)} & \textbf{\scriptsize{}0.007 (0.003)} & {\scriptsize{}0.032 (0.006)} & {\scriptsize{}0.036 (0.006)}\tabularnewline
 & {\scriptsize{}300} & {\scriptsize{}0.065 (0.008)} & {\scriptsize{}0.063 (0.008)} & {\scriptsize{}0.059 (0.007)} & \textbf{\scriptsize{}0.02 (0.004)} & {\scriptsize{}0.052 (0.007)} & {\scriptsize{}0.042 (0.006)}\tabularnewline
\bottomrule
\end{tabular}{\scriptsize\par}

}

{\scriptsize{}Note}\textcolor{black}{\scriptsize{}: Monte Carlo standard
errors are presented in the parentheses. All unac}{\scriptsize{}ceptable
Type I error rates (i.e., <0.025 or >0.075) are bold. ML or REML estimation
sometimes did not have converged results, but the nonconvergence rates
across all conditions were within 0.8\%.}{\scriptsize\par}
\end{table}
{\scriptsize\par}

\subsubsection{\textcolor{black}{5.2.3 Type I error rates in meta-analysis for heterogeneity
magnitude test}}

\textcolor{black}{We explore the Type I error rates of the proposed
methods in testing whether the heterogeneity is larger than a specific
level ($\tau^{2}=\lambda$ verses $\tau^{2}>\lambda$) in this section.
The null hypothesis was $H_{0}:\tau^{2}=\lambda$ and $\lambda$ was
specified as the small level of heterogeneity in Table 2 to mimic
real data. The Type I error rates of B-ML-LRT, B-REML-LRT, and B-Q
are presented in Table 6 with their corresponding Monte Carlo standard
errors. B-ML-LRT and B-REML-LRT could appropriately control the Type
I error rates, except when the study-level sample sizes were small
($SZ=24$) and the effect size was the log odds ratio. B-Q could appropriately
control the Type I error rates across all conditions.}

\textcolor{black}{{[}Table 6{]}}{\scriptsize{}}
\begin{table}
{\scriptsize{}\caption{{\footnotesize{}Type I error rates of the bootstrap based ML LR test
(Q-ML-LRT), the bootstrap based REML LR test (Q-RE}\textcolor{black}{\footnotesize{}ML-LRT),
and the bootstrap based Q test (B-Q) in}{\footnotesize{} meta-analyses
for heterogeneity magnitude test}}
}{\scriptsize\par}

\resizebox{\textwidth}{!}{
\renewcommand{\arraystretch}{0.02}

\textcolor{blue}{\scriptsize{}}%
\begin{tabular}{ccccc}
\toprule 
\textcolor{black}{\scriptsize{}Sample size} & \textcolor{black}{\scriptsize{}Number of} & \textcolor{black}{\scriptsize{}B-ML-LRT} & \textcolor{black}{\scriptsize{}B-REML-LRT} & \textcolor{black}{\scriptsize{}B-Q}\tabularnewline
\cmidrule{3-5} 
\textcolor{black}{\scriptsize{}(SZ)} & \textcolor{black}{\scriptsize{}Studies (K)} & \multicolumn{3}{c}{\textcolor{black}{\scriptsize{}Standardized mean difference}}\tabularnewline
\midrule
\multirow{6}{*}{\textcolor{black}{\scriptsize{}24}} & \textcolor{black}{\scriptsize{}10} & \textcolor{black}{\scriptsize{}0.049 (0.007)} & \textcolor{black}{\scriptsize{}0.048 (0.007)} & \textcolor{black}{\scriptsize{}0.048 (0.007)}\tabularnewline
 & \textcolor{black}{\scriptsize{}20} & \textcolor{black}{\scriptsize{}0.047 (0.007)} & \textcolor{black}{\scriptsize{}0.045 (0.007)} & \textcolor{black}{\scriptsize{}0.048 (0.007)}\tabularnewline
 & \textcolor{black}{\scriptsize{}30} & \textcolor{black}{\scriptsize{}0.057 (0.007)} & \textcolor{black}{\scriptsize{}0.059 (0.007)} & \textcolor{black}{\scriptsize{}0.058 (0.007)}\tabularnewline
 & \textcolor{black}{\scriptsize{}50} & \textcolor{black}{\scriptsize{}0.05 (0.007)} & \textcolor{black}{\scriptsize{}0.05 (0.007)} & \textcolor{black}{\scriptsize{}0.058 (0.007)}\tabularnewline
 & \textcolor{black}{\scriptsize{}100} & \textcolor{black}{\scriptsize{}0.047 (0.007)} & \textcolor{black}{\scriptsize{}0.048 (0.007)} & \textcolor{black}{\scriptsize{}0.048 (0.007)}\tabularnewline
 & \textcolor{black}{\scriptsize{}300} & \textcolor{black}{\scriptsize{}0.056 (0.007)} & \textcolor{black}{\scriptsize{}0.056 (0.007)} & \textcolor{black}{\scriptsize{}0.055 (0.007)}\tabularnewline
\midrule
\multirow{6}{*}{\textcolor{black}{\scriptsize{}91}} & \textcolor{black}{\scriptsize{}10} & \textcolor{black}{\scriptsize{}0.038 (0.006)} & \textcolor{black}{\scriptsize{}0.039 (0.006)} & \textcolor{black}{\scriptsize{}0.04 (0.006)}\tabularnewline
 & \textcolor{black}{\scriptsize{}20} & \textcolor{black}{\scriptsize{}0.053 (0.007)} & \textcolor{black}{\scriptsize{}0.052 (0.007)} & \textcolor{black}{\scriptsize{}0.057 (0.007)}\tabularnewline
 & \textcolor{black}{\scriptsize{}30} & \textcolor{black}{\scriptsize{}0.045 (0.007)} & \textcolor{black}{\scriptsize{}0.044 (0.006)} & \textcolor{black}{\scriptsize{}0.039 (0.006)}\tabularnewline
 & \textcolor{black}{\scriptsize{}50} & \textcolor{black}{\scriptsize{}0.054 (0.007)} & \textcolor{black}{\scriptsize{}0.055 (0.007)} & \textcolor{black}{\scriptsize{}0.046 (0.007)}\tabularnewline
 & \textcolor{black}{\scriptsize{}100} & \textcolor{black}{\scriptsize{}0.061 (0.008)} & \textcolor{black}{\scriptsize{}0.062 (0.008)} & \textcolor{black}{\scriptsize{}0.067 (0.008)}\tabularnewline
 & \textcolor{black}{\scriptsize{}300} & \textcolor{black}{\scriptsize{}0.043 (0.006)} & \textcolor{black}{\scriptsize{}0.043 (0.006)} & \textcolor{black}{\scriptsize{}0.053 (0.007)}\tabularnewline
\midrule
 &  & \multicolumn{3}{c}{\textcolor{black}{\scriptsize{}Fisher-transformed Pearson correlation}}\tabularnewline
\midrule
\multirow{6}{*}{\textcolor{black}{\scriptsize{}24}} & \textcolor{black}{\scriptsize{}10} & \textcolor{black}{\scriptsize{}0.045 (0.007)} & \textcolor{black}{\scriptsize{}0.044 (0.006)} & \textcolor{black}{\scriptsize{}0.041 (0.006)}\tabularnewline
 & \textcolor{black}{\scriptsize{}20} & \textcolor{black}{\scriptsize{}0.061 (0.008)} & \textcolor{black}{\scriptsize{}0.062 (0.008)} & \textcolor{black}{\scriptsize{}0.058 (0.007)}\tabularnewline
 & \textcolor{black}{\scriptsize{}30} & \textcolor{black}{\scriptsize{}0.067 (0.008)} & \textcolor{black}{\scriptsize{}0.066 (0.008)} & \textcolor{black}{\scriptsize{}0.059 (0.007)}\tabularnewline
 & \textcolor{black}{\scriptsize{}50} & \textcolor{black}{\scriptsize{}0.046 (0.007)} & \textcolor{black}{\scriptsize{}0.045 (0.007)} & \textcolor{black}{\scriptsize{}0.051 (0.007)}\tabularnewline
 & \textcolor{black}{\scriptsize{}100} & \textcolor{black}{\scriptsize{}0.059 (0.007)} & \textcolor{black}{\scriptsize{}0.058 (0.007)} & \textcolor{black}{\scriptsize{}0.06 (0.008)}\tabularnewline
 & \textcolor{black}{\scriptsize{}300} & \textcolor{black}{\scriptsize{}0.053 (0.007)} & \textcolor{black}{\scriptsize{}0.053 (0.007)} & \textcolor{black}{\scriptsize{}0.042 (0.006)}\tabularnewline
\midrule
\multirow{6}{*}{\textcolor{black}{\scriptsize{}91}} & \textcolor{black}{\scriptsize{}10} & \textcolor{black}{\scriptsize{}0.05 (0.007)} & \textcolor{black}{\scriptsize{}0.051 (0.007)} & \textcolor{black}{\scriptsize{}0.049 (0.007)}\tabularnewline
 & \textcolor{black}{\scriptsize{}20} & \textcolor{black}{\scriptsize{}0.065 (0.008)} & \textcolor{black}{\scriptsize{}0.064 (0.008)} & \textcolor{black}{\scriptsize{}0.064 (0.008)}\tabularnewline
 & \textcolor{black}{\scriptsize{}30} & \textcolor{black}{\scriptsize{}0.04 (0.006)} & \textcolor{black}{\scriptsize{}0.039 (0.006)} & \textcolor{black}{\scriptsize{}0.039 (0.006)}\tabularnewline
 & \textcolor{black}{\scriptsize{}50} & \textcolor{black}{\scriptsize{}0.059 (0.007)} & \textcolor{black}{\scriptsize{}0.059 (0.007)} & \textcolor{black}{\scriptsize{}0.062 (0.008)}\tabularnewline
 & \textcolor{black}{\scriptsize{}100} & \textcolor{black}{\scriptsize{}0.055 (0.007)} & \textcolor{black}{\scriptsize{}0.055 (0.007)} & \textcolor{black}{\scriptsize{}0.054 (0.007)}\tabularnewline
 & \textcolor{black}{\scriptsize{}300} & \textcolor{black}{\scriptsize{}0.071 (0.008)} & \textcolor{black}{\scriptsize{}0.071 (0.008)} & \textcolor{black}{\scriptsize{}0.069 (0.008)}\tabularnewline
\midrule
 &  & \multicolumn{3}{c}{\textcolor{black}{\scriptsize{}Log odds ratio}}\tabularnewline
\midrule
\multirow{6}{*}{\textcolor{black}{\scriptsize{}24}} & \textcolor{black}{\scriptsize{}10} & \textcolor{black}{\scriptsize{}0.04 (0.006)} & \textcolor{black}{\scriptsize{}0.041 (0.006)} & \textcolor{black}{\scriptsize{}0.046 (0.007)}\tabularnewline
 & \textcolor{black}{\scriptsize{}20} & \textcolor{black}{\scriptsize{}0.025 (0.005)} & \textcolor{black}{\scriptsize{}0.026 (0.005)} & \textcolor{black}{\scriptsize{}0.034 (0.006)}\tabularnewline
 & \textcolor{black}{\scriptsize{}30} & \textcolor{black}{\scriptsize{}0.034 (0.006)} & \textcolor{black}{\scriptsize{}0.033 (0.006)} & \textcolor{black}{\scriptsize{}0.039 (0.006)}\tabularnewline
 & \textcolor{black}{\scriptsize{}50} & \textcolor{black}{\scriptsize{}0.025 (0.005)} & \textcolor{black}{\scriptsize{}0.025 (0.005)} & \textcolor{black}{\scriptsize{}0.035 (0.006)}\tabularnewline
 & \textcolor{black}{\scriptsize{}100} & \textcolor{black}{\scriptsize{}0.031 (0.005)} & \textcolor{black}{\scriptsize{}0.03 (0.005)} & \textcolor{black}{\scriptsize{}0.051 (0.007)}\tabularnewline
 & \textcolor{black}{\scriptsize{}300} & \textbf{\textcolor{black}{\scriptsize{}0.018 (0.004)}} & \textbf{\textcolor{black}{\scriptsize{}0.018 (0.004)}} & \textcolor{black}{\scriptsize{}0.065 (0.008)}\tabularnewline
\midrule
\multirow{6}{*}{\textcolor{black}{\scriptsize{}91}} & \textcolor{black}{\scriptsize{}10} & \textcolor{black}{\scriptsize{}0.051 (0.007)} & \textcolor{black}{\scriptsize{}0.05 (0.007)} & \textcolor{black}{\scriptsize{}0.06 (0.008)}\tabularnewline
 & \textcolor{black}{\scriptsize{}20} & \textcolor{black}{\scriptsize{}0.061 (0.008)} & \textcolor{black}{\scriptsize{}0.062 (0.008)} & \textcolor{black}{\scriptsize{}0.057 (0.007)}\tabularnewline
 & \textcolor{black}{\scriptsize{}30} & \textcolor{black}{\scriptsize{}0.058 (0.007)} & \textcolor{black}{\scriptsize{}0.058 (0.007)} & \textcolor{black}{\scriptsize{}0.058 (0.007)}\tabularnewline
 & \textcolor{black}{\scriptsize{}50} & \textcolor{black}{\scriptsize{}0.043 (0.006)} & \textcolor{black}{\scriptsize{}0.043 (0.006)} & \textcolor{black}{\scriptsize{}0.044 (0.006)}\tabularnewline
 & \textcolor{black}{\scriptsize{}100} & \textcolor{black}{\scriptsize{}0.047 (0.007)} & \textcolor{black}{\scriptsize{}0.047 (0.007)} & \textcolor{black}{\scriptsize{}0.05 (0.007)}\tabularnewline
 & \textcolor{black}{\scriptsize{}300} & \textcolor{black}{\scriptsize{}0.047 (0.007)} & \textcolor{black}{\scriptsize{}0.047 (0.007)} & \textcolor{black}{\scriptsize{}0.049 (0.007)}\tabularnewline
\bottomrule
\end{tabular}{\scriptsize\par}

}

{\scriptsize{}Note: Monte Carlo standard errors are presented in the
parentheses. All unacceptable Type I error rates (i.e., <0.025 or
>0.075) are bold. ML or REML estimation sometimes did not have converged
results, but the nonconvergence rates across all conditions were within
0.3\%.}{\scriptsize\par}
\end{table}
{\scriptsize\par}

\subsubsection{5.2.4 Summary of the performance of the tests on Type I error rates}

\textcolor{black}{In sum, across different conditions and types of
effect sizes, although the performance of REML-LRT and Q test was
better than ML-LRT, the three methods could be too conservative in
the examined conditions and they are not applicable in the heterogeneity
magnitude test. Without covariates, the improved Q test and the Breslow-Day's
test could appropriately control the Type I error rates in the heterogeneity
test but they are not applicable in the heterogeneity magnitude test.
The bootstrap methods are applicable in both the heterogeneity test
and the heterogeneity magnitude test. Among the three bootstrap methods,
B-Q was superior to the other two, and B-REML-LRT could appropriately
control Type I error rates except when the study-level sample sizes
were small ($SZ=24$) and the effect size was the log odds ratio.
B-Q was competitive with the improved Q test and the Breslow-Day test
when the improved Q test and the Breslow-Day test were applicable.}

\subsection{5.3 Statis\textcolor{black}{tical power}}

\textcolor{black}{REML-LRT and ML-LRT will not be discussed further
due to their failure to control Type I error rates. We varied the
between-study heterogeneity ($\tau^{2}$) as mentioned above (see
Table 2 for the small, medium, and large levels of heterogeneity)
and considered the same simulation conditions as in examining Type
I error rates, except the cases where the power values from B-ML-LRT,
B-REML-LRT, B-Q, the improved Q test, the Breslow-Day test, and the
Q test were all nearly 1. }

\subsubsection{\textcolor{black}{5.3.1 Statistical power in meta-analyses for heterogeneity
test}}

\textcolor{black}{In random-effects meta-analyses, the statistical
power values of B-ML-LRT, B-REML-LRT, B-Q, the improved Q test ($Q_{2}$),
the Breslow-Day test, and the Q test are presented in Table 7, with
the highest power values under each condition bolded and the Monte
Carlo standard errors inside the parentheses. More studies, larger
study-level sample sizes, and/or larger $\tau^{2}$ increased power,
which supported the conclusion from \citet{viechtbauer2007hypothesis},
but not \citet{harwell1997empirical}. As a parametric method, the
improved Q test was more powerful than B-REML-LRT and B-ML-LRT in
a number of conditions. In terms of the proposed bootstrap based methods,
when the effect of interest was the standardized mean difference,
compared to the Q test, the maximum power increments of B-ML-LRT,
B-REML-LRT, and B-Q compared to the regular Q test were 0.117, 0.119,
and 0.029 respectively, which were much larger than the Monte Carlo
standard errors. The maximum percentage increments of B-ML-LRT, B-REML-LRT,
and B-Q was 34.0\%, 31.6\%, and 7.5\% respectively. When the effect
size was the Fisher-transformed Pearson correlation, the maximum power
increments of B-ML-LRT, B-REML-LRT, and B-Q were 0.076, 0.076, and
0.003 respectively and the maximum percentage increments of B-ML-LRT,
B-REML-LRT, and B-Q was 17.6\%, 17.6\%, and 1\% respectively. B-Q
barely boosted power, whereas B-ML-LRT and B-REML-LRT boosted power
based on the increased percentages and Monte Carlo standard errors.
When the effect size was the log odds ratio, the maximum power increments
of B-ML-LRT, B-REML-LRT, and B-Q were 0.148, 0.149, and 0.203 respectively,
which were much larger than the Monte Carlo standard errors, and the
maximum percentage increments of B-ML-LRT, B-REML-LRT, and B-Q were
67.1\%, 66.4\%, and 81.9\% respectively. }

\textcolor{black}{{[}Table 7{]}}{\scriptsize{}}
\begin{table}
{\scriptsize{}\caption{{\footnotesize{}Statistical power of the bootstrap based ML LR test
(Q-ML-LRT), the bootstrap based REML LR test}\textcolor{black}{\footnotesize{}
(Q-REML-LRT), the bootstrap based Q test (B-Q), the improved Q test
($Q_{2}$), the Breslow-Day's test (BD), and the regular Q test in
meta-analyses}}
}{\scriptsize\par}

\resizebox{\textwidth}{!}{
\renewcommand{\arraystretch}{0.7}

{\scriptsize{}}%
\begin{tabular}{ccccccccc|cccccc}
\hline 
 &  & \textcolor{black}{\scriptsize{}Number of} & \multirow{2}{*}{\textcolor{black}{\scriptsize{}B-ML-LRT}} & \multirow{2}{*}{\textcolor{black}{\scriptsize{}B-REML-LRT}} & \multirow{2}{*}{\textcolor{black}{\scriptsize{}B-Q}} & \multirow{2}{*}{\textcolor{black}{\scriptsize{}$Q_{2}$}} & \multirow{2}{*}{\textcolor{black}{\scriptsize{}BD}} & \multirow{2}{*}{\textcolor{black}{\scriptsize{}Q}} & \multirow{2}{*}{\textcolor{black}{\scriptsize{}B-ML-LRT}} & \multirow{2}{*}{\textcolor{black}{\scriptsize{}B-REML-LRT}} & \multirow{2}{*}{\textcolor{black}{\scriptsize{}B-Q}} & \multirow{2}{*}{\textcolor{black}{\scriptsize{}$Q_{2}$}} & \multirow{2}{*}{\textcolor{black}{\scriptsize{}BD}} & \multirow{2}{*}{\textcolor{black}{\scriptsize{}Q}}\tabularnewline
 &  & \textcolor{black}{\scriptsize{}Studies} &  &  &  &  &  &  &  &  &  &  &  & \tabularnewline
\hline 
 & \multirow{9}{*}{\textcolor{black}{\scriptsize{}SZ=24}} &  & \multicolumn{6}{c|}{\textcolor{black}{\scriptsize{}$\tau^{2}=0.03$}} & \multicolumn{6}{c}{\textcolor{black}{\scriptsize{}$\tau^{2}=0.1$}}\tabularnewline
\cline{3-15} 
 &  & \textcolor{black}{\scriptsize{}10} & \textcolor{black}{\scriptsize{}0.174 (0.012)} & \textcolor{black}{\scriptsize{}0.174 (0.012)} & \textcolor{black}{\scriptsize{}0.171 (0.012)} & \textbf{\textcolor{black}{\scriptsize{}0.214 (0.013)}} & \textcolor{black}{\scriptsize{}-} & \textcolor{black}{\scriptsize{}0.147 (0.011)} & \textcolor{black}{\scriptsize{}0.542 (0.016)} & \textcolor{black}{\scriptsize{}0.541 (0.016)} & \textbf{\textcolor{black}{\scriptsize{}0.543 (0.016)}} & \textcolor{black}{\scriptsize{}0.53 (0.016)} & \textcolor{black}{\scriptsize{}-} & \textcolor{black}{\scriptsize{}0.514 (0.016)}\tabularnewline
 &  & \textcolor{black}{\scriptsize{}20} & \textcolor{black}{\scriptsize{}0.304 (0.015)} & \textcolor{black}{\scriptsize{}0.305 (0.015)} & \textcolor{black}{\scriptsize{}0.286 (0.014)} & \textbf{\textcolor{black}{\scriptsize{}0.341 (0.015)}} & \textcolor{black}{\scriptsize{}-} & \textcolor{black}{\scriptsize{}0.248 (0.014)} & \textcolor{black}{\scriptsize{}0.795 (0.013)} & \textbf{\textcolor{black}{\scriptsize{}0.796 (0.013)}} & \textcolor{black}{\scriptsize{}0.791 (0.013)} & \textcolor{black}{\scriptsize{}0.78 (0.013)} & \textcolor{black}{\scriptsize{}-} & \textcolor{black}{\scriptsize{}0.766 (0.013)}\tabularnewline
 &  & \textcolor{black}{\scriptsize{}30} & \textcolor{black}{\scriptsize{}0.342 (0.015)} & \textcolor{black}{\scriptsize{}0.342 (0.015)} & \textcolor{black}{\scriptsize{}0.326 (0.015)} & \textbf{\textcolor{black}{\scriptsize{}0.39 (0.015)}} & \textcolor{black}{\scriptsize{}-} & \textcolor{black}{\scriptsize{}0.279 (0.014)} & \textcolor{black}{\scriptsize{}0.885 (0.01)} & \textbf{\textcolor{black}{\scriptsize{}0.886 (0.01)}} & \textcolor{black}{\scriptsize{}0.881 (0.01)} & \textcolor{black}{\scriptsize{}0.875 (0.01)} & \textcolor{black}{\scriptsize{}-} & \textcolor{black}{\scriptsize{}0.852 (0.011)}\tabularnewline
 &  & \textcolor{black}{\scriptsize{}50} & \textcolor{black}{\scriptsize{}0.503 (0.016)} & \textcolor{black}{\scriptsize{}0.507 (0.016)} & \textcolor{black}{\scriptsize{}0.493 (0.016)} & \textbf{\textcolor{black}{\scriptsize{}0.56 (0.016)}} & \textcolor{black}{\scriptsize{}-} & \textcolor{black}{\scriptsize{}0.42 (0.016)} & \textcolor{black}{\scriptsize{}0.981 (0.004)} & \textcolor{black}{\scriptsize{}0.981 (0.004)} & \textbf{\textcolor{black}{\scriptsize{}0.984 (0.004)}} & \textcolor{black}{\scriptsize{}0.982 (0.004)} & \textcolor{black}{\scriptsize{}-} & \textcolor{black}{\scriptsize{}0.975 (0.005)}\tabularnewline
\textcolor{black}{\scriptsize{}Standardized } &  & \textcolor{black}{\scriptsize{}100} & \textcolor{black}{\scriptsize{}0.742 (0.014)} & \textcolor{black}{\scriptsize{}0.743 (0.014)} & \textcolor{black}{\scriptsize{}0.706 (0.014)} & \textbf{\textcolor{black}{\scriptsize{}0.794 (0.013)}} & \textcolor{black}{\scriptsize{}-} & \textcolor{black}{\scriptsize{}0.625 (0.015)} &  &  &  &  &  & \tabularnewline
\cline{3-15} 
\textcolor{black}{\scriptsize{}mean } &  &  & \multicolumn{6}{c|}{\textcolor{black}{\scriptsize{}$\tau^{2}=0.3$}} & \multicolumn{6}{c}{}\tabularnewline
\cline{3-15} 
\textcolor{black}{\scriptsize{}difference} &  & \textcolor{black}{\scriptsize{}10} & \textcolor{black}{\scriptsize{}0.877 (0.01)} & \textcolor{black}{\scriptsize{}0.88 (0.01)} & \textbf{\textcolor{black}{\scriptsize{}0.881 (0.01)}} & \textcolor{black}{\scriptsize{}0.873 (0.011)} & \textcolor{black}{\scriptsize{}-} & \textcolor{black}{\scriptsize{}0.866 (0.011)} &  &  &  &  &  & \tabularnewline
 &  & \textcolor{black}{\scriptsize{}20} & \textcolor{black}{\scriptsize{}0.986 (0.004)} & \textcolor{black}{\scriptsize{}0.986 (0.004)} & \textbf{\textcolor{black}{\scriptsize{}0.988 (0.003)}} & \textcolor{black}{\scriptsize{}0.987 (0.004)} & \textcolor{black}{\scriptsize{}-} & \textcolor{black}{\scriptsize{}0.986 (0.004)} &  &  &  &  &  & \tabularnewline
\cline{2-15} 
 & \multirow{9}{*}{\textcolor{black}{\scriptsize{}SZ=91}} &  & \multicolumn{6}{c|}{\textcolor{black}{\scriptsize{}$\tau^{2}=0.006$}} & \multicolumn{6}{c}{\textcolor{black}{\scriptsize{}$\tau^{2}=0.02$}}\tabularnewline
\cline{3-15} 
 &  & \textcolor{black}{\scriptsize{}10} & \textbf{\textcolor{black}{\scriptsize{}0.153 (0.011)}} & \textbf{\textcolor{black}{\scriptsize{}0.153 (0.011)}} & \textcolor{black}{\scriptsize{}0.133 (0.011)} & \textcolor{black}{\scriptsize{}0.149 (0.011)} & \textcolor{black}{\scriptsize{}-} & \textcolor{black}{\scriptsize{}0.125 (0.01)} & \textcolor{black}{\scriptsize{}0.474 (0.016)} & \textbf{\textcolor{black}{\scriptsize{}0.478 (0.016)}} & \textcolor{black}{\scriptsize{}0.452 (0.016)} & \textcolor{black}{\scriptsize{}0.46 (0.016)} & \textcolor{black}{\scriptsize{}-} & \textcolor{black}{\scriptsize{}0.446 (0.016)}\tabularnewline
 &  & \textcolor{black}{\scriptsize{}20} & \textbf{\textcolor{black}{\scriptsize{}0.284 (0.014)}} & \textcolor{black}{\scriptsize{}0.279 (0.014)} & \textcolor{black}{\scriptsize{}0.228 (0.013)} & \textcolor{black}{\scriptsize{}0.243 (0.014)} & \textcolor{black}{\scriptsize{}-} & \textcolor{black}{\scriptsize{}0.212 (0.013)} & \textbf{\textcolor{black}{\scriptsize{}0.715 (0.014)}} & \textbf{\textcolor{black}{\scriptsize{}0.715 (0.014)}} & \textcolor{black}{\scriptsize{}0.685 (0.015)} & \textcolor{black}{\scriptsize{}0.689 (0.015)} & \textcolor{black}{\scriptsize{}-} & \textcolor{black}{\scriptsize{}0.672 (0.015)}\tabularnewline
 &  & \textcolor{black}{\scriptsize{}30} & \textcolor{black}{\scriptsize{}0.346 (0.015)} & \textbf{\textcolor{black}{\scriptsize{}0.347 (0.015)}} & \textcolor{black}{\scriptsize{}0.317 (0.015)} & \textcolor{black}{\scriptsize{}0.341 (0.015)} & \textcolor{black}{\scriptsize{}-} & \textcolor{black}{\scriptsize{}0.3 (0.014)} & \textcolor{black}{\scriptsize{}0.841 (0.012)} & \textbf{\textcolor{black}{\scriptsize{}0.842 (0.012)}} & \textcolor{black}{\scriptsize{}0.821 (0.012)} & \textcolor{black}{\scriptsize{}0.816 (0.012)} & \textcolor{black}{\scriptsize{}-} & \textcolor{black}{\scriptsize{}0.809 (0.012)}\tabularnewline
 &  & \textcolor{black}{\scriptsize{}50} & \textcolor{black}{\scriptsize{}0.457 (0.016)} & \textbf{\textcolor{black}{\scriptsize{}0.458 (0.016)}} & \textcolor{black}{\scriptsize{}0.415 (0.016)} & \textcolor{black}{\scriptsize{}0.441 (0.016)} & \textcolor{black}{\scriptsize{}-} & \textcolor{black}{\scriptsize{}0.4 (0.015)} & \textcolor{black}{\scriptsize{}0.957 (0.006)} & \textbf{\textcolor{black}{\scriptsize{}0.96 (0.006)}} & \textcolor{black}{\scriptsize{}0.948 (0.007)} & \textcolor{black}{\scriptsize{}0.955 (0.007)} & \textcolor{black}{\scriptsize{}-} & \textcolor{black}{\scriptsize{}0.941 (0.007)}\tabularnewline
 &  & \textcolor{black}{\scriptsize{}100} & \textcolor{black}{\scriptsize{}0.694 (0.015)} & \textbf{\textcolor{black}{\scriptsize{}0.695 (0.015)}} & \textcolor{black}{\scriptsize{}0.606 (0.015)} & \textcolor{black}{\scriptsize{}0.653 (0.015)} & \textcolor{black}{\scriptsize{}-} & \textcolor{black}{\scriptsize{}0.577 (0.016)} & \textcolor{black}{\scriptsize{}0.996 (0.002)} & \textcolor{black}{\scriptsize{}0.996 (0.002)} & \textbf{\textcolor{black}{\scriptsize{}0.998 (0.001)}} & \textcolor{black}{\scriptsize{}0.997 (0.002)} & \textcolor{black}{\scriptsize{}-} & \textcolor{black}{\scriptsize{}0.997 (0.002)}\tabularnewline
\cline{3-15} 
 &  &  & \multicolumn{6}{c|}{\textcolor{black}{\scriptsize{}$\tau^{2}=0.05$}} &  &  &  &  &  & \tabularnewline
\cline{3-15} 
 &  & \textcolor{black}{\scriptsize{}10} & \textbf{\textcolor{black}{\scriptsize{}0.799 (0.013)}} & \textcolor{black}{\scriptsize{}0.798 (0.013)} & \textcolor{black}{\scriptsize{}0.796 (0.013)} & \textcolor{black}{\scriptsize{}0.791 (0.013)} & \textcolor{black}{\scriptsize{}-} & \textcolor{black}{\scriptsize{}0.791 (0.013)} &  &  &  &  &  & \tabularnewline
 &  & \textcolor{black}{\scriptsize{}20} & \textbf{\textcolor{black}{\scriptsize{}0.962 (0.006)}} & \textbf{\textcolor{black}{\scriptsize{}0.962 (0.006)}} & \textcolor{black}{\scriptsize{}0.961 (0.006)} & \textcolor{black}{\scriptsize{}0.961 (0.006)} & \textcolor{black}{\scriptsize{}-} & \textcolor{black}{\scriptsize{}0.959 (0.006)} &  &  &  &  &  & \tabularnewline
\hline 
 & \multirow{9}{*}{\textcolor{black}{\scriptsize{}SZ=24}} &  & \multicolumn{6}{c}{\textcolor{black}{\scriptsize{}$\tau^{2}=0.01$}} & \multicolumn{6}{c}{\textcolor{black}{\scriptsize{}$\tau^{2}=0.03$}}\tabularnewline
\cline{3-15} 
 &  & \textcolor{black}{\scriptsize{}10} & \textbf{\textcolor{black}{\scriptsize{}0.161 (0.012)}} & \textcolor{black}{\scriptsize{}0.158 (0.012)} & \textcolor{black}{\scriptsize{}0.143 (0.011)} & \textcolor{black}{\scriptsize{}-} & \textcolor{black}{\scriptsize{}-} & \textcolor{black}{\scriptsize{}0.145 (0.011)} & \textcolor{black}{\scriptsize{}0.396 (0.015)} & \textbf{\textcolor{black}{\scriptsize{}0.398 (0.015)}} & \textcolor{black}{\scriptsize{}0.368 (0.015)} & \textcolor{black}{\scriptsize{}-} & \textcolor{black}{\scriptsize{}-} & \textcolor{black}{\scriptsize{}0.37 (0.015)}\tabularnewline
 &  & \textcolor{black}{\scriptsize{}20} & \textbf{\textcolor{black}{\scriptsize{}0.225 (0.013)}} & \textcolor{black}{\scriptsize{}0.223 (0.013)} & \textcolor{black}{\scriptsize{}0.197 (0.013)} & \textcolor{black}{\scriptsize{}-} & \textcolor{black}{\scriptsize{}-} & \textcolor{black}{\scriptsize{}0.195 (0.013)} & \textcolor{black}{\scriptsize{}0.575 (0.016)} & \textbf{\textcolor{black}{\scriptsize{}0.578 (0.016)}} & \textcolor{black}{\scriptsize{}0.55 (0.016)} & \textcolor{black}{\scriptsize{}-} & \textcolor{black}{\scriptsize{}-} & \textcolor{black}{\scriptsize{}0.548 (0.016)}\tabularnewline
 &  & \textcolor{black}{\scriptsize{}30} & \textbf{\textcolor{black}{\scriptsize{}0.286 (0.014)}} & \textcolor{black}{\scriptsize{}0.282 (0.014)} & \textcolor{black}{\scriptsize{}0.257 (0.014)} & \textcolor{black}{\scriptsize{}-} & \textcolor{black}{\scriptsize{}-} & \textcolor{black}{\scriptsize{}0.256 (0.014)} & \textcolor{black}{\scriptsize{}0.737 (0.014)} & \textbf{\textcolor{black}{\scriptsize{}0.74 (0.014)}} & \textcolor{black}{\scriptsize{}0.693 (0.015)} & \textcolor{black}{\scriptsize{}-} & \textcolor{black}{\scriptsize{}-} & \textcolor{black}{\scriptsize{}0.69 (0.015)}\tabularnewline
\textcolor{black}{\scriptsize{}Fisher-} &  & \textcolor{black}{\scriptsize{}50} & \textbf{\textcolor{black}{\scriptsize{}0.348 (0.015)}} & \textbf{\textcolor{black}{\scriptsize{}0.348 (0.015)}} & \textcolor{black}{\scriptsize{}0.296 (0.014)} & \textcolor{black}{\scriptsize{}-} & \textcolor{black}{\scriptsize{}-} & \textcolor{black}{\scriptsize{}0.296 (0.014)} & \textbf{\textcolor{black}{\scriptsize{}0.887 (0.01)}} & \textbf{\textcolor{black}{\scriptsize{}0.887 (0.01)}} & \textcolor{black}{\scriptsize{}0.851 (0.011)} & \textcolor{black}{\scriptsize{}-} & \textcolor{black}{\scriptsize{}-} & \textcolor{black}{\scriptsize{}0.849 (0.011)}\tabularnewline
\textcolor{black}{\scriptsize{}transformed } &  & \textcolor{black}{\scriptsize{}100} & \textbf{\textcolor{black}{\scriptsize{}0.615 (0.015)}} & \textbf{\textcolor{black}{\scriptsize{}0.615 (0.015)}} & \textcolor{black}{\scriptsize{}0.539 (0.016)} & \textcolor{black}{\scriptsize{}-} & \textcolor{black}{\scriptsize{}-} & \textcolor{black}{\scriptsize{}0.539 (0.016)} & \textbf{\textcolor{black}{\scriptsize{}0.994 (0.002)}} & \textbf{\textcolor{black}{\scriptsize{}0.994 (0.002)}} & \textcolor{black}{\scriptsize{}0.986 (0.004)} & \textcolor{black}{\scriptsize{}-} & \textcolor{black}{\scriptsize{}-} & \textcolor{black}{\scriptsize{}0.986 (0.004)}\tabularnewline
\cline{3-15} 
\textcolor{black}{\scriptsize{}Pearson} &  &  & \multicolumn{6}{c|}{\textcolor{black}{\scriptsize{}$\tau^{2}=0.1$}} & \multicolumn{6}{c}{}\tabularnewline
\cline{3-15} 
\textcolor{black}{\scriptsize{}correlation} &  & \textcolor{black}{\scriptsize{}10} & \textbf{\textcolor{black}{\scriptsize{}0.815 (0.012)}} & \textbf{\textcolor{black}{\scriptsize{}0.815 (0.012)}} & \textcolor{black}{\scriptsize{}0.804 (0.013)} & \textcolor{black}{\scriptsize{}-} & \textcolor{black}{\scriptsize{}-} & \textcolor{black}{\scriptsize{}0.802 (0.013)} &  &  &  &  &  & \tabularnewline
 &  & \textcolor{black}{\scriptsize{}20} & \textbf{\textcolor{black}{\scriptsize{}0.965 (0.006)}} & \textbf{\textcolor{black}{\scriptsize{}0.965 (0.006)}} & \textcolor{black}{\scriptsize{}0.964 (0.006)} & \textcolor{black}{\scriptsize{}-} & \textcolor{black}{\scriptsize{}-} & \textcolor{black}{\scriptsize{}0.963 (0.006)} &  &  &  &  &  & \tabularnewline
\cline{2-15} 
 & \multirow{8}{*}{\textcolor{black}{\scriptsize{}SZ=91}} &  & \multicolumn{6}{c|}{\textcolor{black}{\scriptsize{}$\tau^{2}=0.06$}} & \multicolumn{6}{c}{\textcolor{black}{\scriptsize{}$\tau^{2}=0.2$}}\tabularnewline
\cline{3-15} 
 &  & \textcolor{black}{\scriptsize{}10} & \textbf{\textcolor{black}{\scriptsize{}0.318 (0.015)}} & \textcolor{black}{\scriptsize{}0.315 (0.015)} & \textcolor{black}{\scriptsize{}0.292 (0.014)} & \textcolor{black}{\scriptsize{}-} & \textcolor{black}{\scriptsize{}-} & \textcolor{black}{\scriptsize{}0.289 (0.014)} & \textcolor{black}{\scriptsize{}0.755 (0.014)} & \textbf{\textcolor{black}{\scriptsize{}0.756 (0.014)}} & \textcolor{black}{\scriptsize{}0.732 (0.014)} & \textcolor{black}{\scriptsize{}-} & \textcolor{black}{\scriptsize{}-} & \textcolor{black}{\scriptsize{}0.735 (0.014)}\tabularnewline
 &  & \textcolor{black}{\scriptsize{}20} & \textbf{\textcolor{black}{\scriptsize{}0.512 (0.016)}} & \textbf{\textcolor{black}{\scriptsize{}0.512 (0.016)}} & \textcolor{black}{\scriptsize{}0.461 (0.016)} & \textcolor{black}{\scriptsize{}-} & \textcolor{black}{\scriptsize{}-} & \textcolor{black}{\scriptsize{}0.461 (0.016)} & \textbf{\textcolor{black}{\scriptsize{}0.942 (0.007)}} & \textbf{\textcolor{black}{\scriptsize{}0.942 (0.007)}} & \textcolor{black}{\scriptsize{}0.931 (0.008)} & \textcolor{black}{\scriptsize{}-} & \textcolor{black}{\scriptsize{}-} & \textcolor{black}{\scriptsize{}0.931 (0.008)}\tabularnewline
 &  & \textcolor{black}{\scriptsize{}30} & \textcolor{black}{\scriptsize{}0.624 (0.015)} & \textbf{\textcolor{black}{\scriptsize{}0.626 (0.015)}} & \textcolor{black}{\scriptsize{}0.586 (0.016)} & \textcolor{black}{\scriptsize{}-} & \textcolor{black}{\scriptsize{}-} & \textcolor{black}{\scriptsize{}0.586 (0.016)} & \textbf{\textcolor{black}{\scriptsize{}0.985 (0.004)}} & \textbf{\textcolor{black}{\scriptsize{}0.985 (0.004)}} & \textcolor{black}{\scriptsize{}0.981 (0.004)} & \textcolor{black}{\scriptsize{}-} & \textcolor{black}{\scriptsize{}-} & \textcolor{black}{\scriptsize{}0.98 (0.004)}\tabularnewline
 &  & \textcolor{black}{\scriptsize{}50} & \textbf{\textcolor{black}{\scriptsize{}0.791 (0.013)}} & \textbf{\textcolor{black}{\scriptsize{}0.791 (0.013)}} & \textcolor{black}{\scriptsize{}0.746 (0.014)} & \textcolor{black}{\scriptsize{}-} & \textcolor{black}{\scriptsize{}-} & \textcolor{black}{\scriptsize{}0.744 (0.014)} &  &  &  &  &  & \tabularnewline
 &  & \textcolor{black}{\scriptsize{}100} & \textbf{\textcolor{black}{\scriptsize{}0.97 (0.005)}} & \textbf{\textcolor{black}{\scriptsize{}0.97 (0.005)}} & \textcolor{black}{\scriptsize{}0.946 (0.007)} & \textcolor{black}{\scriptsize{}-} & \textcolor{black}{\scriptsize{}-} & \textcolor{black}{\scriptsize{}0.945 (0.007)} &  &  &  &  &  & \tabularnewline
\cline{3-15} 
 &  &  & \multicolumn{6}{c|}{\textcolor{black}{\scriptsize{}$\tau^{2}=0.5$}} & \textcolor{black}{\scriptsize{}-} &  &  &  &  & \tabularnewline
\cline{3-15} 
 &  & \textcolor{black}{\scriptsize{}10} & \textcolor{black}{\scriptsize{}0.946 (0.007)} & \textbf{\textcolor{black}{\scriptsize{}0.948 (0.007)}} & \textcolor{black}{\scriptsize{}0.942 (0.007)} &  &  & \textcolor{black}{\scriptsize{}0.942 (0.007)} &  &  &  &  &  & \tabularnewline
\hline 
 & \multirow{11}{*}{\textcolor{black}{\scriptsize{}SZ=24}} &  & \multicolumn{6}{c|}{\textcolor{black}{\scriptsize{}$\tau^{2}=0.1$}} & \multicolumn{6}{c}{\textcolor{black}{\scriptsize{}$\tau^{2}=0.3$}}\tabularnewline
\cline{3-15} 
 &  & \textcolor{black}{\scriptsize{}10} & \textcolor{black}{\scriptsize{}0.115 (0.01)} & \textcolor{black}{\scriptsize{}0.115 (0.01)} & \textcolor{black}{\scriptsize{}0.128 (0.011)} & \textbf{\textcolor{black}{\scriptsize{}0.132 (0.011)}} & \textcolor{black}{\scriptsize{}0.11 (0.01)} & \textcolor{black}{\scriptsize{}0.087 (0.009)} & \textcolor{black}{\scriptsize{}0.308 (0.015)} & \textcolor{black}{\scriptsize{}0.31 (0.015)} & \textcolor{black}{\scriptsize{}0.328 (0.015)} & \textbf{\textcolor{black}{\scriptsize{}0.341 (0.015)}} & \textcolor{black}{\scriptsize{}0.14 (0.011)} & \textcolor{black}{\scriptsize{}0.245 (0.014)}\tabularnewline
 &  & \textcolor{black}{\scriptsize{}20} & \textcolor{black}{\scriptsize{}0.141 (0.011)} & \textcolor{black}{\scriptsize{}0.142 (0.011)} & \textcolor{black}{\scriptsize{}0.168 (0.012)} & \textbf{\textcolor{black}{\scriptsize{}0.182 (0.012)}} & \textcolor{black}{\scriptsize{}0.173 (0.012)} & \textcolor{black}{\scriptsize{}0.11 (0.01)} & \textcolor{black}{\scriptsize{}0.43 (0.016)} & \textcolor{black}{\scriptsize{}0.432 (0.016)} & \textcolor{black}{\scriptsize{}0.459 (0.016)} & \textbf{\textcolor{black}{\scriptsize{}0.48 (0.016)}} & \textcolor{black}{\scriptsize{}0.216 (0.013)} & \textcolor{black}{\scriptsize{}0.355 (0.015)}\tabularnewline
 &  & \textcolor{black}{\scriptsize{}30} & \textcolor{black}{\scriptsize{}0.192 (0.012)} & \textcolor{black}{\scriptsize{}0.19 (0.012)} & \textcolor{black}{\scriptsize{}0.21 (0.013)} & \textbf{\textcolor{black}{\scriptsize{}0.226 (0.013)}} & \textcolor{black}{\scriptsize{}0.196 (0.013)} & \textcolor{black}{\scriptsize{}0.125 (0.01)} & \textcolor{black}{\scriptsize{}0.56 (0.016)} & \textcolor{black}{\scriptsize{}0.565 (0.016)} & \textcolor{black}{\scriptsize{}0.592 (0.016)} & \textbf{\textcolor{black}{\scriptsize{}0.627 (0.015)}} & \textcolor{black}{\scriptsize{}0.256 (0.014)} & \textcolor{black}{\scriptsize{}0.481 (0.016)}\tabularnewline
 &  & \textcolor{black}{\scriptsize{}50} & \textcolor{black}{\scriptsize{}0.249 (0.014)} & \textcolor{black}{\scriptsize{}0.248 (0.014)} & \textcolor{black}{\scriptsize{}0.266 (0.014)} & \textbf{\textcolor{black}{\scriptsize{}0.284 (0.014)}} & \textcolor{black}{\scriptsize{}0.249 (0.014)} & \textcolor{black}{\scriptsize{}0.149 (0.011)} & \textcolor{black}{\scriptsize{}0.725 (0.014)} & \textcolor{black}{\scriptsize{}0.725 (0.014)} & \textcolor{black}{\scriptsize{}0.766 (0.013)} & \textbf{\textcolor{black}{\scriptsize{}0.794 (0.013)}} & \textcolor{black}{\scriptsize{}0.355 (0.015)} & \textcolor{black}{\scriptsize{}0.652 (0.015)}\tabularnewline
 &  & \textcolor{black}{\scriptsize{}100} & \textcolor{black}{\scriptsize{}0.396 (0.015)} & \textcolor{black}{\scriptsize{}0.397 (0.015)} & \textcolor{black}{\scriptsize{}0.451 (0.016)} & \textbf{\textcolor{black}{\scriptsize{}0.463 (0.016)}} & \textcolor{black}{\scriptsize{}0.426 (0.016)} & \textcolor{black}{\scriptsize{}0.248 (0.014)} & \textcolor{black}{\scriptsize{}0.898 (0.01)} & \textcolor{black}{\scriptsize{}0.898 (0.01)} & \textcolor{black}{\scriptsize{}0.909 (0.009)} & \textbf{\textcolor{black}{\scriptsize{}0.938 (0.008)}} & \textcolor{black}{\scriptsize{}0.914 (0.009)} & \textcolor{black}{\scriptsize{}0.818 (0.012)}\tabularnewline
\cline{3-15} 
 &  &  & \multicolumn{6}{c|}{\textcolor{black}{\scriptsize{}$\tau^{2}=0.9$}} & \multicolumn{6}{c}{}\tabularnewline
\cline{3-15} 
\textcolor{black}{\scriptsize{}Natural-} &  & \textcolor{black}{\scriptsize{}10} & \textcolor{black}{\scriptsize{}0.658 (0.015)} & \textcolor{black}{\scriptsize{}0.662 (0.015)} & \textcolor{black}{\scriptsize{}0.689 (0.015)} & \textbf{\textcolor{black}{\scriptsize{}0.718 (0.014)}} & \textcolor{black}{\scriptsize{}0.686 (0.015)} & \textcolor{black}{\scriptsize{}0.601 (0.015)} &  &  &  &  &  & \tabularnewline
\textcolor{black}{\scriptsize{}logarithm- } &  & \textcolor{black}{\scriptsize{}20} & \textcolor{black}{\scriptsize{}0.856 (0.011)} & \textcolor{black}{\scriptsize{}0.854 (0.011)} & \textcolor{black}{\scriptsize{}0.889 (0.01)} & \textbf{\textcolor{black}{\scriptsize{}0.899 (0.01)}} & \textcolor{black}{\scriptsize{}0.881 (0.01)} & \textcolor{black}{\scriptsize{}0.822 (0.012)} &  &  &  &  &  & \tabularnewline
\textcolor{black}{\scriptsize{}transformed} &  & \textcolor{black}{\scriptsize{}30} & \textcolor{black}{\scriptsize{}0.941 (0.007)} & \textcolor{black}{\scriptsize{}0.942 (0.007)} & \textcolor{black}{\scriptsize{}0.955 (0.007)} & \textbf{\textcolor{black}{\scriptsize{}0.963 (0.006)}} & \textcolor{black}{\scriptsize{}0.951 (0.007)} & \textcolor{black}{\scriptsize{}0.915 (0.009)} &  &  &  &  &  & \tabularnewline
\textcolor{black}{\scriptsize{}odds } &  & \textcolor{black}{\scriptsize{}50} & \textcolor{black}{\scriptsize{}0.987 (0.004)} & \textcolor{black}{\scriptsize{}0.987 (0.004)} & \textcolor{black}{\scriptsize{}0.992 (0.003)} & \textbf{\textcolor{black}{\scriptsize{}0.997 (0.002)}} & \textcolor{black}{\scriptsize{}0.993 (0.003)} & \textcolor{black}{\scriptsize{}0.98 (0.004)} &  &  &  &  &  & \tabularnewline
\cline{2-15} 
 & \multirow{9}{*}{\textcolor{black}{\scriptsize{}SZ=91}} &  & \multicolumn{6}{c|}{\textcolor{black}{\scriptsize{}$\tau^{2}=0.03$}} & \multicolumn{6}{c}{\textcolor{black}{\scriptsize{}$\tau^{2}=0.1$}}\tabularnewline
\cline{3-15} 
 &  & \textcolor{black}{\scriptsize{}10} & \textcolor{black}{\scriptsize{}0.176 (0.012)} & \textbf{\textcolor{black}{\scriptsize{}0.177 (0.012)}} & \textcolor{black}{\scriptsize{}0.174 (0.012)} & \textcolor{black}{\scriptsize{}0.174 (0.012)} & \textcolor{black}{\scriptsize{}0.174 (0.012)} & \textcolor{black}{\scriptsize{}0.16 (0.012)} & \textcolor{black}{\scriptsize{}0.457 (0.016)} & \textbf{\textcolor{black}{\scriptsize{}0.46 (0.016)}} & \textcolor{black}{\scriptsize{}0.446 (0.016)} & \textcolor{black}{\scriptsize{}0.449 (0.016)} & \textcolor{black}{\scriptsize{}0.443 (0.016)} & \textcolor{black}{\scriptsize{}0.406 (0.016)}\tabularnewline
 &  & \textcolor{black}{\scriptsize{}20} & \textbf{\textcolor{black}{\scriptsize{}0.269 (0.014)}} & \textcolor{black}{\scriptsize{}0.268 (0.014)} & \textcolor{black}{\scriptsize{}0.227 (0.013)} & \textcolor{black}{\scriptsize{}0.242 (0.014)} & \textcolor{black}{\scriptsize{}0.229 (0.013)} & \textcolor{black}{\scriptsize{}0.196 (0.013)} & \textcolor{black}{\scriptsize{}0.693 (0.015)} & \textbf{\textcolor{black}{\scriptsize{}0.696 (0.015)}} & \textcolor{black}{\scriptsize{}0.686 (0.015)} & \textcolor{black}{\scriptsize{}0.676 (0.015)} & \textcolor{black}{\scriptsize{}0.674 (0.015)} & \textcolor{black}{\scriptsize{}0.649 (0.015)}\tabularnewline
 &  & \textcolor{black}{\scriptsize{}30} & \textcolor{black}{\scriptsize{}0.327 (0.015)} & \textbf{\textcolor{black}{\scriptsize{}0.331 (0.015)}} & \textcolor{black}{\scriptsize{}0.311 (0.015)} & \textcolor{black}{\scriptsize{}0.308 (0.015)} & \textcolor{black}{\scriptsize{}0.297 (0.014)} & \textcolor{black}{\scriptsize{}0.263 (0.014)} & \textbf{\textcolor{black}{\scriptsize{}0.831 (0.012)}} & \textbf{\textcolor{black}{\scriptsize{}0.83 (0.012)}} & \textcolor{black}{\scriptsize{}0.828 (0.012)} & \textcolor{black}{\scriptsize{}0.834 (0.012)} & \textcolor{black}{\scriptsize{}0.821 (0.012)} & \textcolor{black}{\scriptsize{}0.797 (0.013)}\tabularnewline
 &  & \textcolor{black}{\scriptsize{}50} & \textcolor{black}{\scriptsize{}0.45 (0.016)} & \textbf{\textcolor{black}{\scriptsize{}0.453 (0.016)}} & \textcolor{black}{\scriptsize{}0.415 (0.016)} & \textcolor{black}{\scriptsize{}0.41 (0.016)} & \textcolor{black}{\scriptsize{}0.408 (0.016)} & \textcolor{black}{\scriptsize{}0.346 (0.015)} & \textbf{\textcolor{black}{\scriptsize{}0.934 (0.008)}} & \textbf{\textcolor{black}{\scriptsize{}0.934 (0.008)}} & \textcolor{black}{\scriptsize{}0.915 (0.009)} & \textcolor{black}{\scriptsize{}0.922 (0.008)} & \textcolor{black}{\scriptsize{}0.913 (0.009)} & \textcolor{black}{\scriptsize{}0.893 (0.01)}\tabularnewline
 &  & \textcolor{black}{\scriptsize{}100} & \textbf{\textcolor{black}{\scriptsize{}0.652 (0.015)}} & \textbf{\textcolor{black}{\scriptsize{}0.652 (0.015)}} & \textcolor{black}{\scriptsize{}0.602 (0.015)} & \textcolor{black}{\scriptsize{}0.601 (0.015)} & \textcolor{black}{\scriptsize{}0.597 (0.016)} & \textcolor{black}{\scriptsize{}0.521 (0.016)} &  &  &  &  &  & \tabularnewline
\cline{3-15} 
 &  &  & \multicolumn{6}{c|}{\textcolor{black}{\scriptsize{}$\tau^{2}=0.3$}} & \multicolumn{6}{c}{}\tabularnewline
\cline{3-15} 
 &  & \textcolor{black}{\scriptsize{}10} & \textcolor{black}{\scriptsize{}0.833 (0.012)} & \textcolor{black}{\scriptsize{}0.836 (0.012)} & \textbf{\textcolor{black}{\scriptsize{}0.839 (0.012)}} & \textcolor{black}{\scriptsize{}0.834 (0.012)} & \textcolor{black}{\scriptsize{}0.828 (0.012)} & \textcolor{black}{\scriptsize{}0.82 (0.012)} &  &  &  &  &  & \tabularnewline
 &  & \textcolor{black}{\scriptsize{}20} & \textcolor{black}{\scriptsize{}0.965 (0.006)} & \textcolor{black}{\scriptsize{}0.965 (0.006)} & \textcolor{black}{\scriptsize{}0.969 (0.005)} & \textbf{\textcolor{black}{\scriptsize{}0.973 (0.005)}} & \textcolor{black}{\scriptsize{}0.97 (0.005)} & \textcolor{black}{\scriptsize{}0.962 (0.006)} &  &  &  &  &  & \tabularnewline
\hline 
\end{tabular}{\scriptsize\par}

}

{\scriptsize{}Note: Monte Carlo standard errors are presented in the
parentheses. SZ is the average per-group sample sizes. The largest
power values under each condition are highlighted as bold. ML or REML
estimation sometimes did not have converged results, but the nonconvergence
rates across all conditions were within 0.4\%. }{\scriptsize\par}
\end{table}
{\scriptsize{} }{\scriptsize\par}

\subsubsection{5.3.2 Statistical power in mixed-effects \textcolor{black}{models
for heterogeneity test}}

\textcolor{black}{In mixed-effects meta-analyses, the statistical
power values of B-ML-LRT, B-REML-LRT, B-Q, and the Q test with one
covariate are presented in Table 8. The highest power values under
each condition are bolded and the Monte Carlo standard errors are
presented inside the parentheses. Similar to the cases without covariates,
more studies, larger study-level sample sizes, and/or larger $\tau^{2}$
produced higher power. In terms of the proposed bootstrap based methods,
when the effect of interest was the standardized mean difference,
compared to the Q test, the maximum power increments of B-ML-LRT,
B-REML-LRT, and B-Q compared to the Q test were 0.098, 0.098, and
0.082 respectively, and the maximum percentage increments of B-ML-LRT,
B-REML-LRT, and B-Q were 25.8\%, 25.8\%, and 15.7\% respectively.
When the effect size was the Fisher-transformed Pearson correlation,
the maximum power increments of B-ML-LRT, B-REML-LRT, and B-Q were
0.108, 0.108, and 0.004 respectively and the maximum percentage increments
of B-ML-LRT, B-REML-LRT, and B-Q were 23.8\%, 23.8\%, and 1.7\% respectively.
When the effect sizes was the log odds ratio, the maximum power increments
of B-ML-LRT, B-REML-LRT, and B-Q were 0.142, 0.142, and 0.193 respectively,
and the maximum percentage increments of B-ML-LRT, B-REML-LRT, and
B-Q were 65.1\%, 65.1\%, and 88.5\% respectively. Based on both the
percentage increments and Monte Carlo standard errors, the bootstrap
methods increased power compared to the regular Q test except the
case of B-Q with the Fisher-transformed Pearson correlation.}

\textcolor{black}{{[}Table 8{]}}{\scriptsize{}}
\begin{table}
{\scriptsize{}\caption{{\footnotesize{}With one covariate, statistical power of the bootstrap
and ML based LR test (B-ML-LRT), the bootstrap and REML based LR test
}\textcolor{black}{\footnotesize{}(B-REML-LRT), the bootstrap based
Q test (B-Q), the regular ML based LR test (ML-LRT), and the Q test
(Q) in mixed-effects models}}
}{\scriptsize\par}

\resizebox{\textwidth}{!}{
\renewcommand{\arraystretch}{0.52}

{\scriptsize{}}%
\begin{tabular}{ccccccc|cccc}
\hline 
 &  & \textcolor{black}{\scriptsize{}Number of} & \multirow{2}{*}{\textcolor{black}{\scriptsize{}B-ML-LRT}} & \multirow{2}{*}{\textcolor{black}{\scriptsize{}B-REML-LRT}} & \multirow{2}{*}{\textcolor{black}{\scriptsize{}B-Q}} & \multirow{2}{*}{\textcolor{black}{\scriptsize{}Q}} & \multirow{2}{*}{\textcolor{black}{\scriptsize{}B-ML-LRT}} & \multirow{2}{*}{\textcolor{black}{\scriptsize{}B-REML-LRT}} & \multirow{2}{*}{\textcolor{black}{\scriptsize{}B-Q}} & \multirow{2}{*}{\textcolor{black}{\scriptsize{}Q}}\tabularnewline
 &  & \textcolor{black}{\scriptsize{}Studies} &  &  &  &  &  &  &  & \tabularnewline
\hline 
 & \multirow{9}{*}{\textcolor{black}{\scriptsize{}SZ=24}} &  & \multicolumn{4}{c|}{\textcolor{black}{\scriptsize{}$\tau^{2}=0.03$}} & \multicolumn{4}{c}{\textcolor{black}{\scriptsize{}$\tau^{2}=0.1$}}\tabularnewline
\cline{3-11} 
 &  & \textcolor{black}{\scriptsize{}10} & \textbf{\textcolor{black}{\scriptsize{}0.173 (0.012)}} & \textcolor{black}{\scriptsize{}0.169 (0.012)} & \textcolor{black}{\scriptsize{}0.162 (0.012)} & \textcolor{black}{\scriptsize{}0.144 (0.011)} & \textcolor{black}{\scriptsize{}0.457 (0.016)} & \textbf{\textcolor{black}{\scriptsize{}0.459 (0.016)}} & \textcolor{black}{\scriptsize{}0.439 (0.016)} & \textcolor{black}{\scriptsize{}0.425 (0.016)}\tabularnewline
 &  & \textcolor{black}{\scriptsize{}20} & \textbf{\textcolor{black}{\scriptsize{}0.265 (0.014)}} & \textcolor{black}{\scriptsize{}0.263 (0.014)} & \textcolor{black}{\scriptsize{}0.243 (0.014)} & \textcolor{black}{\scriptsize{}0.22 (0.013)} & \textcolor{black}{\scriptsize{}0.72 (0.014)} & \textcolor{black}{\scriptsize{}0.726 (0.014)} & \textbf{\textcolor{black}{\scriptsize{}0.728 (0.014)}} & \textcolor{black}{\scriptsize{}0.696 (0.015)}\tabularnewline
 &  & \textcolor{black}{\scriptsize{}30} & \textbf{\textcolor{black}{\scriptsize{}0.365 (0.015)}} & \textbf{\textcolor{black}{\scriptsize{}0.365 (0.015)}} & \textcolor{black}{\scriptsize{}0.337 (0.015)} & \textcolor{black}{\scriptsize{}0.303 (0.015)} & \textbf{\textcolor{black}{\scriptsize{}0.852 (0.011)}} & \textcolor{black}{\scriptsize{}0.851 (0.011)} & \textcolor{black}{\scriptsize{}0.845 (0.011)} & \textcolor{black}{\scriptsize{}0.832 (0.012)}\tabularnewline
 &  & \textcolor{black}{\scriptsize{}50} & \textcolor{black}{\scriptsize{}0.473 (0.016)} & \textbf{\textcolor{black}{\scriptsize{}0.476 (0.016)}} & \textcolor{black}{\scriptsize{}0.435 (0.016)} & \textcolor{black}{\scriptsize{}0.376 (0.015)} & \textcolor{black}{\scriptsize{}0.976 (0.005)} & \textbf{\textcolor{black}{\scriptsize{}0.979 (0.005)}} & \textcolor{black}{\scriptsize{}0.976 (0.005)} & \textcolor{black}{\scriptsize{}0.971 (0.005)}\tabularnewline
\textcolor{black}{\scriptsize{}Standardized } &  & \textcolor{black}{\scriptsize{}100} & \textbf{\textcolor{black}{\scriptsize{}0.701 (0.014)}} & \textbf{\textcolor{black}{\scriptsize{}0.701 (0.014)}} & \textcolor{black}{\scriptsize{}0.685 (0.015)} & \textcolor{black}{\scriptsize{}0.603 (0.015)} &  &  &  & \tabularnewline
\cline{3-11} 
\textcolor{black}{\scriptsize{}mean } &  &  & \multicolumn{4}{c|}{\textcolor{black}{\scriptsize{}$\tau^{2}=0.3$}} & \multicolumn{4}{c}{}\tabularnewline
\cline{3-11} 
\textcolor{black}{\scriptsize{}difference} &  & \textcolor{black}{\scriptsize{}10} & \textcolor{black}{\scriptsize{}0.867 (0.011)} & \textcolor{black}{\scriptsize{}0.871 (0.011)} & \textbf{\textcolor{black}{\scriptsize{}0.873 (0.011)}} & \textcolor{black}{\scriptsize{}0.856 (0.011)} &  &  &  & \tabularnewline
 &  & \textcolor{black}{\scriptsize{}20} & \textcolor{black}{\scriptsize{}0.984 (0.004)} & \textcolor{black}{\scriptsize{}0.987 (0.004)} & \textbf{\textcolor{black}{\scriptsize{}0.99 (0.003)}} & \textcolor{black}{\scriptsize{}0.99 (0.003)} &  &  &  & \tabularnewline
\cline{2-11} 
 & \multirow{9}{*}{\textcolor{black}{\scriptsize{}SZ=91}} &  & \multicolumn{4}{c|}{\textcolor{black}{\scriptsize{}$\tau^{2}=0.006$}} & \multicolumn{4}{c}{\textcolor{black}{\scriptsize{}$\tau^{2}=0.02$}}\tabularnewline
\cline{3-11} 
 &  & \textcolor{black}{\scriptsize{}10} & \textbf{\textcolor{black}{\scriptsize{}0.157 (0.012)}} & \textcolor{black}{\scriptsize{}0.147 (0.011)} & \textcolor{black}{\scriptsize{}0.144 (0.011)} & \textcolor{black}{\scriptsize{}0.142 (0.011)} & \textcolor{black}{\scriptsize{}0.419 (0.016)} & \textbf{\textcolor{black}{\scriptsize{}0.42 (0.016)}} & \textcolor{black}{\scriptsize{}0.399 (0.015)} & \textcolor{black}{\scriptsize{}0.389 (0.015)}\tabularnewline
 &  & \textcolor{black}{\scriptsize{}20} & \textcolor{black}{\scriptsize{}0.259 (0.014)} & \textbf{\textcolor{black}{\scriptsize{}0.262 (0.014)}} & \textcolor{black}{\scriptsize{}0.246 (0.014)} & \textcolor{black}{\scriptsize{}0.235 (0.013)} & \textcolor{black}{\scriptsize{}0.677 (0.015)} & \textbf{\textcolor{black}{\scriptsize{}0.682 (0.015)}} & \textcolor{black}{\scriptsize{}0.645 (0.015)} & \textcolor{black}{\scriptsize{}0.638 (0.015)}\tabularnewline
 &  & \textcolor{black}{\scriptsize{}30} & \textbf{\textcolor{black}{\scriptsize{}0.302 (0.015)}} & \textcolor{black}{\scriptsize{}0.3 (0.014)} & \textcolor{black}{\scriptsize{}0.257 (0.014)} & \textcolor{black}{\scriptsize{}0.251 (0.014)} & \textcolor{black}{\scriptsize{}0.822 (0.012)} & \textbf{\textcolor{black}{\scriptsize{}0.827 (0.012)}} & \textcolor{black}{\scriptsize{}0.818 (0.012)} & \textcolor{black}{\scriptsize{}0.808 (0.012)}\tabularnewline
 &  & \textcolor{black}{\scriptsize{}50} & \textbf{\textcolor{black}{\scriptsize{}0.462 (0.016)}} & \textcolor{black}{\scriptsize{}0.461 (0.016)} & \textcolor{black}{\scriptsize{}0.424 (0.016)} & \textcolor{black}{\scriptsize{}0.408 (0.016)} & \textcolor{black}{\scriptsize{}0.941 (0.007)} & \textbf{\textcolor{black}{\scriptsize{}0.941 (0.007)}} & \textcolor{black}{\scriptsize{}0.929 (0.008)} & \textcolor{black}{\scriptsize{}0.924 (0.008)}\tabularnewline
 &  & \textcolor{black}{\scriptsize{}100} & \textbf{\textcolor{black}{\scriptsize{}0.631 (0.015)}} & \textbf{\textcolor{black}{\scriptsize{}0.631 (0.015)}} & \textcolor{black}{\scriptsize{}0.572 (0.016)} & \textcolor{black}{\scriptsize{}0.552 (0.016)} &  &  &  & \tabularnewline
\cline{3-11} 
 &  &  & \multicolumn{4}{c|}{\textcolor{black}{\scriptsize{}$\tau^{2}=0.05$}} &  &  &  & \tabularnewline
\cline{3-11} 
 &  & \textcolor{black}{\scriptsize{}10} & \textcolor{black}{\scriptsize{}0.758 (0.014)} & \textbf{\textcolor{black}{\scriptsize{}0.766 (0.013)}} & \textcolor{black}{\scriptsize{}0.763 (0.013)} & \textcolor{black}{\scriptsize{}0.755 (0.014)} &  &  &  & \tabularnewline
 &  & \textcolor{black}{\scriptsize{}20} & \textcolor{black}{\scriptsize{}0.948 (0.007)} & \textbf{\textcolor{black}{\scriptsize{}0.95 (0.007)}} & \textcolor{black}{\scriptsize{}0.949 (0.007)} & \textcolor{black}{\scriptsize{}0.947 (0.007)} &  &  &  & \tabularnewline
\hline 
 & \multirow{10}{*}{\textcolor{black}{\scriptsize{}SZ=24}} &  & \multicolumn{4}{c}{\textcolor{black}{\scriptsize{}$\tau^{2}=0.01$}} & \multicolumn{4}{c}{\textcolor{black}{\scriptsize{}$\tau^{2}=0.03$}}\tabularnewline
\cline{3-11} 
 &  & \textcolor{black}{\scriptsize{}10} & \textcolor{black}{\scriptsize{}0.128 (0.011)} & \textbf{\textcolor{black}{\scriptsize{}0.131 (0.011)}} & \textcolor{black}{\scriptsize{}0.119 (0.01)} & \textcolor{black}{\scriptsize{}0.117 (0.01)} & \textcolor{black}{\scriptsize{}0.359 (0.015)} & \textbf{\textcolor{black}{\scriptsize{}0.367 (0.015)}} & \textcolor{black}{\scriptsize{}0.351 (0.015)} & \textcolor{black}{\scriptsize{}0.348 (0.015)}\tabularnewline
 &  & \textcolor{black}{\scriptsize{}20} & \textbf{\textcolor{black}{\scriptsize{}0.19 (0.012)}} & \textcolor{black}{\scriptsize{}0.188 (0.012)} & \textcolor{black}{\scriptsize{}0.158 (0.012)} & \textcolor{black}{\scriptsize{}0.159 (0.012)} & \textbf{\textcolor{black}{\scriptsize{}0.538 (0.016)}} & \textbf{\textcolor{black}{\scriptsize{}0.538 (0.016)}} & \textcolor{black}{\scriptsize{}0.503 (0.016)} & \textcolor{black}{\scriptsize{}0.503 (0.016)}\tabularnewline
 &  & \textcolor{black}{\scriptsize{}30} & \textcolor{black}{\scriptsize{}0.279 (0.014)} & \textbf{\textcolor{black}{\scriptsize{}0.281 (0.014)}} & \textcolor{black}{\scriptsize{}0.241 (0.014)} & \textcolor{black}{\scriptsize{}0.237 (0.013)} & \textbf{\textcolor{black}{\scriptsize{}0.725 (0.014)}} & \textbf{\textcolor{black}{\scriptsize{}0.725 (0.014)}} & \textcolor{black}{\scriptsize{}0.684 (0.015)} & \textcolor{black}{\scriptsize{}0.688 (0.015)}\tabularnewline
 &  & \textcolor{black}{\scriptsize{}50} & \textbf{\textcolor{black}{\scriptsize{}0.365 (0.015)}} & \textcolor{black}{\scriptsize{}0.364 (0.015)} & \textcolor{black}{\scriptsize{}0.328 (0.015)} & \textcolor{black}{\scriptsize{}0.331 (0.015)} & \textcolor{black}{\scriptsize{}0.881 (0.01)} & \textbf{\textcolor{black}{\scriptsize{}0.884 (0.01)}} & \textcolor{black}{\scriptsize{}0.847 (0.011)} & \textcolor{black}{\scriptsize{}0.847 (0.011)}\tabularnewline
\textcolor{black}{\scriptsize{}Fisher-} &  & \textcolor{black}{\scriptsize{}100} & \textbf{\textcolor{black}{\scriptsize{}0.561 (0.016)}} & \textcolor{black}{\scriptsize{}0.56 (0.016)} & \textcolor{black}{\scriptsize{}0.454 (0.016)} & \textcolor{black}{\scriptsize{}0.453 (0.016)} & \textbf{\textcolor{black}{\scriptsize{}0.983 (0.004)}} & \textbf{\textcolor{black}{\scriptsize{}0.983 (0.004)}} & \textcolor{black}{\scriptsize{}0.974 (0.005)} & \textcolor{black}{\scriptsize{}0.975 (0.005)}\tabularnewline
\cline{3-11} 
\textcolor{black}{\scriptsize{}Pearson} &  &  & \multicolumn{4}{c|}{\textcolor{black}{\scriptsize{}$\tau^{2}=0.1$}} & \multicolumn{4}{c}{}\tabularnewline
\cline{3-11} 
\textcolor{black}{\scriptsize{}correlation} &  & \textcolor{black}{\scriptsize{}10} & \textcolor{black}{\scriptsize{}0.766 (0.013)} & \textbf{\textcolor{black}{\scriptsize{}0.774 (0.013)}} & \textcolor{black}{\scriptsize{}0.769 (0.013)} & \textcolor{black}{\scriptsize{}0.768 (0.013)} &  &  &  & \tabularnewline
 &  & \textcolor{black}{\scriptsize{}20} & \textcolor{black}{\scriptsize{}0.955 (0.007)} & \textbf{\textcolor{black}{\scriptsize{}0.961 (0.006)}} & \textcolor{black}{\scriptsize{}0.956 (0.006)} & \textcolor{black}{\scriptsize{}0.956 (0.006)} &  &  &  & \tabularnewline
 &  & \textcolor{black}{\scriptsize{}30} & \textbf{\textcolor{black}{\scriptsize{}0.991 (0.003)}} & \textbf{\textcolor{black}{\scriptsize{}0.991 (0.003)}} & \textcolor{black}{\scriptsize{}0.988 (0.003)} & \textcolor{black}{\scriptsize{}0.988 (0.003)} &  &  &  & \tabularnewline
\cline{2-11} 
 & \multirow{8}{*}{\textcolor{black}{\scriptsize{}SZ=91}} &  & \multicolumn{4}{c|}{\textcolor{black}{\scriptsize{}$\tau^{2}=0.006$}} & \multicolumn{4}{c}{\textcolor{black}{\scriptsize{}$\tau^{2}=0.02$}}\tabularnewline
\cline{3-11} 
 &  & \textcolor{black}{\scriptsize{}10} & \textcolor{black}{\scriptsize{}0.302 (0.015)} & \textbf{\textcolor{black}{\scriptsize{}0.308 (0.015)}} & \textcolor{black}{\scriptsize{}0.297 (0.014)} & \textcolor{black}{\scriptsize{}0.297 (0.014)} & \textcolor{black}{\scriptsize{}0.696 (0.015)} & \textbf{\textcolor{black}{\scriptsize{}0.699 (0.015)}} & \textcolor{black}{\scriptsize{}0.682 (0.015)} & \textcolor{black}{\scriptsize{}0.681 (0.015)}\tabularnewline
 &  & \textcolor{black}{\scriptsize{}20} & \textbf{\textcolor{black}{\scriptsize{}0.491 (0.016)}} & \textcolor{black}{\scriptsize{}0.49 (0.016)} & \textcolor{black}{\scriptsize{}0.441 (0.016)} & \textcolor{black}{\scriptsize{}0.44 (0.016)} & \textcolor{black}{\scriptsize{}0.932 (0.008)} & \textbf{\textcolor{black}{\scriptsize{}0.934 (0.008)}} & \textcolor{black}{\scriptsize{}0.92 (0.009)} & \textcolor{black}{\scriptsize{}0.917 (0.009)}\tabularnewline
 &  & \textcolor{black}{\scriptsize{}30} & \textbf{\textcolor{black}{\scriptsize{}0.595 (0.016)}} & \textbf{\textcolor{black}{\scriptsize{}0.595 (0.016)}} & \textcolor{black}{\scriptsize{}0.566 (0.016)} & \textcolor{black}{\scriptsize{}0.567 (0.016)} & \textcolor{black}{\scriptsize{}0.977 (0.005)} & \textbf{\textcolor{black}{\scriptsize{}0.978 (0.005)}} & \textcolor{black}{\scriptsize{}0.976 (0.005)} & \textcolor{black}{\scriptsize{}0.976 (0.005)}\tabularnewline
 &  & \textcolor{black}{\scriptsize{}50} & \textcolor{black}{\scriptsize{}0.805 (0.013)} & \textbf{\textcolor{black}{\scriptsize{}0.806 (0.013)}} & \textcolor{black}{\scriptsize{}0.763 (0.013)} & \textcolor{black}{\scriptsize{}0.764 (0.013)} &  &  &  & \tabularnewline
 &  & \textcolor{black}{\scriptsize{}100} & \textbf{\textcolor{black}{\scriptsize{}0.952 (0.007)}} & \textbf{\textcolor{black}{\scriptsize{}0.952 (0.007)}} & \textcolor{black}{\scriptsize{}0.933 (0.008)} & \textcolor{black}{\scriptsize{}0.934 (0.008)} &  &  &  & \tabularnewline
\cline{3-11} 
 &  &  & \multicolumn{4}{c|}{\textcolor{black}{\scriptsize{}$\tau^{2}=0.05$}} &  &  &  & \tabularnewline
\cline{3-11} 
 &  & \textcolor{black}{\scriptsize{}10} & \textcolor{black}{\scriptsize{}0.909 (0.009)} & \textbf{\textcolor{black}{\scriptsize{}0.917 (0.009)}} & \textcolor{black}{\scriptsize{}0.915 (0.009)} & \textcolor{black}{\scriptsize{}0.916 (0.009)} &  &  &  & \tabularnewline
\hline 
 & \multirow{11}{*}{\textcolor{black}{\scriptsize{}SZ=24}} &  & \multicolumn{4}{c|}{\textcolor{black}{\scriptsize{}$\tau^{2}=0.1$}} & \multicolumn{4}{c}{\textcolor{black}{\scriptsize{}$\tau^{2}=0.3$}}\tabularnewline
\cline{3-11} 
 &  & \textcolor{black}{\scriptsize{}10} & \textbf{\textcolor{black}{\scriptsize{}0.102 (0.01)}} & \textcolor{black}{\scriptsize{}0.101 (0.01)} & \textcolor{black}{\scriptsize{}0.099 (0.009)} & \textcolor{black}{\scriptsize{}0.07 (0.008)} & \textcolor{black}{\scriptsize{}0.254 (0.014)} & \textcolor{black}{\scriptsize{}0.252 (0.014)} & \textbf{\textcolor{black}{\scriptsize{}0.278 (0.014)}} & \textcolor{black}{\scriptsize{}0.205 (0.013)}\tabularnewline
 &  & \textcolor{black}{\scriptsize{}20} & \textcolor{black}{\scriptsize{}0.138 (0.011)} & \textcolor{black}{\scriptsize{}0.139 (0.011)} & \textbf{\textcolor{black}{\scriptsize{}0.143 (0.011)}} & \textcolor{black}{\scriptsize{}0.084 (0.009)} & \textcolor{black}{\scriptsize{}0.443 (0.016)} & \textcolor{black}{\scriptsize{}0.445 (0.016)} & \textcolor{black}{\scriptsize{}0.441 (0.016)} & \textcolor{black}{\scriptsize{}0.359 (0.015)}\tabularnewline
 &  & \textcolor{black}{\scriptsize{}30} & \textcolor{black}{\scriptsize{}0.174 (0.012)} & \textcolor{black}{\scriptsize{}0.177 (0.012)} & \textbf{\textcolor{black}{\scriptsize{}0.184 (0.012)}} & \textcolor{black}{\scriptsize{}0.108 (0.01)} & \textcolor{black}{\scriptsize{}0.549 (0.016)} & \textcolor{black}{\scriptsize{}0.55 (0.016)} & \textbf{\textcolor{black}{\scriptsize{}0.574 (0.016)}} & \textcolor{black}{\scriptsize{}0.431 (0.016)}\tabularnewline
 &  & \textcolor{black}{\scriptsize{}50} & \textcolor{black}{\scriptsize{}0.221 (0.013)} & \textcolor{black}{\scriptsize{}0.22 (0.013)} & \textbf{\textcolor{black}{\scriptsize{}0.25 (0.014)}} & \textcolor{black}{\scriptsize{}0.14 (0.011)} & \textcolor{black}{\scriptsize{}0.71 (0.014)} & \textcolor{black}{\scriptsize{}0.711 (0.014)} & \textbf{\textcolor{black}{\scriptsize{}0.752 (0.014)}} & \textcolor{black}{\scriptsize{}0.609 (0.015)}\tabularnewline
 &  & \textcolor{black}{\scriptsize{}100} & \textcolor{black}{\scriptsize{}0.36 (0.015)} & \textcolor{black}{\scriptsize{}0.359 (0.015)} & \textbf{\textcolor{black}{\scriptsize{}0.411 (0.016)}} & \textcolor{black}{\scriptsize{}0.218 (0.013)} & \textbf{\textcolor{black}{\scriptsize{}0.902 (0.009)}} & \textbf{\textcolor{black}{\scriptsize{}0.902 (0.009)}} & \textcolor{black}{\scriptsize{}0.893 (0.01)} & \textcolor{black}{\scriptsize{}0.79 (0.013)}\tabularnewline
\cline{3-11} 
 &  &  & \multicolumn{4}{c|}{\textcolor{black}{\scriptsize{}$\tau^{2}=0.9$}} & \multicolumn{4}{c}{}\tabularnewline
\cline{3-11} 
\textcolor{black}{\scriptsize{}Natural-} &  & \textcolor{black}{\scriptsize{}10} & \textcolor{black}{\scriptsize{}0.607 (0.015)} & \textcolor{black}{\scriptsize{}0.619 (0.015)} & \textbf{\textcolor{black}{\scriptsize{}0.64 (0.015)}} & \textcolor{black}{\scriptsize{}0.566 (0.016)} &  &  &  & \tabularnewline
\textcolor{black}{\scriptsize{}logarithm- } &  & \textcolor{black}{\scriptsize{}20} & \textcolor{black}{\scriptsize{}0.851 (0.011)} & \textcolor{black}{\scriptsize{}0.854 (0.011)} & \textbf{\textcolor{black}{\scriptsize{}0.881 (0.01)}} & \textcolor{black}{\scriptsize{}0.826 (0.012)} &  &  &  & \tabularnewline
\textcolor{black}{\scriptsize{}transformed} &  & \textcolor{black}{\scriptsize{}30} & \textcolor{black}{\scriptsize{}0.923 (0.008)} & \textcolor{black}{\scriptsize{}0.924 (0.008)} & \textbf{\textcolor{black}{\scriptsize{}0.947 (0.007)}} & \textcolor{black}{\scriptsize{}0.909 (0.009)} &  &  &  & \tabularnewline
\textcolor{black}{\scriptsize{}odds } &  & \textcolor{black}{\scriptsize{}50} & \textcolor{black}{\scriptsize{}0.987 (0.004)} & \textcolor{black}{\scriptsize{}0.988 (0.003)} & \textbf{\textcolor{black}{\scriptsize{}0.989 (0.003)}} & \textcolor{black}{\scriptsize{}0.974 (0.005)} &  &  &  & \tabularnewline
\cline{2-11} 
 & \multirow{9}{*}{\textcolor{black}{\scriptsize{}SZ=91}} &  & \multicolumn{4}{c|}{\textcolor{black}{\scriptsize{}$\tau^{2}=0.03$}} & \multicolumn{4}{c}{\textcolor{black}{\scriptsize{}$\tau^{2}=0.1$}}\tabularnewline
\cline{3-11} 
 &  & \textcolor{black}{\scriptsize{}10} & \textbf{\textcolor{black}{\scriptsize{}0.16 (0.012)}} & \textcolor{black}{\scriptsize{}0.157 (0.012)} & \textcolor{black}{\scriptsize{}0.144 (0.011)} & \textcolor{black}{\scriptsize{}0.119 (0.01)} & \textcolor{black}{\scriptsize{}0.412 (0.016)} & \textbf{\textcolor{black}{\scriptsize{}0.416 (0.016)}} & \textcolor{black}{\scriptsize{}0.392 (0.015)} & \textcolor{black}{\scriptsize{}0.372 (0.015)}\tabularnewline
 &  & \textcolor{black}{\scriptsize{}20} & \textbf{\textcolor{black}{\scriptsize{}0.228 (0.013)}} & \textcolor{black}{\scriptsize{}0.224 (0.013)} & \textcolor{black}{\scriptsize{}0.215 (0.013)} & \textcolor{black}{\scriptsize{}0.182 (0.012)} & \textbf{\textcolor{black}{\scriptsize{}0.652 (0.015)}} & \textcolor{black}{\scriptsize{}0.651 (0.015)} & \textcolor{black}{\scriptsize{}0.627 (0.015)} & \textcolor{black}{\scriptsize{}0.587 (0.016)}\tabularnewline
 &  & \textcolor{black}{\scriptsize{}30} & \textbf{\textcolor{black}{\scriptsize{}0.314 (0.015)}} & \textbf{\textcolor{black}{\scriptsize{}0.314 (0.015)}} & \textcolor{black}{\scriptsize{}0.297 (0.014)} & \textcolor{black}{\scriptsize{}0.245 (0.014)} & \textcolor{black}{\scriptsize{}0.767 (0.013)} & \textbf{\textcolor{black}{\scriptsize{}0.77 (0.013)}} & \textcolor{black}{\scriptsize{}0.744 (0.014)} & \textcolor{black}{\scriptsize{}0.703 (0.014)}\tabularnewline
 &  & \textcolor{black}{\scriptsize{}50} & \textbf{\textcolor{black}{\scriptsize{}0.42 (0.016)}} & \textbf{\textcolor{black}{\scriptsize{}0.42 (0.016)}} & \textcolor{black}{\scriptsize{}0.388 (0.015)} & \textcolor{black}{\scriptsize{}0.333 (0.015)} & \textcolor{black}{\scriptsize{}0.922 (0.008)} & \textbf{\textcolor{black}{\scriptsize{}0.923 (0.008)}} & \textcolor{black}{\scriptsize{}0.913 (0.009)} & \textcolor{black}{\scriptsize{}0.885 (0.01)}\tabularnewline
 &  & \textcolor{black}{\scriptsize{}100} & \textcolor{black}{\scriptsize{}0.622 (0.015)} & \textbf{\textcolor{black}{\scriptsize{}0.623 (0.015)}} & \textcolor{black}{\scriptsize{}0.574 (0.016)} & \textcolor{black}{\scriptsize{}0.494 (0.016)} &  &  &  & \tabularnewline
\cline{3-11} 
 &  &  & \multicolumn{4}{c|}{\textcolor{black}{\scriptsize{}$\tau^{2}=0.3$}} & \multicolumn{4}{c}{}\tabularnewline
\cline{3-11} 
 &  & \textcolor{black}{\scriptsize{}10} & \textcolor{black}{\scriptsize{}0.789 (0.013)} & \textbf{\textcolor{black}{\scriptsize{}0.795 (0.013)}} & \textcolor{black}{\scriptsize{}0.809 (0.012)} & \textcolor{black}{\scriptsize{}0.784 (0.013)} &  &  &  & \tabularnewline
 &  & \textcolor{black}{\scriptsize{}20} & \textcolor{black}{\scriptsize{}0.955 (0.007)} & \textbf{\textcolor{black}{\scriptsize{}0.957 (0.006)}} & \textcolor{black}{\scriptsize{}0.967 (0.006)} & \textcolor{black}{\scriptsize{}0.954 (0.007)} &  &  &  & \tabularnewline
\hline 
\end{tabular}{\scriptsize\par}

}

{\scriptsize{}Note: Monte Carlo standard errors are presented in the
parentheses. SZ is the average per-group sample sizes. The largest
power values under each condition are highlighted as bold. ML or REML
estimation sometimes did not have converged results, but the nonconvergence
rates across all conditions were within 0.4\%. }{\scriptsize\par}
\end{table}
{\scriptsize\par}

\subsubsection{\textcolor{black}{5.3.3 Statistical power in meta-analysis for heterogeneity
magnitude test}}

\textcolor{black}{$\tau^{2}$ under the null hypothesis was specified
as the small level of heterogeneity in Table 2 and the true $\tau^{2}$
which generated data was specified as the medium level of heterogeneity.
We explore the power of the proposed methods in testing whether the
heterogeneity is larger than a specific level ($\tau^{2}=\lambda$
verses $\tau^{2}>\lambda$) in this section. The power values are
presented in Table 9 with the highest power values under each condition
bolded and the Monte Carlo standard errors inside the parentheses.
Based on Monte Carlo standard errors, different bootstrap methods
did not differ noticeably.}

\textcolor{black}{{[}Table 9{]}}{\scriptsize{}}
\begin{table}
{\scriptsize{}\caption{{\footnotesize{}Statistical power of the bootstrap based ML LR test
(Q-ML-LRT), the bootstrap based REML LR test (Q-RE}\textcolor{black}{\footnotesize{}ML-LRT),
and the bootstrap based Q test (B-Q) in }{\footnotesize{}meta-analyses
for heterogeneity magnitude test}}
}{\scriptsize\par}

\resizebox{\textwidth}{!}{
\renewcommand{\arraystretch}{0.45}

{\scriptsize{}}%
\begin{tabular}{cccccc}
\hline 
 &  & \textcolor{black}{\scriptsize{}Number of} & \multirow{2}{*}{\textcolor{black}{\scriptsize{}B-ML-LRT}} & \multirow{2}{*}{\textcolor{black}{\scriptsize{}B-REML-LRT}} & \multirow{2}{*}{\textcolor{black}{\scriptsize{}B-Q}}\tabularnewline
 &  & \textcolor{black}{\scriptsize{}Studies} &  &  & \tabularnewline
\hline 
 & \multirow{6}{*}{\textcolor{black}{\scriptsize{}SZ=24}} &  & \multicolumn{3}{c}{\textcolor{black}{\scriptsize{}$\lambda=0.03$, $\tau^{2}=0.1$}}\tabularnewline
\cline{3-6} 
 &  & \textcolor{black}{\scriptsize{}10} & \textcolor{black}{\scriptsize{}0.284 (0.014)} & \textcolor{black}{\scriptsize{}0.285 (0.014)} & \textbf{\textcolor{black}{\scriptsize{}0.286 (0.014)}}\tabularnewline
 &  & \textcolor{black}{\scriptsize{}20} & \textcolor{black}{\scriptsize{}0.476 (0.016)} & \textbf{\textcolor{black}{\scriptsize{}0.477 (0.016)}} & \textcolor{black}{\scriptsize{}0.471 (0.016)}\tabularnewline
 &  & \textcolor{black}{\scriptsize{}30} & \textbf{\textcolor{black}{\scriptsize{}0.566 (0.016)}} & \textcolor{black}{\scriptsize{}0.565 (0.016)} & \textcolor{black}{\scriptsize{}0.565 (0.016)}\tabularnewline
 &  & \textcolor{black}{\scriptsize{}50} & \textcolor{black}{\scriptsize{}0.75 (0.014)} & \textcolor{black}{\scriptsize{}0.751 (0.014)} & \textbf{\textcolor{black}{\scriptsize{}0.756 (0.014)}}\tabularnewline
\textcolor{black}{\scriptsize{}Standardized } &  & \textcolor{black}{\scriptsize{}100} & \textbf{\textcolor{black}{\scriptsize{}0.94 (0.008)}} & \textbf{\textcolor{black}{\scriptsize{}0.94 (0.008)}} & \textbf{\textcolor{black}{\scriptsize{}0.94 (0.008)}}\tabularnewline
\cline{2-6} 
\textcolor{black}{\scriptsize{}mean } & \multirow{6}{*}{\textcolor{black}{\scriptsize{}SZ=91}} &  & \multicolumn{3}{c}{\textcolor{black}{\scriptsize{}$\lambda=0.006$, $\tau^{2}=0.02$}}\tabularnewline
\cline{3-6} 
\textcolor{black}{\scriptsize{}difference} &  & \textcolor{black}{\scriptsize{}10} & \textcolor{black}{\scriptsize{}0.243 (0.014)} & \textbf{\textcolor{black}{\scriptsize{}0.247 (0.014)}} & \textcolor{black}{\scriptsize{}0.242 (0.014)}\tabularnewline
 &  & \textcolor{black}{\scriptsize{}20} & \textbf{\textcolor{black}{\scriptsize{}0.418 (0.016)}} & \textbf{\textcolor{black}{\scriptsize{}0.418 (0.016)}} & \textcolor{black}{\scriptsize{}0.409 (0.016)}\tabularnewline
 &  & \textcolor{black}{\scriptsize{}30} & \textbf{\textcolor{black}{\scriptsize{}0.528 (0.016)}} & \textbf{\textcolor{black}{\scriptsize{}0.528 (0.016)}} & \textcolor{black}{\scriptsize{}0.505 (0.016)}\tabularnewline
 &  & \textcolor{black}{\scriptsize{}50} & \textcolor{black}{\scriptsize{}0.702 (0.014)} & \textbf{\textcolor{black}{\scriptsize{}0.704 (0.014)}} & \textcolor{black}{\scriptsize{}0.698 (0.015)}\tabularnewline
 &  & \textcolor{black}{\scriptsize{}100} & \textbf{\textcolor{black}{\scriptsize{}0.911 (0.009)}} & \textbf{\textcolor{black}{\scriptsize{}0.911 (0.009)}} & \textcolor{black}{\scriptsize{}0.902 (0.009)}\tabularnewline
\hline 
 & \multirow{6}{*}{\textcolor{black}{\scriptsize{}SZ=24}} &  & \multicolumn{3}{c}{\textcolor{black}{\scriptsize{}$\lambda=0.1$, $\tau^{2}=0.3$}}\tabularnewline
\cline{3-6} 
 &  & \textcolor{black}{\scriptsize{}10} & \textcolor{black}{\scriptsize{}0.211 (0.013)} & \textcolor{black}{\scriptsize{}0.213 (0.013)} & \textbf{\textcolor{black}{\scriptsize{}0.215 (0.013)}}\tabularnewline
 &  & \textcolor{black}{\scriptsize{}20} & \textcolor{black}{\scriptsize{}0.314 (0.015)} & \textbf{\textcolor{black}{\scriptsize{}0.315 (0.015)}} & \textcolor{black}{\scriptsize{}0.301 (0.015)}\tabularnewline
 &  & \textcolor{black}{\scriptsize{}30} & \textbf{\textcolor{black}{\scriptsize{}0.425 (0.016)}} & \textcolor{black}{\scriptsize{}0.422 (0.016)} & \textcolor{black}{\scriptsize{}0.414 (0.016)}\tabularnewline
\textcolor{black}{\scriptsize{}Fisher-} &  & \textcolor{black}{\scriptsize{}50} & \textbf{\textcolor{black}{\scriptsize{}0.535 (0.016)}} & \textbf{\textcolor{black}{\scriptsize{}0.535 (0.016)}} & \textcolor{black}{\scriptsize{}0.513 (0.016)}\tabularnewline
\textcolor{black}{\scriptsize{}transformed } &  & \textcolor{black}{\scriptsize{}100} & \textbf{\textcolor{black}{\scriptsize{}0.815 (0.012)}} & \textbf{\textcolor{black}{\scriptsize{}0.815 (0.012)}} & \textcolor{black}{\scriptsize{}0.807 (0.012)}\tabularnewline
\cline{2-6} 
\textcolor{black}{\scriptsize{}Pearson} & \multirow{5}{*}{\textcolor{black}{\scriptsize{}SZ=91}} &  & \multicolumn{3}{c}{\textcolor{black}{\scriptsize{}$\lambda=0.03$, $\tau^{2}=0.1$}}\tabularnewline
\cline{3-6} 
\textcolor{black}{\scriptsize{}correlation} &  & \textcolor{black}{\scriptsize{}10} & \textbf{\textcolor{black}{\scriptsize{}0.391 (0.015)}} & \textbf{\textcolor{black}{\scriptsize{}0.391 (0.015)}} & \textcolor{black}{\scriptsize{}0.388 (0.015)}\tabularnewline
 &  & \textcolor{black}{\scriptsize{}20} & \textcolor{black}{\scriptsize{}0.613 (0.015)} & \textbf{\textcolor{black}{\scriptsize{}0.617 (0.015)}} & \textcolor{black}{\scriptsize{}0.609 (0.015)}\tabularnewline
 &  & \textcolor{black}{\scriptsize{}30} & \textcolor{black}{\scriptsize{}0.753 (0.014)} & \textbf{\textcolor{black}{\scriptsize{}0.754 (0.014)}} & \textcolor{black}{\scriptsize{}0.751 (0.014)}\tabularnewline
 &  & \textcolor{black}{\scriptsize{}50} & \textcolor{black}{\scriptsize{}0.909 (0.009)} & \textbf{\textcolor{black}{\scriptsize{}0.91 (0.009)}} & \textcolor{black}{\scriptsize{}0.907 (0.009)}\tabularnewline
\hline 
 & \multirow{6}{*}{\textcolor{black}{\scriptsize{}SZ=24}} &  & \multicolumn{3}{c}{\textcolor{black}{\scriptsize{}$\lambda=0.01$, $\tau^{2}=0.03$}}\tabularnewline
\cline{3-6} 
 &  & \textcolor{black}{\scriptsize{}10} & \textcolor{black}{\scriptsize{}0.157 (0.012)} & \textcolor{black}{\scriptsize{}0.158 (0.012)} & \textbf{\textcolor{black}{\scriptsize{}0.175 (0.012)}}\tabularnewline
 &  & \textcolor{black}{\scriptsize{}20} & \textcolor{black}{\scriptsize{}0.202 (0.013)} & \textcolor{black}{\scriptsize{}0.203 (0.013)} & \textbf{\textcolor{black}{\scriptsize{}0.236 (0.013)}}\tabularnewline
 &  & \textcolor{black}{\scriptsize{}30} & \textcolor{black}{\scriptsize{}0.286 (0.014)} & \textcolor{black}{\scriptsize{}0.286 (0.014)} & \textbf{\textcolor{black}{\scriptsize{}0.326 (0.015)}}\tabularnewline
\textcolor{black}{\scriptsize{}Natural-} &  & \textcolor{black}{\scriptsize{}50} & \textcolor{black}{\scriptsize{}0.394 (0.015)} & \textcolor{black}{\scriptsize{}0.391 (0.015)} & \textbf{\textcolor{black}{\scriptsize{}0.468 (0.016)}}\tabularnewline
\textcolor{black}{\scriptsize{}logarithm- } &  & \textcolor{black}{\scriptsize{}100} & \textcolor{black}{\scriptsize{}0.569 (0.016)} & \textcolor{black}{\scriptsize{}0.568 (0.016)} & \textbf{\textcolor{black}{\scriptsize{}0.648 (0.015)}}\tabularnewline
\cline{2-6} 
\textcolor{black}{\scriptsize{}transformed} & \multirow{6}{*}{\textcolor{black}{\scriptsize{}SZ=91}} &  & \multicolumn{3}{c}{\textcolor{black}{\scriptsize{}$\lambda=0.006$, $\tau^{2}=0.02$}}\tabularnewline
\cline{3-6} 
\textcolor{black}{\scriptsize{}odds } &  & \textcolor{black}{\scriptsize{}10} & \textcolor{black}{\scriptsize{}0.286 (0.014)} & \textcolor{black}{\scriptsize{}0.285 (0.014)} & \textbf{\textcolor{black}{\scriptsize{}0.291 (0.014)}}\tabularnewline
 &  & \textcolor{black}{\scriptsize{}20} & \textcolor{black}{\scriptsize{}0.389 (0.015)} & \textcolor{black}{\scriptsize{}0.388 (0.015)} & \textbf{\textcolor{black}{\scriptsize{}0.395 (0.015)}}\tabularnewline
 &  & \textcolor{black}{\scriptsize{}30} & \textcolor{black}{\scriptsize{}0.519 (0.016)} & \textcolor{black}{\scriptsize{}0.516 (0.016)} & \textbf{\textcolor{black}{\scriptsize{}0.526 (0.016)}}\tabularnewline
 &  & \textcolor{black}{\scriptsize{}50} & \textcolor{black}{\scriptsize{}0.668 (0.015)} & \textcolor{black}{\scriptsize{}0.667 (0.015)} & \textbf{\textcolor{black}{\scriptsize{}0.67 (0.015)}}\tabularnewline
 &  & \textcolor{black}{\scriptsize{}100} & \textcolor{black}{\scriptsize{}0.877 (0.01)} & \textcolor{black}{\scriptsize{}0.877 (0.01)} & \textbf{\textcolor{black}{\scriptsize{}0.889 (0.01)}}\tabularnewline
\hline 
\end{tabular}{\scriptsize\par}

}

{\scriptsize{}Note: Monte Carlo standard errors are presented in the
parentheses. $\lambda$ indicates the heterogeneity level under the
null hypothesis. $\tau^{2}$ is the true data generating heterogeneity
level. SZ is the average per-group sample sizes. The largest power
values under each condition are highlighted as bold. ML or REML estimation
sometimes did not have converged results, but the nonconvergence rates
across all conditions were within 0.4\%. }{\scriptsize\par}
\end{table}
{\scriptsize\par}

\subsubsection{5.3.3 Summary of the performance of the tests on statistical power}

In sum, regardless of the method and type of effect\textcolor{black}{{}
size, more studies, larger study-level sample sizes, and/or larger
$\tau^{2}$ provided higher power. The bootstrap based methods boosted
power compared to the regular Q test. The increments of power were
the highest when the log odds ratio was of interest. The improved
Q test demonstrated the highest power in a number of conditions when
it was applicable. }

\section{\textcolor{black}{6. Empirical Illustration}}

\textcolor{black}{We applied the proposed bootstrap based heterogeneity
test in three real meta-analyses. Specifically, we considered the
bootstrap based REML LR test (B-REML-LRT) and bootstrap based Q test
(B-Q) which generally appropriately controlled Type I error rates.
Datasets used in the three examples along with the R codes are provided
in the supplemental materials. The first meta-analysis consists of
13 studies that studied the correlation between sensation seeking
scores and levels of monoamine oxidase \citep{zuckerman1994}. The
sample sizes range from 10 to 125 with a median of 40. In this example,
the REML estimate of the overall Fisher-transformed Pearson correlation
is -0.26 and the REML estimate of the between-study heterogeneity
is 0.03. The Q test result is $Q(df=12)=29.06,\;p=0.004$. Similarly,
B-REML-LRT and B-Q also reject the assumption of homogeneity with
$p=0.004$ and 0.002 respectively when $\alpha=0.05$. }

\textcolor{black}{The second meta-analysis consists of 18 studies
in which the effect of open versus traditional education on students'
self-concept was studied \citep{Hedges1981}. The per-group sample
sizes range from 13.5 to 140 with a median of 64.5. In this example,
the REML estimate of the overall standardized mean difference is 0.01
and the REML estimate of the between-study heterogeneity is 0.02.
The Q test result is $Q(df=17)=23.39,\;p=0.137$, and thus the test
fails to reject the assumption of homogeneity when $\alpha=0.05$
or $0.1$. B-REML-LRT and B-Q fail to reject the assumption of homogeneity
when $\alpha=0.05$ with $p=0.053$. The improved Q test is not applicable
because the sample mean and sample variance information is not available
in this meta-analysis.}

\textcolor{black}{The third meta-analysis consists of 26 studies on
nicotine replacement therapy (nicotine chewing gum) for smoking cessation
\citep{silagy2003nicotine}. The numbers of participants in the control
group and treatment group (i.e., nicotine replacement therapy) who
stopped smoking for at least 6 months after treatment were recorded.
The total sample sizes for the control and treatment groups range
from 47 to 1217 with a median of 175. The REML estimate of the overall
log odds ratio is 0.56 and the REML estimate of the between-study
heterogeneity is 0.05. The Q test result is $Q(df=25)=34.87,\;p=0.091$,
and thus the test fails to reject the assumption of homogeneity when
$\alpha=0.05$. The improved Q test and B-Q also fail to reject the
assumption of homogeneity with $p=0.073$ and 0.088, respectively.
In contrast, B-REML-LRT rejects the assumption of homogeneity with
$p=0.037$. This real data analysis result is consistent with the
simulation results: when the effect size was the log odds ratio and
$SZ=91$, B-REML-LRT was the most powerful method.}

\section{\textcolor{black}{7. Discussion}}

\textcolor{black}{Testing the between-study heterogeneity is the research
of interest in some studies and widely used in practice. We considered
two types of between-study heterogeneity tests. In the heterogeneity
test, we test $\tau^{2}=0$ versus $\tau^{2}\neq0$. In the heterogeneity
magnitude test, we test $\tau^{2}=\lambda$ versus $\tau^{2}>\lambda$.
The heterogeneity test can be viewed as the a special case of heterogeneity
magnitude test with $\lambda=0$. Researchers may be interested in
the heterogeneity test and the heterogeneity magnitude test for three
reasons. First, with a large between-study heterogeneity, interpreting
study-specific effect size is more practically meaningful than the
overall effect size. Second, researchers may also be interested in
exploring factors that can explain the systematic between-study heterogeneity.
The relevant test can be viewed as the goodness of fit test. If the
unexplained heterogeneity in mixed-effects model is significantly,
we can consider interactive combinations of moderators by creating
product terms. Third, in replication research, the heterogeneity test
can be used in exact replication test and the heterogeneity magnitude
test can be used in approximate replication test. For both heterogeneity
test and heterogeneity magnitude test, failure of controlling Type
I error rates and low power will lead to misleading testing results.
Hence, the goal of the current student is to propose a powerful and
reliable test for both hypothesis testings.}

\textcolor{black}{The existing methods, such as the Q test \citep{hedges1985statistical,hedges1983random}
and likelihood ratio test \citep{viechtbauer2007hypothesis}, are
criticized by their failure to control the Type I error rate and/or
failure to attain enough statistical power. Kulinskaya\textquoteright s
improved Q test, as a more recently proposed test could better control
Type I error rates and improve power, but so far it is limited to
standardized mean differences, risk differences, and log odds ratios.
Additionally, few studies have examined the performance of the heterogeneity
test in the mixed-effect meta-analysis and the R functions for the
improved Q test cannot accommodate covariates yet. To the best of
our knowledge, no existing methods can handle heterogeneity magnitude
test so far. Consistent with the findings from \citet{harwell1997empirical},
\citet{hedges1985statistical}, and \citet{viechtbauer2007hypothesis},
we found that the regular ML based LR test (ML-LRT), the regular REML
based LR test (REML-LRT), and the Q test could be too conservative
with the examined types of effect sizes. We have summarized the strengths
and limitations of the existing methods in Table 1.}

\textcolor{black}{We propose a bootstrap based heterogeneity test
combining the likelihood ratio test (ML- and REML based) and the Q
test with parametric bootstrap procedures. The proposed method has
several advantages over the existing approaches. First, the proposed
bootstrap based heterogeneity test can be applied to heterogeneity
magnitude test while other methods cannot. Second, based on our simulation
results, B-REML-LRT and B-Q outperformed ML-LRT, REML-LRT, and the
Q test in terms of controlling the Type I error rate. B-REML-LRT and
B-Q appropriately controlled the Type I error rate and were competitive
with the improved Q test, except a few cases (B-REML-LRT was too conservative
when the study-level sample sizes were small and the effect size was
the log odds ratio). }

\textcolor{black}{Third, the proposed bootstrap based heterogeneity
test increased statistical power compared to the Q test, especially
when the study-level sample sizes were small. The increment in power
was the largest with the log odds ratio. As a parametric method, the
improved Q test had the highest power in a number of conditions when
it was applicable. We also found that for both the bootstrap methods
and Q test, more studies, larger study-level sample sizes, and/or
larger $\tau^{2}$ provided higher power.}

\textcolor{black}{Another strength of the proposed method is its flexibility.
Unlike the improved Q test which requests a sophisticated analytical
form for each type of effect size and the information of raw data,
the bootstrap based heterogeneity test can be easily extended to cases
where effect sizes are not the types of effect size discussed in the
current paper, such as standardized mean change. The only place that
would need modification is the analytical form of the sampling distribution.
We have summarized the strengths and limitations of the proposed bootstrap
methods in Table 1.}

\textcolor{black}{Overall, based on both the Type I error rate and
statistical power, we recommend B-Q for the heterogeneity test and
heterogeneity magnitude test. If one wants to pursue higher power,
B-REML-LRT is also recommended as long as study-level sample sizes
are not too small. One also can consider the improved Q test when
it is applicable. Future work could look into developing the bootstrap
based heterogeneity test for multivariate meta-analysis. The generalization
is feasible when the analytical form of the sampling distribution
of multivariate meta-analysis is given. Furthermore, published studies
may not be truly representative of all valid studies undertaken on
a research topic which is known as publication bias or the file-drawer
problem \citep{hunter2004methods,iyengar1988selection,rosenthal1979file}.
Conclusions based on the meta-analyses of published studies without
correcting for publication bias compromise the validity of a systematic
review \citep{du2017bayesian,klein2018many,kvarven2020comparing,open2015estimating,sutton2000modelling}.
Future research should look into developing the bootstrap based heterogeneity
test while reducing the bias caused by the file-drawer problem.}

\bibliographystyle{apalike}
\bibliography{test_of_random_effects}

\end{document}